%% file: Main.tex
\documentclass[12pt,a4paper,twoside,openright]{report}
\input{include/settings/Settings}

\begin{document} 

% COVER PAGE, TITLE PAGE AND IMPRINT PAGE
\pagenumbering{roman}			% Roman numbering (starting with i (one)) until first main chapter
\input{include/frontmatter/Titlepage}

% ABSTRACT
\newpage
\input{include/frontmatter/Abstract}

% ACKNOWLEDGEMENTS
\newpage
\input{include/frontmatter/Acknowledgements}

% TABLE OF CONTENTS
\newpage
\tableofcontents

% OTHER FRONTMATTER
% List of figures (add to table of contents)
\cleardoublepage
\addcontentsline{toc}{chapter}{\listfigurename} 
\listoffigures
% List of tables (add to table of contents)
\cleardoublepage
\addcontentsline{toc}{chapter}{\listtablename}  
\listoftables

% START OF MAIN DOCUMENT
\cleardoublepage
\setcounter{page}{1}
\pagenumbering{arabic}			% Arabic numbering starting from 1 (one)
\setlength{\parskip}{0pt plus 1pt}

% INTRODUCTION
\input{include/Introduction}

% BACKGROUND
\input{include/Background}

% RELATED_WORK
\input{include/RelatedWork}

% METHODS
\input{include/Methods}

% RESULTS
\input{include/Results}

% DISCUSSION
\input{include/Discussion}

% CONCLUSION
\input{include/Conclusion}

% REFERENCES / BIBLIOGRAPHY
\cleardoublepage

% BIBLIOGRAPHY
\phantomsection
\addcontentsline{toc}{chapter}{Bibliography}
\bibliographystyle{ieeetr}
\bibliography{Main}

% APPENDICES
\let
\cleardoublepage
\clearpage
\appendix
\setcounter{page}{1}
\pagenumbering{Roman}			% Capitalized roman numbering starting from I (one)
\titleformat{\chapter}[display]{\bfseries\filcenter}{\huge\appendixname~\thechapter}{2ex}{\LARGE}
\begin{appendices}
    \input{include/backmatter/Appendix_1}
    \input{include/backmatter/Appendix_2}
    \input{include/backmatter/Appendix_3}
    \input{include/backmatter/Appendix_10}

    \input{include/backmatter/Appendix_4}
    \input{include/backmatter/Appendix_5}
    \input{include/backmatter/Appendix_6}
    \input{include/backmatter/Appendix_7}
    \input{include/backmatter/Appendix_8}
    \input{include/backmatter/Appendix_9}
    \input{include/backmatter/Appendix_11}
\end{appendices}

\end{document}

%% file: include/settings/Settings.tex
\usepackage{geometry}
\usepackage{moreverb}								% List settings
\usepackage{textcomp}								% Fonts, symbols etc.
\usepackage{lmodern}								% Latin modern font
\usepackage{helvet}	
\usepackage[T1]{fontenc}							% Output settings
\usepackage[english]{babel}							% Language settings
\usepackage[utf8]{inputenc}							% Input settings
\usepackage{amsmath}								% Mathematical expressions (American mathematical society)
\usepackage{amssymb}								% Mathematical symbols (American mathematical society)
\usepackage{graphicx}								% Figures
\usepackage{subfig}									% Enables subfigures
\numberwithin{equation}{chapter}					% Numbering order for equations
\numberwithin{figure}{chapter}						% Numbering order for figures
\numberwithin{table}{chapter}						% Numbering order for tables
\usepackage{eso-pic}								% Create cover page background
\newcommand{\backgroundpic}[3]{
	\put(#1,#2){
	\parbox[b][\paperheight]{\paperwidth}{
	\centering
	\includegraphics[width=\paperwidth,height=\paperheight,keepaspectratio]{#3}}}}
\usepackage{float} 									% Enables object position enforcement using [H]
\usepackage{parskip}								% Enables vertical spaces correctly 
\usepackage{array}
\usepackage{rotating}
\usepackage{makecell}
\usepackage{pgfplots}
\usepackage[titletoc]{appendix}

\usepackage[labelfont=bf, textfont=normal,
			justification=centering,
			singlelinecheck=false]{caption}

\usepackage{hyperref}								
\hypersetup{colorlinks, citecolor=black,
   		 	filecolor=black, linkcolor=black,
    		urlcolor=black}

\usepackage{titlesec}		
\titleformat{\chapter}[display]
  {\Huge\bfseries\filcenter}
  {{\fontsize{50pt}{1em}\vspace{-4.2ex}\selectfont \textnormal{\thechapter}}}{1ex}{}[]

\usepackage{fancyhdr}								
\def\layout{2}	% Choose 1 for one-sided or 2 for two-sided layout
\ifnum\layout=2	% Two-sided
	\fancypagestyle{plain}{			% Redefine the plain page style
	\fancyhf{}
	 		
	\fancyfoot[LE,RO]{\thepage}}	
\else			% One-sided  	
\fi

\usepackage[textsize=tiny]{todonotes}   % Include the option "disable" to hide all notes
\usepackage{verbatim} 
\usepackage[framemethod=TikZ]{mdframed}
\mdfdefinestyle{MyFrame}{%
    linecolor=black,
    outerlinewidth=0.5pt,
    roundcorner=0.5pt,
    innertopmargin=\baselineskip,
    innerbottommargin=\baselineskip,
    innerrightmargin=20pt,
    innerleftmargin=20pt,
    backgroundcolor=white}

%% file: include/frontmatter/Titlepage.tex
% CREATED BY DAVID FRISK, 2016
% MODIFIED BY JAKOB JARMAR, 2016
% A few changes by Birgit Grohe, 2017 and 2018

% COVER PAGE
\begin{titlepage}
\newgeometry{top=3cm, bottom=3cm,
			left=2.25 cm, right=2.25cm}	% Temporarily change margins		
			
% Cover page background 
\AddToShipoutPicture*{\backgroundpic{-4}{56.7}{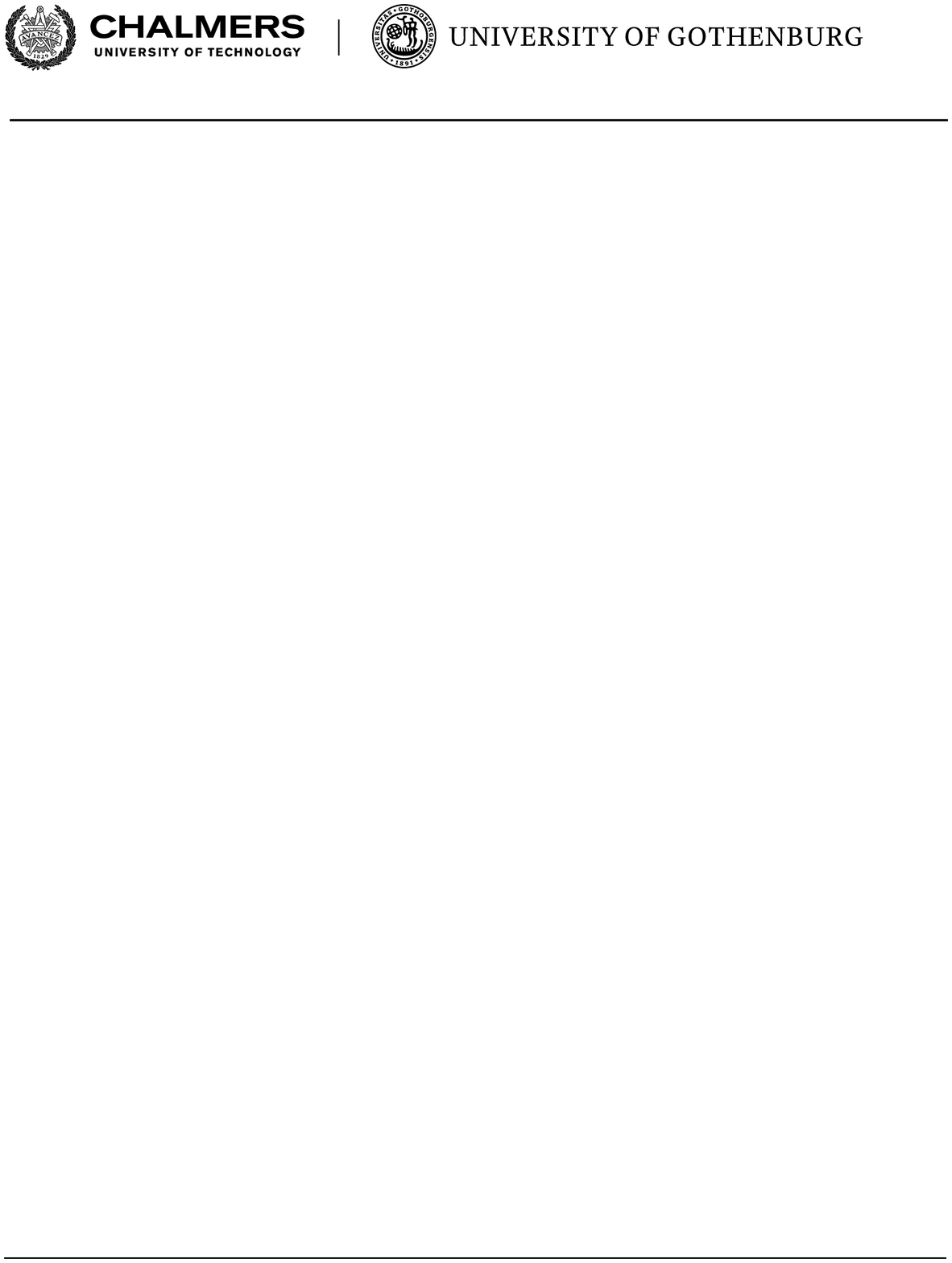}}
\addtolength{\voffset}{2cm}

% Cover picture (replace with your own or delete)		
\begin{figure}[H]
\centering
\vspace{1cm}	% Adjust vertical spacing here
\includegraphics[width=0.9\linewidth]{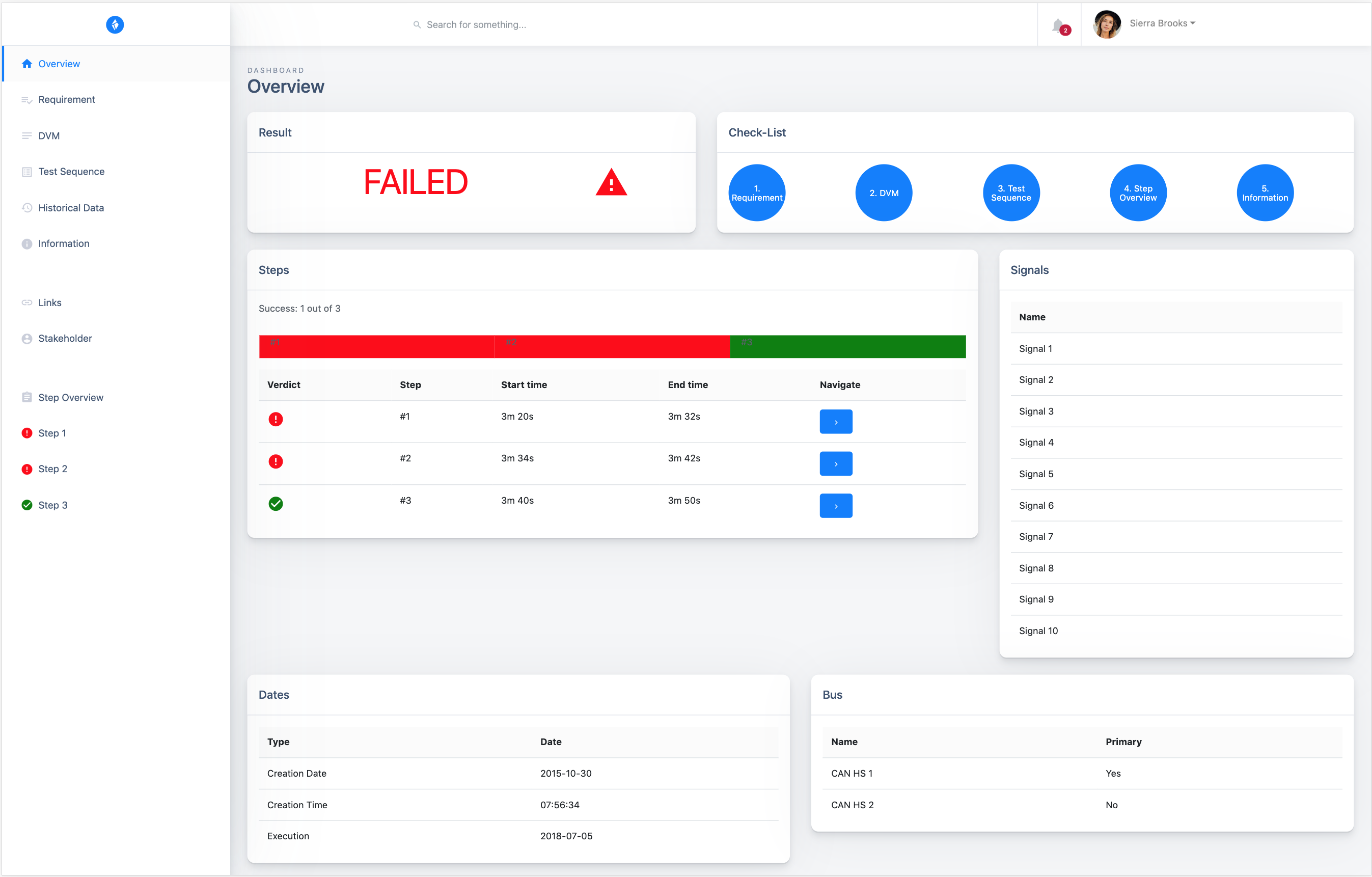}
\end{figure}

% Cover text
\mbox{}
\vfill
\renewcommand{\familydefault}{\sfdefault} \normalfont % Set cover page font
\textbf{{\Huge 	Structuring and presenting data for testing of automotive electronics to reduce effort during decision making }} 	\\[0.5cm]
Master's thesis in Computer Science and Engineering \setlength{\parskip}{1cm}

{\Large MARTIN TRAN} \setlength{\parskip}{2.9cm} \\
{\Large ASHISH M HUSAIN} \setlength{\parskip}{2.9cm}

Department of Computer Science and Engineering \\
\textsc{Chalmers University of Technology} \\
\textsc{University of Gothenburg} \\
Gothenburg, Sweden 2020

\renewcommand{\familydefault}{\rmdefault} \normalfont % Reset standard font
\end{titlepage}

% BACK OF COVER PAGE (BLANK PAGE)
\newpage
\restoregeometry
\thispagestyle{empty}
\mbox{}

% TITLE PAGE
\newpage
\thispagestyle{empty}
\begin{center}
	\textsc{\large Master's thesis 2020}\\[4cm]		% Report number is currently not in use
	\textbf{\Large Structuring and presenting data for testing of automotive electronics to reduce effort during decision making } \\[1cm]

	{\large MARTIN TRAN}\\
    {\large ASHISH M HUSAIN}
	
	\vfill	
	% Logotype on titlepage	
	\begin{figure}[H]
	\centering
	% Remove the following line to remove the titlepage logotype
	\includegraphics[width=0.25\pdfpagewidth]{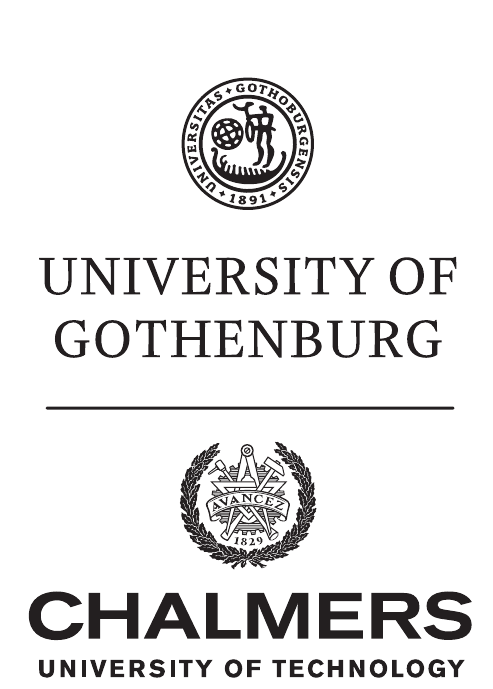}
	\end{figure}	\vspace{5mm}	
	
	Department of Computer Science and Engineering\\
	%\emph{Division of Division name}\\
	%Name of research group (if applicable)\\
	\textsc{Chalmers University of Technology} \\
	\textsc{University of Gothenburg} \\
	Gothenburg, Sweden 2020 \\
\end{center}

% IMPRINT PAGE (BACK OF TITLE PAGE)
\newpage
\thispagestyle{plain}
\vspace*{4.5cm}
A case-study for structuring and presenting data for testing of automotive electronics, to reduce effort and aid decision making. \\
Martin Tran\\
Ashish Husain \setlength{\parskip}{1cm}

\copyright ~ MARTIN TRAN, 2020. \\
\copyright ~ ASHISH M HUSAIN, 2020. \setlength{\parskip}{1cm}

Academic Supervisor: Gul Calikli, Department of Computer Science and Engineering \& Gregory Gay, Department of Computer Science and Engineering\\
Industry Supervisor: Jurij Sivakov, Volvo Cars AB\\
Examiner: Jennifer Horkoff, Department of Computer Science and Engineering \setlength{\parskip}{1cm}

Master's Thesis 2020\\	% Report number currently not in use 
Department of Computer Science and Engineering\\
%Division of Division name\\
%Name of research group (if applicable)\\
Chalmers University of Technology and University of Gothenburg\\
SE-412 96 Gothenburg\\
Telephone +46 31 772 1000 \setlength{\parskip}{0.5cm}

\vfill
% Caption for cover page figure if used, possibly with reference to further information in the report
Cover: The picture on the cover page is an alternative system prototype developed in this study to improve data structure and presentation for test diagnostics of automotive electronics. The prototype was designed and evaluated with stakeholders in the automotive domain. Evaluation of the prototype was used for this study's research.

Typeset in \LaTeX \\
%Printed by [Name of printing company]\\
Gothenburg, Sweden 2020

%% file: include/frontmatter/Abstract.tex
\thispagestyle{plain}			% Supress header 
\setlength{\parskip}{0pt plus 1.0pt}
\section*{Abstract}
% Abstract text about your project in Computer Science and Engineering.
% WHAT TO INCLUDE HERE: The aim, statement of the problem, methods used, the results and the conclusion.

%----Version 2
Automotive engineering is recognized as a combination of software and mechanical engineering due to the ever-increasing number of software-based components in vehicles. Since vehicles have become more sophisticated than before to ensure robustness, testing of automotive electronics is performed in high volume, producing immense test-related data.  

\hfill

This study investigates how unstructured and decentralized test-related data from testing of automotive electronics creates issues in decision making during the testing and analysis process of test artifacts by performing an exploratory case-study at one of the leading automotive companies, Volvo Cars. From the findings of the exploratory study, a prototype was designed to improve the data and information structure and presentation for test analysis and diagnostics for automotive electronics. The prototype's results showed that providing better data and information structure significantly increases the efficiency and reduces the workload for testers when conducting test analysis and diagnostics. Testers showed a decrease in task load for tasks related to testing due to better information structure, presentation, correctness and accessibility. Hence, the improvements aided the testers to arrive at decisions regarding root cause analysis of failed tests efficiently. The findings of this study can assist automotive companies in systematically investigating and improving the testing process of automotive electronics in regards to managing and structuring test-related data.

% KEYWORDS (MAXIMUM 10 WORDS)
\vfill
Keywords: Testing, Automotive Electronics, Electronic Control Unit, ECU, Unstructured Data.

\newpage				% Create empty back of side
\thispagestyle{empty}
\mbox{}

%% file: include/frontmatter/Acknowledgements.tex
% CREATED BY DAVID FRISK, 2016
\thispagestyle{plain}			% Supress header
\section*{Acknowledgements}
%Here, you can say thank you to your supervisor(s), company advisors and other people that supported you during your project.
The foundations of this thesis were built upon the help, guidance, and support of many people, and for that, we are truly in debt.

\hfill

Foremost, we would like to express our sincere gratitude to our academic supervisor Asst Prof Gul Calikili for her continuous support of our Master's thesis and research. Her patience, knowledge, and guidance helped us immensely to perform and complete this thesis in the utmost professional manner. We could not have imagined a better academic supervisor and mentor for ourselves.

\hfill

We would also like to thank Asst Prof Gregory Gay for helping us finish the thesis with his extraordinary insights and feedback on the thesis', especially the academic aspects of the thesis regarding the writing and structure. We are grateful to Gregory as, without his guidance, we would not have been able to finish the thesis while maintaining the standards and quality.

\hfill

Besides our supervisor, we would like to thank our industrial supervisor Jurij Sivakov at Volvo Cars, for motivating us through every step of the way. He brought the best out of us and guided us through his knowledge, experience, and charisma.

\hfill

We like to thank the various team members we collaborated with, in Volvo Cars, especially Albin Johannsson, Petter Emanuelsson, and Martin Turesson for their unconditional help and support. Without their guidance, many aspects of this thesis would not see fruition. The effort spent and the time they spent on us truly deserves an ovation.
We would also like to share our humblest of gratitude to everyone in the Power \& Energy Department at Volvo Cars AB for their time and support.

\hfill

Finally, we would like to thank our examiner, Jennifer Horkoff for guiding us especially during challenging times in our thesis. Her advice and feedbacks have had an extraordinary impact on our thesis.

\vspace{1.5cm}
\hfill
Martin Tran \& Ashish M Husain, Gothenburg, June 2020

\newpage				% Create empty back of side
\thispagestyle{empty}
\mbox{}

%% file: include/Introduction.tex
\chapter{Introduction} 
In recent years the advancement in the automotive industry has increased significantly. The advancement is so great that Bringmann and Krämer \cite{bringmann2008model} suggested that development in the automotive industry can be recognized as a combination of software, electrical and mechanical engineering due to the ever-increasing number of software-based components in vehicles. Hence, the importance of software grew as it became one of the fundamental sources of innovation \cite{grimm2003software}. The effects of this trend also demand frequent testing of automotive electronics, as vehicles have become more sophisticated than before, and testing automotive electronics is done in high volume to ensure robustness and better quality. 

\hfill

There can be numerous Electronic Control Units (ECU) in a vehicle. The functionalities of each ECU are requirement specific. As vehicle designs have become more sophisticated, their functionality has increased significantly \cite{Ecu}. ECUs communicate with various sensors in a vehicle to determine specific actions. For example, the ECU can take input from sensors inside the vehicle. Based on the sensor's value, an ECU can determine output signals that would control functionalities like fuel injection or internal throttling. Testers inspect the signal values and decide if sensor and ECU behavior match the expectations set by functional and test requirements.

\hfill

Testers frequently make decisions during the testing process based on data gathered while testing. An example of such decision making is when testers evaluate sensor signal values from an ECU. Testers make judgments on whether those values meet the thresholds set by the ECU's functional requirements. A tester could strengthen their claims by analyzing more signal data or using their experience from similar test cases. Having a lot of data and information to work with can have a negative impact on testers. If the data is not organized in a structured manner or if it is stored in multiple places. Having unstructured data makes relevant information harder to locate, and irrelevant information can cause inaccuracy in judgment \cite{jorgensen2008avoiding}. In general, people do not want to spend more time than necessary dealing with unstructured data \cite{rao2003Unstructured}. 

\hfill

This study investigates the testing process, data from the tests (e.g., test case information, signals, metadata), and how its unstructured nature and lack of presentability can affect decision making and effort. For simplicity, ECUs will be referred to as test objects in the upcoming chapters. An iterative process was conducted with the partnering company, Volvo Cars AB, specifically in the Power and Energy department. The department consisted of various stakeholders, including testers of automotive electronics, developers who worked on features for ECUs, product owners who were in charge of the feature deliveries and test environment providers who focused on test environment development. In the department, elicitation methods were used in the beginning to understand and gather data on the challenges they faced. The data was then used to create a prototype with an interface that would improve test case data presentation and structure. Finally, evaluations were conducted on the prototype to understand how structure and presentation could solve a few of their challenges. The result of this study showed that by providing a tool for the testers with a better data and information structure reduced the workload, increase the efficiency and aided in the decision making process by increasing the accessibility of information during test diagnostics and analysis.

%Such data can be used to derive decisions and having a larger set of it can affect data-driven decision making \cite{russom2011big}.

% Problem Description Start
\section{Problem Description}
\label{section:problemDescription}
There were three major types of issues that this study uncovered in the partnering company, Volvo Cars. They were about storing and maintaining testing related files, retrieving relevant data from log files consisting of a large amount of unstructured sensor signal data, and conducting the testing of a test object. Combining those three issues, the required effort to conduct test diagnostics on testers' ECUs was high. It would take a lot of time and effort for a tester to gather all necessary data and information before starting their test diagnostics task. If more data and information were needed to support decision making during test diagnostics, it would also take much effort to find the required data. The effort needed to conduct test diagnostics also differed significantly between new and experienced testers. In particular, inexperienced testers were challenged by working with unstructured data.

\subsection{Storing and Maintaining Files}
The files required to drive the testing process were not organized and kept in a well-defined location. Instead of opting to keep the files in a centrally accessible place, they were instead kept in multiple locations, including individual testers' local drives, storage servers, or the private company repositories. Test objects were tested frequently with specific test requirements using the standard testing practices in the automotive industry \cite{colleaux2009bridging}, with the help of testing tools to perform tests and analyze test output data. Files such as configuration, calibration, database, and master database files were required to configure the testing tools and allow the testers to decrypt and understand the values from the sensor signals and ECUs. These files were often stored in differing and inconsistent locations.

\subsection{Retrieval of Information and Data}
Not having these files readily accessible to the testers in a well-defined location made it extremely difficult and time-consuming for the testers to begin the testing process. They needed to look for the files manually in different storage locations. Version control of the files was also not practiced in a standard way, making it difficult for testers to differentiate outdated versions of files from newer versions. Having the correct files was crucial to the correctness of the tests' outcome, and often, testers had to perform trial and error to find the right files for initiating the test tools. There were situations when testers would proceed with incorrect test tool configuration files, which would affect the test output results. As a result, they would deem a correctly functioning test object as faulty. 

\hfill

Outputs from a test run could consist of sensor signal log data to and from a test object. As testing was done in real-time, large log files would be created. Testers found it very difficult to retrieve relevant information from the log files as data was represented in an unstructured manner. For example, if specific signals were not behaving as they should during the test run, testers manually had to look for the signal data, amongst a large data-set in the log files. Making sure they analyzed the correct signal became extremely difficult and time-consuming due to continually having to eliminate irrelevant data in the log files. Identification of the root causes of faults was affected by these issues. As the testers have no overview of the test results in a precise and concise manner, they always divert their attention, time, and effort to traversing and locating the required signal data. Making decisions from unstructured data
has been shown to be difficult \cite{blumberg2003problem}. To help testers make faster, less mentally taxing, and correct decisions based on faulty sensor signals, the data need to be structured or represented in a way that prioritizes relevant data over irrelevant data. 

\subsection{Process}
When the testers figured out the issue with the test object, they did not have well-documented guidelines on how to contact the relevant stakeholders responsible for fixing the problems. Most of the time, they would have to ask their colleagues to identify the person responsible. They would use messaging applications to broaden the search for the responsible person. Still, they would sometimes end up with no exact answers due to multiple teams or individuals having ownership over the test object or lack of documentation specifying ownership. It was essential to quickly delegate the issues to the correct team or individual to mitigate the problems found during testing and provide continuous integration and deployment within the organization and avoid delaying business plans and deliveries to the concerning stakeholders.

% Problem Description End

% Purpose Of Study And Research Questions start
\section{Purpose of Study}
\label{section:purposeOfTheStudy}
The purpose of the case study was to identify the root causes of the problems mentioned earlier and help solve them by iteratively developing a testing-framework that improved the practices and processes of testing automotive electronics. 

\hfill

One of the study's purposes was managing various testing-related files efficiently to perform the test initiation quickly. The goal was to better map the files with the test object, allowing testers to perform and reproduce the tests with consistent configurations and decreasing the time and effort required to collect all-important test initiation files from various storages.

\hfill
    
Another purpose was to aid testers to visualize relevant data amongst the large unstructured data logs. Testers needed to make decisions about whether the test object is faulty or not by analyzing the log files' signal, as explained earlier. The goal was to present the unstructured log data to the testers in a structured way, allowing them to focus on the data they want to investigate rather than finding it from the messy log files always, reducing the time and effort needed throughout the investigation process.

\hfill
    
Another objective was to improve testers' communication with stakeholders to convey the verdicts of the test outcome efficiently. The goal was to minimize the time and effort spent on unnecessarily looking for the right person to contact and to have better documentation of which stakeholders own which test objects and the issues that may surface after testing. The following research questions were investigated and answered by the end of this exploratory case-study:

\hfill

\begin{description}
    \item[RQ1:]What aspects deteriorate decision making during test analysis of automotive electronics?
    \item[RQ2:]How should the design of a test analysis framework for automotive testing reflect the decision making models followed by testers?
    \item[RQ3:]How should data be structured, presented, and documented in this test analysis framework to improve the quality of information available during test analysis?
    \item[RQ4:]How can this test analysis framework be designed to support ease of use and accessibility of test related data during test analysis?
\end{description}

\hfill

To investigate effort spent during collecting the test files from various locations, investigating sensor signal data from the unstructured log files and communication about the verdict of the test to the responsible stakeholder, a workload index assessment tool called NASA TLX \footnote{https://humansystems.arc.nasa.gov/groups/TLX/downloads/TLXScale.pdf} was used to measure subjective workloads and performance. NASA TLX is a subjective multi-dimensional assessment tool that rates perceived workload in tasks to access performance.

\hfill

The study shows the variation of workload index with data structure and presentation of test-related data. The findings from the study suggest that data structure and presentation improved testers workload index results. In the automotive testing domain, information presentation had a positive impact on testers, aiding their decision making abilities. As the testers had to work with a lot of unstructured test data and files, improving presentation and navigation efficiency of those data and file improved the workload index scores, time efficiency and decision making satisfaction of the testers.
% Purpose Of Study end

% Significance of the study Start
\section{Significance of the study}
\label{section:significaneOfTheStudy}
Overall, this study addresses effort and process efficiency from an industrial standpoint, focusing on decision making, which has relevance for both practitioners and researchers. This study investigates the automotive industry's current practices in managing and presenting data and information from automotive electronics tests and how to support decision making efficiency by using techniques based on data visualization and presentation. The findings in this study contribute to solving data visualization and presentation challenges in the context of the current advancement in the automotive industry. The contributions of this study include the following:

\hfill

\begin{itemize}
    \item Identification of the issues surrounding unstructured data and its presentation for testing automotive electronics.
    \item A user-centered framework to address the issues, developed using techniques and literature from psychology, design science, and software engineering.
    \item A system developed using the proposed framework to mitigate the identified problems.
    \item Providing observational and experimental data on using the framework and system and showing relationships and trends regarding its usage from the stakeholder's perspective.
    \item A generalized set of guidelines to help mitigate the problems with unstructured data for automotive electronics.
\end{itemize}

\hfill

The beneficiaries of this study will be the stakeholders in the automotive industry, particularly the testers. Furthermore, researchers in the field of data-driven decision making will have further data from an industrial case.
% Significance of the study End

% The outline basically conatins the whole structure of the study.

% briefly describe the background for the reseach and the domains it will cover (DDD, Testing of automotive, Unstrctured data)

% Briefly explain how the related work section will look and what it conatins

% Breifly describe the initial planning of how the different methodologies will be used and the section in general

% describe the discussion and results section
% Thesis Outline Start
\label{section:thesisOutline}
\section{Thesis Outline}
Chapter 2 presents relevant research topics and terminology relevant to this study. Electronic Control Units are explained in general, as well as how they are tested, along with the current challenges seen in other studies. Decision making is described in the context of how it relates to data management and task complexity. 

\hfill

Chapter 3 describes findings and concepts from other studies which are expanded on in this study. It contains an overview of current research for testing of automotive electronics and ECUs. Decision making models and data visualization findings from previous studies are described, as well as how they are used by this study to investigate our research questions.

\hfill

Chapter 4 describes the methods used in this study, following the Design Science Methodology. Our study is divided into three phases: problem identification, artifact design, and design evaluation. As the Design Science Methodology is an iterative process where the phases are conducted multiple times, this chapter is structured chronologically. The problem identification phase describes methods used to understand the problem and challenges currently in the partner company. The artifact design phase describes the development of a new system for test analysis and diagnostics, designed to solve the identified issues and challenges. Finally, the evaluation phase assesses the new system's performance. 

\hfill

Chapter 5 contains the results of the study. The results are presented for each of the three phases from Chapter 4. The challenges and issues of unstructured data are presented in the problem identification section. The chapter also presents the artifact, and its evaluation results.

\hfill

Chapter 6 contains the discussion of the results from Chapter 5, where the research questions of this study are answered. The chapter also discusses the key contributions, benefits, threats to the validity of this study. Chapter 7 summarizes and concludes the findings of the study.

% Thesis Outline End

%% file: include/Background.tex
\chapter{Background}
This chapter aims to explain the terminology and background knowledge required to understand this study. 

% https://corescholar.libraries.wright.edu/cgi/viewcontent.cgi?article=3296&context=etd_all
% https://www.ecutesting.com/categories/ecu-explained/
\section{Electronic Control Units}
\label{section:electronicControlUnit}
Electronic Control Units (ECU) are embedded systems that control one or more systems in a vehicle. An ECU works as a switching system that toggles flags and sets values for the different electronic modules and manages power distribution in the vehicle. An ECU provides automotive engines with inputs and outputs and communicates with the electronic modules through an internal computer network (e.g., Controller Area Network (CAN bus)). ECU inputs can include sensor input, such as temperature and pressure values, as well as data from other electronic modules. The input information is processed and used to make decisions, for example, performing an action on the engine or controlling the correct amount of power provided to actuators, such as ignition system timing. The amount of ECUs in a vehicle can be up to 80, and this number has been growing over time \cite{ebert2009embedded}. As the software complexity of vehicles increases \cite{grimm2003software}, as does the quantity of ECUs in a vehicle as well \cite{ebert2009embedded}.

% Testing Start
% Testing in the field of Automotive
% How do the Automotive Industry do tests -> Electronics -> ECU -> Outputs
\section{Testing of Electronic Control Units}
\label{section:testingOfAutomotiveElectronics}
The development approach within automotive electronics has taken the trend of model-based development, using modeling, simulation, and code generation tools as standard practices \cite{bringmann2008model, lamberg2004model}. Testing is affected by adapting model-based testing, wherein the earlier phases contain systematic automated testing and ECU testing activities in later phases. Hence, model-based testing enables testing the system in the earlier phases of development, in a test environment, before implementing the final hardware. The model and functional code are first tested and implemented, then integrated into the ECU. The ECUs are first tested in a black box environment where inputs are simulated, and outputs are monitored, followed by integration testing to test the ECU's integration with other ECUs and electronic modules. During the ECU tests, the ECU inputs and outputs are monitored to ensure the data is within the thresholds specified by requirements.

\hfill

Input and output data from the tests of an ECU are sensor signals that contain values and information from other ECUs and electronic modules. The tester's task is to decide if the test has passed or not by analyzing the sensor signal's values from the system and comparing it to the expected value provided in the requirements. An example of a test case could be if the engine in a vehicle had started its ignition, then ECUs responsible for the vehicle lighting system should receive sensor signals values that the vehicle has started and later process them accordingly. As requirements can be vague, confusing, and even missing some information, it can be challenging for the tester to judge the verdict of the test. Up to 40\% of embedded-software defects in ECUs come from insufficient requirements with testing analysis activities\cite{ebert2009embedded}. Testing of ECUs can be highly subjective due to factors such as requirement vagueness and large test sizes or quantity, which can lead to judgments errors, can be time-consuming, and requires testing experience \cite{conrad2005automatic}.

\hfill

ECUs are monitored through the network buses in the vehicle such as CAN buses. CAN buses serve as the communication bus technology to broadcast the sensor signals and are used for diagnostic services. The data gathered through the CAN bus in a test vehicle is estimated as 4500 megabytes per day \cite{johanson2014big}. This means that there may be many sensor signals to analyze during a test. Testing of ECUs can take up to 30-40\% of the embedded-development resources \cite{ebert2009embedded}. 30\% of test cases are redundant as embedded software development tends to do tests without taking into account coverage or related effectiveness criteria when assessing the completeness of testing \cite{ebert2009embedded}.
%Testing End

\section{Decision Making in Testing Electronic Control Units}
\label{section:DecisionMakingInTestingElectronicControlUntis}
Testing of ECUs requires testers to make various decisions and judgments during the testing process. When testers receive a test case, they will have to read the test case's test requirements and the functional requirements to understand the context of what is being tested. The testers have to set up their testing tools with the correct files and interpreter, such as a database file, to make sense of the test's output. There could be multiple data interpreters the testers have to choose from to initiate the testing. If files and interpreters are hard to find because of the unstructured nature of how they are stored, it can pose a challenge for the testers to pick the correct ones. The next step is to compare the outputs to see if they meet the test requirements' threshold values and if the functional requirements are still fulfilled. If the test's output was not satisfactory enough, the tester has to find a responsible stakeholder to contact who would continue the investigation. As the requirements can contain insufficient information and be vague \cite{ebert2009embedded}, it can be challenging for the testers to make judgments and decisions such as testing accordingly, with respect to the correct requirements or interpreting the acceptable thresholds during the test process \cite{conrad2005automatic}. Mismanaging requirements during test analysis activities for embedded-software systems can result up to 40\% of the defects to incorrect reviewing and analysis of requirements, missing and vague requirements or confusing needs of requirements \cite{ebert2009embedded}.

%write about how in general complex tasks makes decision making harder for subjects. Talk about some models of decision making.
\label{section:taksComplexityAndDecisionMaking}
\section{Task Complexity and Decision Making}
Task complexity and decision making are related to each other and are considered a widespread problem in psychology \cite{payne1976task}. Payne \cite{payne1976task} focused on how variation in task complexity affected decision making. Their findings suggested that when performing complex tasks, individuals performed different heuristics, keeping information processing demands of the situation within the boundaries of their capacities \cite{payne1976task}. When humans were required to perform a task, the information processing leading to a preferred choice would vary as a function of task complexity. When subjected to many alternatives for a specific task, humans tend to quickly eliminate some of the alternatives to avoid building up stress progressively \cite{payne1976task}. People tend to resort to choice-based heuristics to reduce cognitive strains when faced with deriving decisions from tasks involving increased complexity. Findings from the study by Payne confirmed the issues that lead up to decision making that the testers went through. Testers in the current testing process had to find test initiation files from different locations. Due to them being unorganized, they had to locate and decide to proceed with the correct files for their test cases, amongst an abundance of records. To quickly start the testing process, the testers sometimes had to eliminate outdated test initiation files to get the correct ones rapidly. Even then, there was a chance that they might end up with the incorrect initiation files and find it out at a later stage of testing. In the testing process, testers visualized all the test data (sensor signals, files, test cases, test requirements) in multiple test suites' windows, which made it increasingly difficult for them to maintain focus and maintain information processing demands.

%% file: include/RelatedWork.tex
\chapter{Related Work}
\label{chapter:relatedWork}
This chapter will explore and present related work. The findings from previous research can be investigated further and extended upon by our industrial case-study. 

% Testing Processes -> Different Approaches -> They all get data, then what? -> We focus on what comes next
% Companies, use methods based on their business goals/self-tailored
% Most of the researches/related works are about optimizing the ECU output, we want to work on the next step
\section{Test Analysis in Automotive Electronics}
\label{section:testAnalysisInAutomotiveElectronics}

There has been work done related to improving the process of testing automotive electronics. There are proposed methods such as robust engineering techniques \cite{leaphart2006application} and model-based testing approaches \cite{bringmann2008model, lamberg2004model} to improve the process. Automation of sensor signal comparison \cite{conrad2005automatic} has been conducted to reduce time and cost compared to manually reviewing sensor signals. A diagnostic system \cite{foran2005intelligent} to intelligently and automatically diagnose faults in multiple ECUs dependencies when one ECU can cause a fault to another ECU has been investigated. Automating the testing process to reduce the influence of human factors, resulting in improved quality of test data has also been investigated \cite{novak2005automated}. 

\hfill

Existing research and innovation have primarily focused on improving the overall testing procedures and increasing the quality of test analysis. Metadata, such as test requirements, information, and data from test procedures, has been overlooked. The actual structure and presentation of the metadata and its impact on test analysis quality has been given little to no attention. Our study focused on how test-related data could be stored, structured, and presented to improve the testing analysis and diagnostics quality. 

\section{Decision Making Models}
\label{section:decisionMakingModel}
Relationships between task complexity and decision making have been researched extensively. Earlier studies have suggested that increasing information increases the variability of the responses \cite{hayes1962human, jacoby1974brand, hendrick1968decision} and decreases the quality of choices while increasing the confidence decision-makers have in their judgment \cite{lichtenstein1973response}. The study by Payne \cite{payne1976task} further discussed various models of decision making that represented the process of making choices. Here, alternative-choices have been defined as something leading to the completion of an overall problem or task:

\hfill

\begin{description}
\item[Additive/Linear Model:] Each alternative in a set of choices was evaluated separately. A perceived subjective/objective value was arrived at for each component/dimension of the alternative. The components were blended in additively, forming an overall value for the alternative. After comparisons between the alternatives, the one with the highest value were chosen. The study also mentioned that the additive model's nature was compensatory. The subject would trade-off between a high value in one dimension of an alternative and a low value on another dimension.
 
\item[Conjunctive Model:] An alternative must have a certain minimum value on all the relevant dimensions in order to be chosen. 

\item[Elimination-by-aspects (EBA) model] As the name suggests, based on a selection of dimension or aspect, all the other alternatives that do not possess that dimension were eliminated. This process was repeated until one alternative remains. The probability of selecting an aspect or dimension was assumed to be proportional to its weight or importance.
\end{description}

\hfill

In an experiment done in the same study by Payne \cite{payne1976task}, it was observed that subjects would employ decision making strategies in challenges with two-alternative choices that involved searching the same amount of information for each alternative. This behavior changed when the alternatives were increased, and subjects adopted decision making strategies that would eliminate some of the available alternatives as quickly as possible based on a limited amount of information searching and evaluation. Furthermore, practices done in the models such as conjunctive and elimination-by-aspects were used early in the decision making process to simplify the task's decisions by eliminating alternatives quickly.

\hfill

The findings from the study by Payne \cite{payne1976task} were used in our study to understand how the tester's decision making process related to the three decision making models. The approach proposed to improve the testing procedure uses the three models to introduce improvements concerning information structuring and presentation. 

\section{Data Visualization}
\label{section:dataVisualization}
Visualization is important for software testing as it is a long-term and complex process based on significant quantities of data. Better visibility of the results and outputs of tests benefits testers with a quicker general perspective and leads to a better understanding of the test \cite{wang2012mavis}. 
This study investigated the following data visualization factors to solve the challenges.

\subsection{Information Quality and Presentation}
The study performed by Bharati and Chaudhury \cite{bharati2004empirical} investigated decision making satisfaction, which was the ability to support decision making and problem-solving activities. The study used a web-based decision support system for the investigation with the motivation that web-based technologies were among the leading approaches for decision making systems \cite{bhargava2001decision}. An investigation on how the web-based system's characteristics affect the decision making could provide useful insights for development. The study created three hypotheses to investigate the characteristics: System quality, information quality, and information presentation. The results in the study showed that:

\hfill

\begin{itemize}
    \item System quality has a positive correlation to decision making satisfaction. Access and system reliability enables ease of use, which had a positive effect on decision making satisfaction.
    \item Information quality has a positive correlation to decision making satisfaction. Providing relevant, accurate, and complete information resulted in a positive effect on decision making satisfaction. 
    \item Information presentation did not have a positive correlation to decision making satisfaction. The graphics, color, presentation style, and navigation efficiency did not impact decision making satisfaction.
\end{itemize}

\hfill

This study used the system quality, information quality and information presentation criteria to design a solution with improvements to the challenges with unstructured data and information, using a proposed web-based tool. This study also investigated whether the presentation of information aided in decision making satisfaction in the testing of automotive electronics and working with large amount of unstructured data. 

\subsection{Impact of Irrelevant Information} 
The study done by Jørgensen and Grimstad \cite{jorgensen2008avoiding} explained how working with estimation and judgment-based tasks, irrelevant information, and variation of wording could affect understanding of requirements. One source of inaccuracy in judgment, decision, and estimation comes from irrelevant and misleading information, in which unconscious cognitive processes have a big factor \cite{jorgensen2008avoiding}. The study's findings indicate that highlighting relevant or crossing out irrelevant information did not significantly reduce the impact, but completely removing the irrelevant information did. In general, people always move on when they do not find useful information quickly. Therefore, categorizing information, codified as documents and collection of meta-data, made it practical to help solve issues with unstructured data, which can remove much irrelevant information \cite{rao2003Unstructured}. In our study, the impact of irrelevant information and the effort for decision making were investigated further by examining approaches for handling irrelevant data.  

\section{Applying the Existing Related Work} 
\label{section:applyingTheExistingWork}
It was vital to understand how the existing testing process influenced the testers' decision making. The current testing process had issues in the storage of test-related files, finding relevant information from massive unstructured test data, needing to traverse between multiple testing tools to get an overview of the test data, and locating the responsible stakeholder for the test object. Those issues have implications on the testers regarding the time and effort testers spend on testing and efficiently locating faults found during testing. The existing research regarding the decision making models' practices helped inform how testers come to conclusions or make decisions during the testing process. This past research provides rationale and context for how the drawbacks of how the current system provided information affect the testers.   

\begin{table}[h!]
\caption{Criteria used to investigate and answer research questions.}
\label{table:criteria}
\centering
    \begin{tabular}{ |m{8em}|m{25em}| } 
        \hline
            \textbf{Criteria} & \textbf{How it was used} \\ 
        \hline
        \hline
            Decision Making Models & To investigate how testers reach conclusions during testing while having to face problems due to poor system and information quality.  \\
        \hline
            System Quality & To investigate ease of use and accessibility of test related data in the existing testing process and to improve them in the new artifact. \\ 
        \hline
            Information Quality & To investigate information completeness, relevancy and accuracy for test related data in the existing testing process and to imporve them in the new artifact.   \\
        \hline
    \end{tabular}
\end{table}
\hfill

%% file: include/Methods.tex
% Write briefly about the how the problem statement came to be, How the initial stage of the problem were discovered. 
% Explain that the study follows design Science, and that there are three iteration that will be discussed throughout the thesis. 
% Explain that there will be an ovelap between sections
%conclude this paragraph, by stating the course of the events, how the study went from -> Undertstanding the problem -> Design stage -> Evaluation/Validation and finally the experiment.
\chapter{Methodology}
This chapter will describe the methodology used in the study to answer the following research questions:

\hfill

\begin{description}
    \item[RQ1:]What aspects deteriorate decision making during test analysis of automotive electronics?
    \item[RQ2:]How should the design of a test analysis framework for automotive testing reflect the decision making models followed by testers?
    \item[RQ3:]How should data be structured, presented, and documented in this test analysis framework to improve the quality of information available during test analysis?
    \item[RQ4:]How can this test analysis framework be designed to support ease of use and accessibility of test related data during test analysis?
\end{description}

%\begin{mdframed}[style=MyFrame]
%\end{mdframed}

\hfill

This study followed the Design Science methodology  \cite{von2004design} to answer the research questions. Design Science is used to design and evaluate an artifact intended to solve the identified problems. For this study, the artifact was used as a new testing system for the testers that, incorporating improvements found during the iteration of design and evaluation. The Design Science approach consists of iterative development phases\cite{von2004design}. The phases include:

\hfill

\begin{enumerate}
    \item \textbf{Problem Identification}. This phase involved acquiring knowledge and understanding the problem and environment.
    \item \textbf{Artifact Design}. An artifact was created to address the problem. We applied existing knowledge and research to propose a solution to the problem.
    \item \textbf{Design Evaluation}. We evaluated the utility, quality, and efficiency of the artifact. 
\end{enumerate}

\hfill

The problem identification phase (fully described in Section \ref{section:understandingTheProblem}) consisted of investigating the existing testing process and the stakeholders' involvement using data collection methods, including semi-structured interviews and questionnaires. Participant observation sessions with the testers were conducted to practically witness the testers perform tasks in their natural environment. The findings from interviews, questionnaires, and observations were categorized using the affinity diagramming approach. The affinity diagram helped organize and categorize the findings based on their natural relationships, which were then coded using the Nvivo tool to understand the requirements that the stakeholders needed to see in an improved system for preparing and performing the test cases. The requirements consisted of the stakeholders' wish-list of improvements, based on their overall testing process experience. The observations collected during this phase provided a foundation to answer the first research question.

\hfill

The artifact design phase (fully described in Section \ref{section:designNewSystem}) was used to develop an improved testing system using prototypes based on the requirements gathered from the problem identification phase. After each iteration of the prototype, new requirements emerged and improved the prototype iteratively through the design evaluation phase, where cognitive walkthrough and semi-structured interviews were used to evaluate the prototype. Three iterations of the prototype were conducted in this study. Requirements were mapped with criteria from previous work described in Section \ref{section:applyingTheExistingWork}. Finally, a high fidelity prototype was designed that the testers could use interactively to perform real testing. This high fidelity prototype was evaluated by the testers using the usability testing technique with the think-aloud protocol. The usability testing consisted of tasks that the testers would do to perform tests on a test object. The tasks included gathering all the necessary files to initiate the test environment, performing steps as defined in test cases, find issues in the sensor signals wherever necessary, and convey the verdict to the responsible stakeholders. The testers' verbal and usability feedback was essential to justify how well the improved system fared in undertaking the issues highlighted in the second research question. In total, the prototypes went through three iterations.

\hfill

The final evaluation of the improved testing system (described fully in Section \ref{section:evaluationOfTheNewSystem}) was done to measure effort, workload, and time spent performing tests using NASA TLX, a subjective multi-dimensional assessment tool that rates perceived workload in tasks to access performance. NASA TLX also incorporates a multi-dimensional rating procedure to derive scores for workload in a task. The goal was to compare the performance between the existing testing system and the new, improved system by providing scores based on the perceived effort and workload during the testing process. An open-ended interview followed the experiment to discuss the scores of the NASA TLX. The observations made in the evaluation phase provided a foundation to answer the second research question.

\hfill

\begin{figure}[h!]
  \centering
  \includegraphics[scale=0.3]{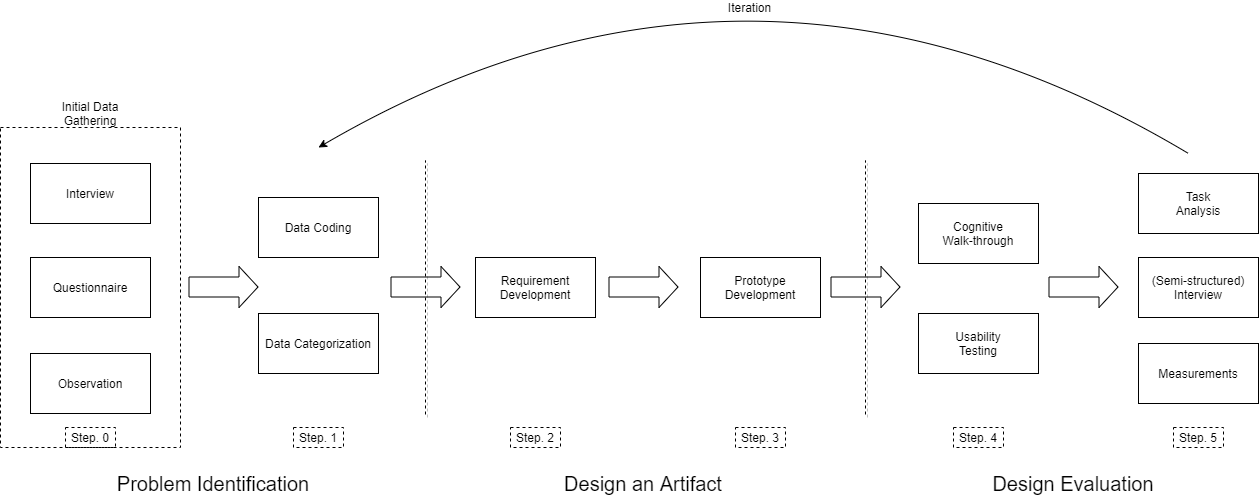}
  \centering
  \caption{Activity timeline for the exploratory case study.}
  \label{fig:activityInMethods}
\end{figure}

Activities done during this study were iterated upon multiple times and will be described in this chapter chronologically. The following sections will explain the phases of problem identification, artifact design, and evaluation. The activities and methods in this study can be seen in Figure \ref{fig:activityInMethods}. 

\newpage
\section{Participants in the Partner Company}
\label{section:participantsInThePartnerCompany}
The participants for this study were stakeholders involved in the testing process. The stakeholders' role were product owners, testers and test environment providers. In the initial problem identification phase, two of each stakeholder were participating for interviews. Two testers were used for participant observation. A questionnaire was sent out to the department, and it resulted in 13 respondents. 

\hfill

For the evaluation phases, only testers were the participants and they not involved or participated in the problem identification phase. Two other testers were used for the iterative evaluation of the new system. Six other testers were used for a usability test, for the final evaluation of the new system. One tester was also used for the pilot test of the usability test and did not participate in the actual usability test. 

% Short description of the initial problem description (transition of man to auto and unstructure data, hard to make decsion from it)
% Interview 1. Environment
% Questionnaire: Environment & Testing
% Interview 2. Testing process
% What are we going to do with the data gathered?
\section{Understanding the Problem}
\label{section:understandingTheProblem}
The goal was to investigate problems with the existing testing process in detail while exploring broader issues concerning testing. A vital area studied was how testers made decisions during the testing process. The testers' decision making strategies were compared and contrasted with the three decision making models (Additive/Linear, Conjunction, and Elimination by Aspects) discussed in Chapter \ref{chapter:relatedWork}. A combination of data collection and categorization methods were used, such as interviews, questionnaires \cite[Chapter~7.4]{preece2015human}, observation \cite{johnson2003data} and affinity diagram \cite{takai2010use}, to investigate the existing testing process and to identify problems.  

\hfill

During the initial data gathering stage, interviews were done with different stakeholders, including the testers, product owners, and test environment providers to understand and find issues with the existing testing process. Each stakeholder was interviewed twice in the problem identification stage. The product owners in the company were in charge of delivering the features that were implemented in vehicles (involving ECUs). They were familiar with the development and testing life-cycle of the test objects. The testers performed the testing on the test object to find issues. The test environment providers had detailed knowledge of the testing tools and their configurations. Questionnaires were used to reach out to more stakeholders and departments involved with testing. After that, observation techniques (participant observation) were used to see how testers tested a test object. Two experienced testers were observed for a duration of 30 minutes. The goal was to reflect upon the interviews and questionnaires' findings and investigate visual ques of stress and effort during testing. One observation session with the experienced testers was conducted because they were accustomed to the testing process. Having one session was enough to identify the problems for further discussion about the problems during the interviews. Once enough data to understand the existing testing process and its problems was collected, the findings were then categorized using the affinity diagram technique, described further in Section \ref{findingsCategorized}. 

%write about how different data collection methods were chosen for this case-study --> Briefly talk about (interviews, questionnaire, affinity diagram)  --> why use such methods

% what will the output of the data collection method be used for
\subsection{Applying data collection and categorization methods}
Throughout the case-study, semi-structured interviews were used to investigate the problems in the existing process of testing. Such interviewing techniques were well suited for their exploratory \cite{runeson2012case} and flexible nature. Both researchers were in the interview room, where one acted as the interviewer while the other took notes that would be recapped with the interviewee.

\hfill

Interviews were conducted with various stakeholders, including the testers, product owners, and test environment providers. Testers could provide information on how they initiated and performed the tests and found problematic issues during the tests. Product owners could provide an overview of the entire testing process of a test object and the importance of mitigating problems encountered from testing the test object quickly for their continuous integration and deployment strategy. The test environment providers could share the information on test initiation files needed to start testing and describe their importance for determining the correctness of testing outcomes. 

\hfill

Each session was transcribed using NVivo \footnote{https://www.qsrinternational.com/nvivo/home}, which is a qualitative data analysis tool. The tool made it possible to categorize the interviews' transcriptions and find similarities and differences between the interviewee's answers. NVivo aided in identifying trends from the observations collected above and representing them visually using various data categorization charts, making it easier to understand and pinpoint variations. The goals of the interviews were as follow: 

\hfill

\begin{enumerate} \label{investigationPoints1}
    \item Investigate the existing testing process.
    \item Identify problems in the current process.
    \item Understand stakeholder involvement in the process.
\end{enumerate}

\hfill

The interview with the product owner established the basis for the existing test process as an overview. The interview guide for product owners can be found in Appendix \ref{appendix:interviewProductOwner}. During this interview, the product owner highlighted the testing process, protocols, and involved stakeholders in the testing domain. Interviews with various testers and test-environment providers were conducted, with varying experience levels to investigate practical aspects of testing. The interview guide for testers can be found in Appendix \ref{appendix:interviewTester}. An introduction to the various issues in the testing process was given beforehand by the product owners before conducting the problem identification phase. Hence, this made it easier to develop the questions for the semi-structured interviews. The primary goals for the interview included the following:

\hfill
\begin{enumerate} \label{investigationPoints2}
    \item How were test requirements associated with the test object?
    \item How do testers initiate the testing process?
        \begin{itemize}
            \item What kind of files were needed to initiate the testing process.
            \item Where are these files stored, and are they easily accessible to testers?
        \end{itemize}
    \item What test-related outputs were produced while the test object was being tested, and where were they stored?
    \begin{itemize}
            \item Were the test output files easy to locate if tests needed to be revised later?
            
        \end{itemize}
    \item How did testers attempt to locate faults found during the test object's signal analysis?
     \begin{itemize}
            \item What were the challenges in finding the faults?
            \item Did the testers get overwhelmed by the quantity of sensor signal data? 
        \end{itemize}
    \item How did testers find and collaborate with stakeholders responsible for mitigating the issues identified during testing? What were the challenges the testers faced?

\end{enumerate}

\hfill

%\subsection{Open Ended Questionnaire}
Once the interviews helped to illustrate the testing process, an open-ended questionnaire was made to target a broader audience in the testing domain to gain further insights. Due to the legal obligations of the industrial partner data policies, the questionnaire was made using the Microsoft Forms platform \cite{MicrosoftForms}. The reason for opting to choose an open-ended questionnaire was because it allowed respondents to express an opinion without being influenced by the researcher \cite{foddy1994constructing}, while also helping gain valuable insights based on the respondents' perception and experience. The questionnaire design was focused on the aspects of the testing process that the audience considered problematic and to which extent such issues hampered them during the testing process. The questionnaire also gave the respondents the flexibility to mention further problematic issues in the testing process. The questionnaire can be found in Appendix \ref{appendix:questionnaireProblemIdentification}. In total, thirteen responses were received for the questionnaire. All thirteen respondents had prior experience within the testing process in the department. There was no overlap between the respondents from the questionnaire and the initial interviewees. That is because the questionnaire was sent to everyone other than the initial interviewees.

\newpage
\subsection{Participant Observation}
Participant observation is a method in which a researcher can participate in the daily activities, rituals, interactions, and events of a group of people to learn about their implicit and explicit aspects of their routines \cite{musante2010participant}. The participant observation method was used to observe the entire testing process and ensure consistency between the interviews and questionnaires' findings. According to Johnson and Turner, people might not always practice what they claim. Hence it became essential to practically witness the testing process and validate the information from interviews and questionnaires \cite{johnson2003data}.

\hfill

There were two participant observation sessions focused on viewing the entire testing process for a test object.
Two expert testers with significant working experience were selected randomly from the previous interview sessions for the observation. The reason for only including two experienced testers was to help identify every detail in the investigation and avoid biasses between the testers' choices. The sessions were booked in advance, on different working days, based on the testers' schedules. The researcher's participation was kept to a minimum by preventing interruption in the testers' work. The sessions were performed in the tester's actual working environment. Due to legal obligations, recording the sessions was prohibited; thus, notes had to be taken to document the observation. The observation was concluded by conducting a brief interview with the participants. During the interview, the test process was summarized and discussed to mitigate any confusion or misunderstandings.

\hfill
\subsection{Coding Qualitative Data}
\label{findingsCategorized}
The results from the data collection methods explained in Section \ref{section:understandingTheProblem} were coded by grouping the data into various categories using NVIVO. Initially, the verbatim transcriptions of the recorded interviews were done manually by listening to record and writing it down in the NVIVO tool. Then, a deductive coding approach was used with predefined themes to categorize the transcripts were identified \cite{rivas2012coding}. The themes were made using the following key points:

\hfill

\begin{enumerate}
    \item Existing testing and test-output analysis process.
    \item Identifying associations for the test object concerning testing related files, testing outputs.
    \item The problems in the testing and analysis process.
    \item Involvement of stakeholders.
\end{enumerate}

\hfill

Further categories/subcategories were made during the coding process by considering the points mentioned in \ref{investigationPoints1}, and by analyzing the transcriptions with which identified further essential findings. 

\hfill

From here onward, the respondents' findings were associated with similar grouping needs to represent standard criteria by using the affinity diagram approach in NVivo. An affinity diagram is a tool that gathers large amounts of language data (ideas, opinions, issues) and organizes them into groupings based on their natural relationships \cite{affinity}. The categorization helped understand the various aspects involved with the testing process in a detailed and concise manner. The diagram also helped in coding the data in NVIVO by creating nodes for the categories. A Node in NVIVO represented the categories consisting of references and vital information from the data collection methods. Findings were segmented hierarchically and organized to identify similarities and differences between respondents. To eliminate biases, we opted to perform the coding together. Requirements for improving the existing testing process were developed and improved upon with the stakeholders until a common consensus was established.   

% The findings from Understanding the Problem will be mapped to studies/research from the related work to build a foundation when designing the new system. 
% Discuss about the existing techniques in the various industry/academic related to the problem statement
% Decision Making "Guidelines, model"
% Information Presentation / Data Structure
% Testing Techniques in Automotive Electronic

%\section{Mapping Decision Making Models} % rename this criteria to some thing else!
%\label{section:criteriaFromPreviousWork}
%HAVE TAKEN A BACKUP OF PREVIOUS TEXT! They are being re-distributed where ever necessary.

% We have data, we have criteria -> Affinity Diagram / Themeification -> Create Requirements
% Having the knowledge of Data From Questionaire, interviews
%How criterias will be connected
% Affinity diagram for the identified themes
%requirement creation
\section{Designing the New System}
\label{section:designNewSystem}

%Refactoring..
After establishing the answer to the first research question, a concrete understanding of the existing testing process's problems was formed. In order to answer the second question, we designed and evaluated a new testing system that incorporates improvements and requirements from the problem identification phase. 
\hfill

\subsection{Areas Improved By  the Artifact}
An artifact was iteratively designed based on the knowledge and findings from the problem identification phase. The artifact is an improved testing system that was prototyped and evaluated multiple times with the stakeholders. The key areas of improvement targeted by the artifact are as follows:

\hfill
% these are basically the extension of the bold bullets points on page 38(Chapter 5, 5.2.1), but now they are being used to map and possibly inspire to derive requirements.
\begin{itemize}
    	\item \textbf{Data Presentation: }The artifact improves how testers visualize and find all the test related data and files without needing to jump between different testing tools and storage.

    	\item \textbf{Data Structuring: }The artifact improves accessibility and readability of test-related information (test cases, testing requirements, and initiation files) and test output files (log files) by efficiently organizing and structuring them. 
    	\item \textbf{Testing Process: }The artifact attempts to improve the overall testing process by decreasing the time spent doing unnecessary work that does not add value to the testing process. Unnecessary work, in this case, refers to finding files from different storage locations to initiate the testing process, as well as locating and traversing the large logs for sensor signals essential for evaluating the correctness of the test object.
 	\item \textbf{Communication: }The artifact improves the process of communicating with stakeholders responsible for mitigating the issues found during the testing.  
\end{itemize}

\hfill

The requirements were iteratively developed from the observation made during the problem identification phase and focused on delivering the improvements listed above. During the prototyping process, each iteration was validated to ensure that the areas of improvement were appropriately considered. Further requirements emerged during the interview sessions after the prototypes' usability testing, which will be further discussed in the Section \ref{prototypeDevelopment}.
%affinity stuff written before

\hfill 

The criteria from Section \ref{section:applyingTheExistingWork} were used as a reference during the creation of requirements. The requirements aimed to improve the problems in the existing system, and incorporating them would enrich the functionality and design of the new system. The requirements were iterated multiple times and later used to develop the artifact, which is further explained in Section \ref{section:evaluationOfTheNewSystem}. 

% Mention about the requirements from the affinity diagrams corresponded to making a low fi prototype
% How all the themes in the affinity were used as features
% Prototype methodology was "T-prototyping", read more in the UX book, and it is also in our shared doc where we read both the books
% Talk about the two iteration of the high fidelity prototype using figma and later React
\subsection{Prototype Development}
\label{prototypeDevelopment}
Prototyping was used for developing the artifact to investigate, evaluate, and think of new ideas. Prototypes were useful for evaluating the stakeholders' requirements and worked as a communication device to discuss and explore ideas \cite[Chapter~11.2]{preece2015human} effectively. As low-fidelity prototyping was simple and quick to modify, it was an important tool during the early stages of the requirements development. The exploratory nature of prototyping made it easier to try out and generate new ideas with the design. The low-fidelity prototypes were sketched using pen and paper.

\hfill

The first set of requirements for the new artifact were created by validating the low-fidelity prototypes \cite[Chapter~11.2.3]{preece2015human}, using a cognitive walkthrough approach followed by open-ended interviews with the testers are described further in this chapter. User requirements 1 through 10 were identified after validating with the testers. The requirements are listed in Table \ref{table:requirementsInLowFid}. The overview of the low-fidelity prototype of the artifact is shown in Figure \ref{appendix:lowPrototype:test_subject}.

\begin{table}[h!]
\caption{User requirements for the artifact found after validating Low-fidelity prototype.}
\label{table:requirementsInLowFid}
\centering
  \begin{tabular}{ |m{7em}|m{27em}| } 
    \hline
      \textbf{Requirement} & \textbf{Name} \\ 
    \hline
    \hline
      UR1 & View an overview of the test case. \\
    \hline
      UR2 & View and download (if available) all the test initiation and test output files related to the test object. \\ 
    \hline
      UR3 & View information on a particular step in the test case. \\ 
    \hline
      UR4 & View graphs of the signals in the test case. \\
    \hline
      UR5 & Search for signals. \\
    \hline
      UR6 & View the stakeholder for the test object. \\
    \hline
      UR7 & View information of the test performers. \\
    \hline
      UR8 & View the Design Verification Method (DVM) for the test object. \\
    \hline
      UR9 & View the requirements for the test object. \\
    \hline
      UR10 & View the ECUs and its associated Bus of the test object used in the test case. \\
    \hline
  \end{tabular}
\end{table}

\hfill

Additional details and rationale for the requirements are discussed in Appendix \ref{appendix:requirements}. Each requirement laid a foundation to create a better artifact, compared to the existing testing system. Hence, the requirements were validated and mapped against the criteria(Decision Models, Information, and System Quality) discussed in Table \ref{table:criteria}, and the targeted areas of improvement. 

\hfill

During the cognitive walkthrough, testers' actions and decision making styles were observed with regard to the decision making models and the system and information quality. The requirements were mapped with the criteria throughout the observation of the cognitive walkthrough. Subsequently, the mappings were discussed and verified with the testers during the interview session. The testers were informed about the mappings, and, based on their feedback and explanation, the mappings were improved. The mappings were essential for improving the prototype because they help ensure improvements in system and information quality and testing efficiency.

Each of the requirements was mapped to the targeted areas for improvement to understand the importance and rationality of building a better testing system. The mappings are listed in the Table \ref{table:mappingRequirementsWithImprovmentAreas}.

\hfill

\begin{table}[h!]
\caption{User requirements for the artifact found after validating High-fidelity prototype.}
\label{table:requirementsInHighFid}
\centering
  \begin{tabular}{ |m{7em}|m{27em}| } 
    \hline
      UR11 & View the details of the software and hardware read-outs. \\
    \hline
      UR12 & View the the calibration and configuration of the test object for the test case. \\
    \hline
      UR13 & View historical data for a test object. \\
    \hline
      UR14 & View and add favorite signals. \\
    \hline
      UR15 & View and add comments on signals. \\
    \hline
  \end{tabular}
\end{table}

\hfill

The next step for the prototype development was to create a high-fidelity prototype \cite[Chapter~11.2.4]{preece2015human} that would look more like the final product. Hence, this provided more functionality than the low-fidelity prototype, and the stakeholders could interact with it more freely. The high-fidelity prototype was developed in an interface design application called Figma\footnote{https://www.figma.com} as it was quick to modify after each evaluation. Once the Figma representation of the prototype was completed, it was later transformed into a web-based system for the final evaluation, developed in a JavaScript library called React\footnote{https://reactjs.org} to ensure usability. The artifact was validated once again with the testers for feedback on how well the artifact represented the existing requirements (see Table \ref{table:requirementsInLowFid}) and the improvement areas. The testers' feedback included a few more requirements that focused on working with the sensor signals and keeping documentation on them to archive important findings. The requirements from the second validation have been listed in Table \ref{table:requirementsInHighFid}. Additional details and rationality for the requirements are mentioned in Appendix \ref{appendix:requirements}. Those requirements were also mapped with the decision making models, information quality concerns, and system quality concerns, and during the interview session, the requirements were validated. 

\begin{table}[h!]
\caption{Mapping of User Requirements with Improvement Areas.}
\label{table:mappingRequirementsWithImprovmentAreas}
\centering
  \begin{tabular}{ |m{14em}|m{20em}| } 
    \hline
      \textbf{Improvement Area} & \textbf{Requirements} \\ 
    \hline
      Data Presentation  & UR1, UR3, UR4, UR5, UR10-UR13 \\
    \hline
    \hline
      Data Structuring  & UR2, UR8, UR9, UR14, UR15 \\
    \hline
    \hline
    Testing Process  &UR1-UR10 \\
    \hline
    \hline
       Communication  & UR6, UR7 \\
    \hline
  \end{tabular}
\end{table}

\hfill

% Adapt the requirements we got to Volvo -> Signals, DVM etc.
\subsection{Applying the New System}
\label{section:applyNewSystem}

%The requirements defined in Section \ref{section:designPhase}
The artifact had to be adapted to the requirements to ensure practical usability. As the artifact was designed to be a tool for the testers to perform testing on test objects, it required a representation of a test case as a reference point. 

\hfill

The test case included data such as information about the test and the results and data related to the test object and its requirements. Availability for the required and related data for performing the test diagnostic was mandatory in the new system. Hence they were gathered from different repositories and storage within the partnering company. The adaptation of the old system followed this process:

\hfill

\begin{enumerate}
    \item Choose a test case as a reference point.
    \item Collect test related files (configuration information, log-files, test requirements) and post-test data from the test case.
    \item Collect similar test cases manually, by referring to Elektra as historical data.
    \item Collect information about the stakeholders related to the test case and test object.
    \item Adapt the collected data and information to the new system.
\end{enumerate}

\hfill

\iffalse
\begin{table}[h!]
\caption{Data aspects to adapt for the artifact.}
\label{table:aspect}
\centering
    \begin{tabular}{ |m{16em}|m{18em}| } 
        \hline
        \textbf{Aspect} & \textbf{Description} \\ 
        \hline
        \hline
        Requirements & Test requirements. \\ 
        \hline
        Configuration Information & Tool specific setting preferences and files for Test object, car, database, software and hardware. \\ 
        \hline
        Sensor Signal & The sensor signal outputs from the ECUs \\ 
        \hline 
        Historical Test Data & Data and test results for similar test cases. \\ 
        \hline
        Related Stakeholders & Stakeholders that have direct/indirect involvement with the test object. \\ 
        \hline
    \end{tabular}
\end{table}
\fi

% Iterative talk about the different evaluation phases for the system (Feedback from two testers, *), how was the evaluation done (Describe in general)
% Cognitive Walk-through, Feedback
\newpage
\section{Evaluation of the New System}
\label{section:evaluationOfTheNewSystem}
The new system was evaluated iteratively in each of the design science phases. Testers were considered the ideal choice for assessment due to their substantial involvement in the testing process. Two testers were selected for the evaluation. One was a highly experienced tester who had been working with testing for many years, and the latter was a fairly inexperienced tester who had been in this role for a year. The two testers were used strictly for evaluation purposes only and did not participate in the usability test described in Section \ref{section:usabilityTesting}, to avoid biases. The evaluation of the new system was reviewed with them using two methods, involving a cognitive walkthrough \cite[Chapter~15.2]{preece2015human} for the system's usability and semi-structured interviews for feedback. The semi-structured interview was conducted directly after the cognitive walkthrough to investigate if the new system was useful for test analysis. The interview guide can be found in Appendix \ref{appendix:interviewEvaluationPhase}.

\hfill

The cognitive walkthrough involved conducting tasks by the participants in the new system. The method was used to find issues in the design, structure, and presentation of the new system. The tasks were designed to help find the necessary information and data in the prototype that would help test analysis and diagnostics. 

The tasks and their steps that were used in the evaluation phase can be found in Table \ref{table:cognitiveWalkthroughTasksForLowFiPrototypes} for the low-fidelity prototype and in Table \ref{table:cognitiveWalkthroughTasksForHighFiPrototypes} for the high-fidelity prototype. For the task analysis, four common cognitive walkthrough questions were used. They were as follows \cite{blackmon2002cognitive}:

\hfill

\begin{itemize}
    \item Will the user try to successfully achieve the outcome for a specific task (testing tasks)?
    \item Will the user notice the correct action available for selection?
    \item Will the user understand the action and its outcome as expected?
    \item Will the user get the feedback they expected?
\end{itemize}

\hfill

The evaluation feedback was used to improve the existing requirements while adding new ones that were used to improve the new system. The methods described in Section \ref{section:designNewSystem} and \ref{section:applyNewSystem} were performed three times in different design phases for evaluation and development of the new system, which was enough to achieve a consensus between the two testers. 

\newpage
\begin{table}[H]
\caption{Cognitive Walk-through Tasks For Low-Fidelity Prototypes.}
\label{table:cognitiveWalkthroughTasksForLowFiPrototypes}
\centering
    \begin{tabular}{ |m{14em}|m{19em}| } 
        \hline
            \textbf{Task} & \textbf{Task Break Down} \\ 
        \hline
        \hline
            Get information about the test case. & \hfill
            \begin{enumerate}
                \item Open the system.
                \item Go to the Overview.
                \item Get general information.
                \item Get Stakeholder and Tester information.
                \item Get hardware and software information.
            \end{enumerate}  \\
            
        \hline
           Get information about the requirement and DVM of the test case. & \hfill
            \begin{enumerate}
                \item Open the system.
                \item Go to DVM.
                \item Get description.
                \item Get test sequence.
            \end{enumerate} \\ 
        \hline
            Get information about a specific step. & \hfill 
            
            \begin{enumerate}
                \item Open the system.
                \item Go to a step number.
                \item Get the sensor signals involved in that specific step.
                \item Read the description of the specific step.
                \item Compare the sensor signals outputs in the graph.
            \end{enumerate}   \\
         \hline
            Get information about test cases which are similar with the current test case. & \hfill
            
             \begin{enumerate}
                \item Open the system.
                \item Go to Historical Data.
                \item Get the list of similar test cases.
            \end{enumerate} \\
        \hline
        
    \end{tabular}
\end{table}

\newpage
\begin{table}[H]
\caption{Cognitive Walk-through Tasks For High-Fidelity Prototypes.}
\label{table:cognitiveWalkthroughTasksForHighFiPrototypes}
\centering
    \begin{tabular}{ |m{14em}|m{19em}| } 
        \hline
            \textbf{Task} & \textbf{Task Break Down} \\ 
        \hline
        \hline
            Get an overview of the test case. & \hfill
           \begin{enumerate}
            \item Open the system.
            \item Go to Overview.
            \item Get the results of the test case.
            \item Get an overview of the steps in the test case.
            \item Get the sensor signals used in the test case.
            \item Get the buses involved in the test case.
        \end{enumerate}  \\
            
        \hline
           Get information about the requirement and DVM of the test case. & \hfill
            \begin{enumerate}
                \item Open the system.
                \item Go to DVM.
                \item Get description.
                \item Get test sequence.
            \end{enumerate} \\ 
        \hline
            Get information about test cases which are similar with the current test case. & \hfill
            
             \begin{enumerate}
                \item Open the system.
                \item Go to Historical Data.
                \item Get the list of similar test cases.
            \end{enumerate} \\
        \hline
        
            Get the information related to the test case. & \hfill
            
            \begin{enumerate}
            \item Open the system.
            \item Get the software and hardware information.
            \item Get the car information, such as database and configurations.
            \item Get the tester information.
            \item Get the stakeholder information.
        \end{enumerate} \\
        \hline
         Get the links to the file-repository of the log-files used in the test case. & \hfill
            
            \begin{enumerate}
            \item Open the system.
            \item Go to link to log files.
        \end{enumerate} \\
        \hline
            Get information about a specific step. & \hfill 
            
            \begin{enumerate}
                \item Open the system.
                \item Go to a step number.
                \item Get the sensor signals involved in that specific step.
                \item Read the description of the specific step.
                \item Compare the sensor signals outputs in the graph.
            \end{enumerate}   \\
         \hline
    \end{tabular}
\end{table}

% https://link.springer.com/content/pdf/10.1007/BF02299683.pdf
\section{Usability Testing}
\label{section:usabilityTesting}
A usability test was conducted as a final evaluation for the artifact. A usability test was described as a process of involving users to evaluate a system to ensure that it meets usability criteria \cite{barnum2010usability}. Dumas and Redish described usability as a systematic way of observing actual users trying out a product and collecting information about how the product was easy or difficult for them to use \cite{dumas1993practical}. The following characteristics, as explained by Dumas and Redish were taken into consideration when performing the usability test \cite{dumas1993practical}.

\hfill

\begin{itemize}
    \item The primary goal was to improve the usability of the artifact.
    \item The participants represented real users.
    \item The participants performed real tasks.
    \item We observed and recorded what the participants did.
    \item We analyzed the data, diagnosed the real problems, and recommended changes to fix those problems.
\end{itemize}

\hfill

The goal of the usability test was to measure how much improvement was achieved with the artifact over the existing process of testing. A usability test using the think-aloud protocol \cite[Chapter~7.6]{preece2015human} was conducted on six testers where they were required to perform tasks related to testing using the new system and the old system. These six testers were different from the ones in the cognitive walkthrough, which was conducted in the previous evaluations. The think-aloud protocol involves participants verbally saying what they thought while performing specific tasks. The think-aloud protocol was used as it allowed us to understand the participants' way of thinking and the actions they took during the usability test.  
% describe briefly what levels of tasks in general testers solve for
% testers definition of easy medium and hard level task
\subsection{Defining Tasks}
\label{definingTasks}
Tasks from the perspective of test diagnostics and post-test-data analysis centered around the necessary work done when analyzing failed test cases. Hence, according to the testers, the common tasks that needed to be performed consisted of the following: 

\hfill

\begin{enumerate}
    \item \textbf{Understanding the issue}: Read and understand the requirements and define DVM for the test case.
    \item \textbf{Gathering information and data}: Collecting all the necessary output files generated from the test case after completion and the configurations information to be able to analyze the files (i.e., log files, database, configuration and setup information). 
    \item \textbf{Finding the issue}:
        \begin{enumerate}
            \item Readying the tools with the data found from task 2.
            \item Creating an overview of all the data, in the tools/tool, to find the root cause.
            \item Finding the root cause for the failing test case. Analyzing output files and its sensor signals. Figuring out which faulty sensor signals are associate with the failing test case. Identifying any other sensor signals that might have caused additional issues.
        \end{enumerate}
    \item \textbf{Reporting the issue}: Reporting the issues to the correct test object owner or stakeholder.
\end{enumerate}

\hfill

For usability testing, the tasks mentioned above were performed on failing test cases. To reduce learning effects, the test cases were different for each system but same in difficulty and complexity level. The performance of the tester was measured in regard to task completion in both the existing testing system and the new system. The tasks were associated with the requirements of the new system (see Section \ref{section:designNewSystem}), which were used for evaluation of the requirements of the new system. 

%needs rephrasing... old content below
% Why selecting medium level tasks
%collection of all medium level tasks
%verifying wether they have equal difficulty 
% selecting one from the collection randomly using a random number generator using background noise.
\subsection{Test Case Selection Strategy}
\label{testCaseSelectionStrategy}

For the usability test, failing test cases were selected to portray a real problem that needed to be solved. The test cases chosen represented various difficulty and complexity levels. The difficulty was determined based on an increasing number of steps in the test case and how complicated each step was. For example, the quantity of test initiation steps, values from the sensor signal data, mathematical calculations to derive sensor values, were all added to the difficulty of solving test cases. The complexity was defined by the increase in difficulty caused by the involvement of multiple test objects in test cases.

\hfill

According to the testers, test cases were categorized into three sets of difficulty: easy, medium, and hard. This research focused on medium difficulty test cases only during usability testing because, on average, testers faced medium level difficulty test cases the most. A guideline for selecting medium levels test cases was created with the testers to negate its influence on the usability test. A compendium of medium level test cases was made, and from that, two tests were randomly chosen to be used for the usability test. The reason for choosing two test cases was to minimize learning effects from completing tasks on different systems. The two randomly selected test case was verified for its difficulty level during the pilot test.

%talk about how testers were found 
% reason for grouping them
% experience level considered
% strategy for grouping, maintaining equality of both types of testers
\subsection{Test Subjects}
The test subjects were testers with different testing domain experiences, ranging from less than a year to more than two years. A survey helped determine the prospective subjects and their experience level. Two groups of testers were made from the total respondents, having assigned an equal number of experienced and less experienced testers randomly. Three testers were randomly selected from each of the two groups, making six test subjects in total. The overall selection process can be seen in figure \ref{fig:test_subject}. The survey can be found in Appendix \ref{appendix:questionnaireSubjectUsabilityTest}.

\hfill

The rationality behind having the number of subjects was because, in previous studies, 80\% of the products usability problems could be identified by having 4 to 5 test subjects \cite{turner2006determining}. 

\begin{figure}[ht!]
  \centering
  \includegraphics[scale=0.5]{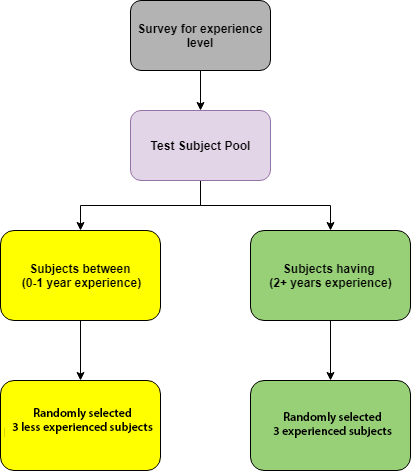}
  \centering
  \caption{Test subject selection methodology.}
  \label{fig:test_subject}
\end{figure}

\subsection{Measurements}
To evaluate the artifact, subjective and objective measuring techniques were applied. The subjective measure was commonly associated with benchmarking tasks, while the objective measure was commonly associated with a user questionnaire \cite{hartson2012ux}. 

\subsubsection{Subjective Measures}
\label{subjectiveMeasures}
To investigate both the old system and the new artifact, in regards to the effort while performing tasks, NASA Task Load Index\footnote{https://humansystems.arc.nasa.gov/groups/TLX/downloads/TLXScale.pdf} was used. The NASA TLX helped to identify workload for the tasks \cite{hart1988development}. The think-aloud protocol was also used to identify problem areas when performing the tasks followed by interviews after completing the test for further feedback. The interview strategy employed was to gather data about what aspects of the new system led to improvements in efficiency and effort. The interview guide can be found in Appendix \ref{appendix:interviewUsabilityTest}. The subjective measurements can be seen in Table \ref{table:subjectiveMeasurment}.

\hfill

As the NASA Task load index was primarily a subjective measure, it focused on calculating a workload index based on different factors. According to the NASA Task Load Index manual, previous studies revealed that workload patterns depended on different tasks and the test subjects' unique experience. Effort, stress, and frustrations were some of the highlighted factors and based on these factors, the workload was determined and measured on a linear scale. The factors, according to the NASA Technical Report \cite{NASATLX_1986} were as follows:

\hfill

\begin{itemize}
    \item \textbf{Mental Demand: } The amount of mental activity required for the task (Thinking, deciding, calculating, and searching). 
    
    \item \textbf{Physical Demand: } How much physical activity was required while performing the task. 
    
    \item \textbf{Temporal Demand} How much time pressure was the person under while performing the task (slow-paced or rapid).
    
    \item \textbf{Performance: } How successful the person was in accomplishing the goals of the task.
    
    \item \textbf{Effort: } How difficult it was to complete the task in terms of mental and physical work.
        
    \item \textbf{Frustration Level: }  How insecure, discouraged, irritated, stressed, and annoyed versus secure, gratified, content, relaxed, and complacent the person felt while performing the task.
    
\end{itemize}

The goals of using the NASA Task Load Index were to identify the impact of such factors on the testers' performance of tasks using both the old and the new system. Such findings would help evaluate the impact of the new system on the efficiency of the testing process.
\hfill

\begin{table}[h!]
\caption{Subjective measures and techniques for the usability test.}
\label{table:subjectiveMeasurment}
\centering
    \begin{tabular}{ |m{14em}|m{20em}| } 
        \hline
        \textbf{Subjective Measure} & \textbf{Techniques} \\ 
        \hline
        \hline
        Task performance and effort index &  NASA Task Load Index and Interview\\ 
        \hline
        Usability & Think-aloud protocol and Interview \\ 
        \hline
    \end{tabular}
\end{table}

\hfill

\subsubsection{Objective Measures}
\label{objectiveMeasures}
An objective measurement was used to measure benchmarks regarding the tasks the subjects performed in the usability test. The time needed to complete the task (time-on-task) was taken into account.

\begin{table}[h!]
\caption{Goals to investigate in the experiment}
\label{table:objectiveMeasurment}
\centering
    \begin{tabular}{ |m{14em}|m{20em}| } 
        \hline
        \textbf{Objective Measure} & \textbf{Techniques} \\ 
        \hline
        Time on task  & Time in seconds to complete the task \\ 
        \hline
    \end{tabular}
\end{table}

\subsection{Usability Test Design}
Test subjects were allocated individually in an office room consisting of two moderators. One moderator hosted the test and recorded error rates while the other served as a timekeeper. The moderator explained that the task would be needed to be completed using both systems by using the think-aloud protocol throughout the completion of the task. The think-aloud protocol was breifly explained to the testers.

\hfill

\begin{figure}[ht!]
  \centering
  \includegraphics[width=\textwidth]{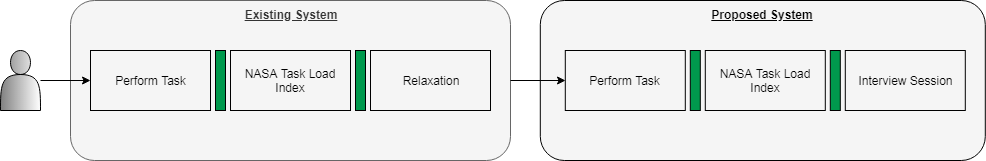}
  \centering
  \caption{The usability test design.}
  \label{fig:usabilityDesignMethod}
\end{figure}

The subject was given a randomly selected test case and was asked to perform the four tasks, mentioned in Section \ref{definingTasks} that testers did in the testing process on both the systems. The timekeeper started the timer when the subject initiated the task. In case the subject forgot to talk while they performed, the moderator reminded them to do so. When the task was completed, the subject was asked to fill the NASA Task Load Index form. After a relaxation break, the subject was asked to perform the series of tasks again, but on a different test case using the other system, which was followed by filling the NASA task load index form. A  semi-structured interview was done to gain insights on their experience with the systems when the usability test was finished. Details regarding the interview design are described below in Section \ref{userInterviewsUsability} The entire usability process is depicted in Figure \ref{fig:usabilityDesignMethod}.

\subsection{User Interviews}
\label{userInterviewsUsability}

The interviews were designed to address the test-subjects' feedback and answer the research questions through the qualitative explorative manner. These semi-structured questions were designed using the research questions as the focus. The interview question guide can be seen in Appendix \ref{appendix:interviewUsabilityTest}. The criteria addressed in the interviews were as follows:

\hfill

\begin{itemize}
    \item Comparison of both the old and the new system regarding usefulness, efficiency, and effort spent while performing the tasks.
    \item Data structure and presentation for diagnosing test-related data.
    \item Impact on decision making capabilities.
    \item Ease of use of the new system.
\end{itemize}

\subsection{Pilot Tests}
A pilot test had to be done to confirm if the usability test design and the process would work with the test subjects in the actual test. Furthermore, it was vital to validate if it was feasible to solve the task and correctly measure the measurements. One experience test subject was selected for the pilot test. Both the test environment and the process were mimicked to feel like the real test. The insights and feedback collected during the pilot test were used to improve the usability test design and process and reduce inaccuracy and flaws.

%% file: include/Results.tex
% CREATED BY DAVID FRISK, 2016
\chapter{Results}
\label{chapter:results}
%Describe you results. Use tables, diagrams etc. for illustration.
Based on the Design Science Research methodology, the results in this chapter have been divided along with the three phases. The problem identification phase results were described first, followed by the design phase, and finally, the evaluation phase presents the design's validation. The discussion of each phase consists of results found using various data collection methods such as interviews, questionnaires, observation, and methodologies such as affinity diagramming, prototyping, and a usability test.

\section{Problem Identification}
\label{section:problemIdentification}
This section will go through the results found using interviews and questionnaires. During the initial phase of the study, multiple semi-structured interviews were performed with stakeholders, including a product owner, tester, test environment provider, and a comprehensive questionnaire to target a larger group of testers. The aim was to understand the existing process of testing and analyzing test output data. 

\begin{figure}[H]
 \centering
 \includegraphics[width=\textwidth]{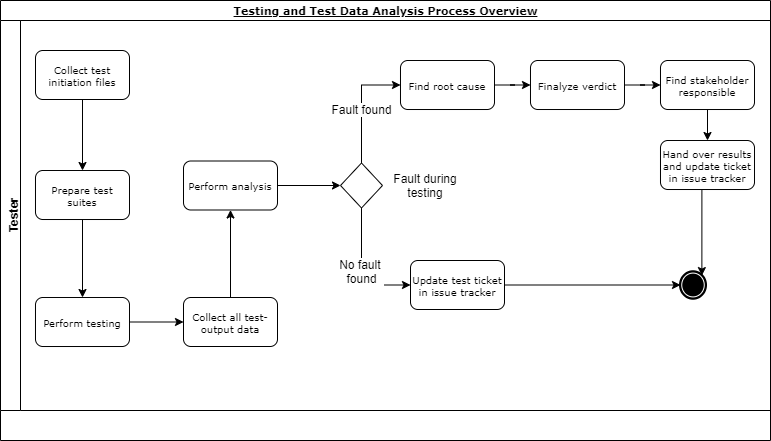}
 \centering
 \caption{Existing testing and test output data analysis process overview.}
 \label{fig:existingTestingProcessOverview}
\end{figure}

\newpage
\subsection{Interviews with Product Owners}
During the initial interview with the product owners, the existing process and protocols for testing and test output data analysis were discussed. The product owner highlighted the issues with the current process and how the problems began emerging. The product owner explained how the testing process was conducted in the partnering company specifically because he wanted to distinguish between some of the company's unique practices compared to the traditional way of managing test-related data.

\begin{table}[H]
\caption{test output files and data}
\label{table:testOuputDataFiles}
\centering
  \begin{tabular}{{ | p{2.5cm} | p{9cm} | p{2.5cm} |}} 
    \hline
      \textbf{File Name} & \textbf{Description} & \textbf{Storage } \\ 
    \hline
    \hline
    Topology File & Contained the schematics of the test object and its connection in modules of a car.  It was used during test analysis. & Shared Disk and Local Repository \\ 
    \hline
    Binary Logging File(BLF) & Contained data for events that were captured during testing of the test object. Data logged from test objects were stored here, consisting of messages from CAN busses. It is used during test analysis. & Local Repository \\ 
    \hline
    Master Database File(MDF) & Mostly contained logged signals with time-stamp from the test object and other devices attached to the bus. It was used during test analysis. & Local Repository \\ 
    \hline
    Logging File(MF4) & Similar to MDF but redundantly kept as a log. & Local Repository \\ 
    \hline
    Configuration File(CFG) & Contained configuration settings of test objects for conducting tests. CFG files were vital for performing tests. It was used during test initiation. & Storage Server \\ 
    \hline
    Calibration File(CLG) & It consisted of settings for the test bench environment and held vital information of the test object and other objects in the test environment. The calibration file was used to initiate the test environment for testing along with the CFG file. & Local Repository \\ 
    \hline
    Database File & Consisted of the decrypted signal name for the signal database. Used by testers when reading the MDF and BLF files using the analysis tools. & Local Repository \\ 
    \hline
    Design Verification Method (DVM) & Consisted of the detailed test cases for test runs. The file also held crucial information about each step in the test-run and how it was performed during testing. & ELEKTRA (Third Party Software Suite) \\ 
    \hline
    Requirement & The file contained all the requirements for the test object's functionalities and was used during the testing process & ELEKTRA (Third-party software suite)\\ 
    \hline
    
  \end{tabular}
\end{table}

\newpage
Testing ECUs in the department's context produced a large quantity of data as testing devices continuously monitored the ECUs. There were cases when multiple ECUs were unnecessarily logged, creating an abundance of data that was later used to perform analysis. Such data was not structured efficiently and kept in varying file formats, which are explained in Table \ref{table:testOuputDataFiles}, it became harder to locate faults, efficiently make decisions that mitigate problems and share knowledge with the involved stakeholders. 

\hfill

\begin{mdframed}[style=MyFrame]
\begin{quote}
    \begin{itshape}
    "In our department, testing produces an immense amount of logging of data, and you know what, sometimes, un-necessary ECUs are logged to make sure that they were not part of the problem when testing the test object. All this produces huge data, and it is being stored in so many different places."
    \end{itshape}
    - Product Owner.
\end{quote}
\end{mdframed}

\hfill

Each test object had many files associated with it, and during the testing phase, it was crucial to have most of the files to perform testing. Files such as a CFG and a database were essential to initiate the testing process. They were not stored in a centralized system, making it inconvenient for testers to find and collect them for analysis.  There was a significant amount of time spent doing this, and this task had no tangible benefit in making the testing process better. 

\hfill

\begin{mdframed}[style=MyFrame]
\begin{quote}
    \begin{itshape}
    "Testers need to find all the config, database, and other information to initiate the testing, which is a hassle, as all the files are not exactly in the same place, making it harder for the testers to collect all the required files quickly."
    \end{itshape}
    - Product Owner.
\end{quote}
\end{mdframed}

\hfill

The product owner described test analysis as a hectic situation. As all the information and data describing the test object were stored in various files, they needed to be collectively visualized using different tools during the analysis process. MDF and BLF files (produced from testing) were the primary files required for analysis, and CFG and Database files were needed to set-up the tools used to perform analysis (see Table \ref{table:testOuputDataFiles}). Furthermore, testers had to work with at least three different tools to work with additional files. They had to frequently navigate between the tools during test data/information analysis, which was time-consuming and prone to introducing mistakes. Testers admitted to missing out on information due to the constant navigation.

\newpage
\begin{mdframed}[style=MyFrame]
\begin{quote}
    \begin{itshape}
    "Testing tools like Elektra, CANape, CANalyzer are needed to be open all the time.  I have seen testers jumping from screen to screen to find the relevant information for the failed test cases. Jumping so much can also make them miss out on certain information and loose concentration."
    \end{itshape}
    - Product Owner.
\end{quote}
\end{mdframed}

\hfill

Many stakeholders had ownership of a single test object. Ownership can be described as functional responsibility for the software or hardware of the ECU. After testing, the testers were responsible for conveying the test results to the correct stakeholder to solve problems with the test object. As there were no guidelines in the testing process for locating the proper person to get the solution to the problem quickly, testers ended up spending much time looking for them either physically, asking around within or between teams and departments, or using messaging.

\subsection{Interviews with testers} \label{testerInterview}
Upon interviewing the testers, more in-depth details of the process and its problems were identified. The tester mentioned not having a standard for storing the various files for the test initiation and test output data analysis. 
As different files held different types of data required for analysis, having all of them at the tester's disposal was vital. However, due to them not being stored in a centralized storage system with a standard method of storage, it took much time and effort to collect them. The files and their storage locations have been described in Table \ref{table:testOuputDataFiles}. 

\hfill

\begin{mdframed}[style=MyFrame]
\begin{quote}
    \begin{itshape}
    "Time spent collecting all the test initiation and later, test output files is time-consuming and does not deliver value to our work. We want to analyze and find issues in the tests because then, we actually do something productive."
    \end{itshape}
    - Tester.
\end{quote}
\end{mdframed}

\hfill

%this passage talks about the test output files, where they are stored and what are they.
% also talks about how abundance of data takes time to work with and makes the analysis error prone. as in tester may make mistakes by disregarding data which was vital.
As the test objects were primarily ECUs, they produced a large amount of data. In the partnering company's context, the testers had to monitor multiple ECUs despite them not being part of the test process. In the worst-case scenarios, additional ECUs and the test object became part of the tests. The testers emphasized that logging multiple ECUs and the test object produced many log data and files. Logs were kept in different file formats and under various storage systems, making retrieval of the files for analysis time-consuming and challenging job. The storage systems for storing data included repositories, servers, shared disks, and local storage like personal computers, loosely associated with their corresponding test objects. Apart from the DVM file which contained the test cases and steps, the Requirement file, and the Topology file, data in all the other files were large in quantity and not user-friendly in terms of readability, requiring significant effort to find the relevant data needed for test analysis. The testers also mentioned that problem identification became overwhelming and challenging due to not focusing on the essential aspects of the test analysis because of the abundance of data. The analysis became time-consuming and would occasionally cause testers to disregard essential and relevant information mistakenly. 

\hfill

\begin{mdframed}[style=MyFrame]
\begin{quote}
    \begin{itshape}
    "The data scattered around in front of you, constantly looking for things in different files, causes an expenditure in terms of time. It becomes tedious to skim through so much unstructured data. To come up with a conclusion(issues found in testing) on the problem is not impossible for the testers because they are capable of their job, but this makes it tedious us."
    \end{itshape}
    - Testers.
\end{quote}
\end{mdframed}

\hfill

%talks about tools, issues with having many tools
Another issue identified was that the testers used multiple tools to analyze the test output data, making test analysis cumbersome. Their explanation noted that different test-data files had different sets of information, having different standards and notations. Hence, different tools were needed. The tools used to process the files mentioned in Table \ref{table:testOuputDataFiles} have been described below in Table \ref{table:interviewFindings1}.

\begin{table}[H]
\caption{Software/Tools used for test output data diagnosis}
\label{table:interviewFindings1}
\centering
  \begin{tabular}{ |m{14em}|m{20em}| } 
    \hline
      \textbf{Software/Tool} & \textbf{Usage} \\ 
    \hline
    \hline
      CANoe \cite{vector} & Used to read BLF and MDF files. It also has advanced features like simulation for an entire CAN/Flexray bus missing in CANalyzer. \\
    \hline
    \hline
      CANalyzer \cite{vector} & Used to read BLF and MDF files. The tool was used by both experienced and inexperienced testers, however experienced testers prefered CANoe due to its simulation features.\\ 
    \hline
    \hline
      ELEKTRA \cite{vector} & Used to read the DVM and its details. It was also used to manually log findings related to test analysis.\\ 
    \hline
  \end{tabular}
\end{table}

\hfill

Each file held different, yet vital, information for the test object, the files needed to be visualized through various tools. ELEKTRA was used to visualize information from the DVM, test requirements, and test cases. The tool did not help visualize all of this information optimally and usually required testers to spend effort and time managing the tool's complicated settings and interfaces. CANoe and CANalyzer were used to read the BLF/MDF files to visualize the sensor signal data.

\hfill

The use of multiple tools led to challenges in how to make sense of the data. Testers noted that their concentration was interrupted when they had to relate data from the various files while working with all the tools simultaneously. It was standard practice for testers to have all the tools open in different windows in their computer while navigating between them to get the required information.

\hfill

\begin{mdframed}[style=MyFrame]
\begin{quote}
    \begin{itshape}
    "Decision making regarding finding issues in tests gets delayed when testers need to find the right things to look at and prepare for everything before it can be visually represented properly. Unless you know what to look for, especially for new testers, the task will be challenging."
    \end{itshape}
    - Testers on getting an overview from tools.
\end{quote}
\end{mdframed}

\hfill

The initial coding of the tester interview with all the major findings are categorized and coded in Appendix \ref{appendix:codedInterviews:testerGroup1Interview} and \ref{appendix:codedInterviews:formerTestersInterview}.

\subsection{Interviews with Test Environment Providers}
Two test environment providers were interviewed. They also confirmed the problems with managing unstructured data. The lack of refinement of the existing testing process and its implementation caused technical debts, which caused issues led to data storage, structure, and management. They also noted that, in the existing testing process, test cases were written without maintaining a strict standard which made it difficult for testers to perceive the test cases in the correct way. Depending on the experience of testers, test cases were written in varying ways. Sometimes, experience testers omitted to write certain steps because, for them, they were unnecessary. The omission caused the inexperienced testers difficulties in understanding the test cases and their scenarios and made it difficult for them to run the test cases and perform test analysis. No additional findings emerged from the interview with the test environment providers.

\hfill

\begin{mdframed}[style=MyFrame]
\begin{quote}
    \begin{itshape}
    "In most cases testers belong to teams, and when developers create the functionality and they then say that as this functionality is made, can the tester test it? Then the testers in a team writes the DVM(test cases), and then when the test is performed, there are lost of files and testers have to find them all and find a verdict. Basically, we are trying to automate this so that the testers no longer need to manually test them."
    
    \hfill
    \end{itshape}
    -  Feedback on existing testing process.
    
\hfill

\end{quote}
\begin{quote}
    \begin{itshape}
    "Also DVM is written by humans and it is written with some slack/fussiness which makes tester try some things like figuring out the intention behind the test case. DVM is written manually, hence it is interpreted different by different testers and in worse way, the test might pass, but as testers interpreted the DVM differently, in reality the test should have failed. So the existing testing system is not strict in terms or process."
    
    \hfill
    \end{itshape}
    - Feedback on existing testing process.
\end{quote}

\end{mdframed}

\hfill

\subsection{Summary of the Interviews} \label{summaryOfInterviews}
A summary of the main findings from the initial round of semi-structured interviews is briefly described below. The entire testing process specific to the partnering company was depicted in Figure \ref{fig:existingTestingProcessOverview}. The diagram depicts the testing process the testers perform based on the information collected from the interviews, which is as following: 

\hfill

\begin{itemize}
  \item Product Owners
    \begin{itemize}
      \item The existing testing process results in unstructured data and leads to the logging of unnecessary data.
      \item The files that described and held information for the test object were scattered across various storage systems.
      \item Accumulation of all the associated files for the test object took time and effort.
      \item Testers were required to navigate between testing tools as not all information could be found using just one, which interrupted their work and information flow.
      \item It was difficult to share analysis results with all the respective stakeholders.
    \end{itemize}
    
  \item Testers
    \begin{itemize}
      \item Too many files are associated that associate with the test object (see Table
      \ref{table:testOuputDataFiles}).
      \item Accumulating all the test object files was time-consuming and error-prone due to files being stored in various storage systems.
      \item A large quantity of sub-optimally structured data was stored in the files, preventing testers from finding the correct or important information quickly.
      \item Different tools were simultaneously used to analyze the test output data, causing testers to constantly shift attention in order to process, understand, and make decisions from the information. 
      \item Test output, especially important findings from the testers, were not archived systematically, causing loss of vital information.
      \item It was not easy to locate or discuss test analysis results with stakeholders.
    \end{itemize}
    
  \item Test Environment Providers
    \begin{itemize}
       \item The testing process led to unstructured data, causing difficulties.
       \item DVMs are not written in a standard way, which makes testers misinterpret the test cases. 
    \end{itemize}
\end{itemize}

\newpage
\subsection{Tester Questionnaire} \label{questionnaire}
A questionnaire was prepared to investigate further the claims made in the interviews at a larger scale and can be seen in Appendix \ref{appendix:questionnaireProblemIdentification}. The questionnaire was organized in the following manner:

\hfill

\begin{itemize}
  \item General questions involving respondent's role
  \item The testing process
  \item Unstructured files and data
  \item Testing process and test output data analyzing issues
  \item Analytic tools or software used by testers
  \item Demographics
\end{itemize}

\hfill

In total, thirteen responses were received for the questionnaire. All thirteen respondents had prior experience within the testing process in the department. 

%Regarding the migration of testing process from manual to automated, none of the respondents preferred the manual testing process. Nine out of the thirteen however, shared the limiting factors of the automated test system in regards to being able to conduct tests. Not all functionalities can be performed using manual testing, making certain testing jobs difficult. 

Regarding the issues related to unstructured data and files related to the test object, the responses were diverse. Seven respondents suggested that the information that needs to be used for data analysis was difficult to find due to the unstructured data. All respondents suggested that the files' unstructured nature affected their test-analysis and diagnostics process.

\hfill

Regarding issues with test output data analysis, more than two-thirds of the respondents suggested that they required significant time to perform their analysis due to unstructured data and files. The respondents also indicated that finding the right stakeholders to aid with fault mitigation was challenging as many stakeholders were involved with each test object. The respondents also discussed the tools they used for performing diagnosis. All the tools were used in combination, during diagnosis. Navigating between the tools to extract the required information to continue the test analysis was time-consuming and strenuous. Locating faulty signals ended up taking significant time and effort. The main findings from the questionnaire are briefly described below:

\hfill

\begin{itemize}
  \item Unstructured files and data
    \begin{itemize}
      \item Required information about the test object and test output was difficult to find due to the abundance of files scattered in various storage systems.
      \item Gathering all of the required files for the test object took a lot of effort and time. 
    \end{itemize}
  \item Analyzing test output data
    \begin{itemize}
      \item Analysis of test output data tool took a lot of time and effort due to the files' unstructured nature and abundance of data.
       \item Different tools needed to be opened and navigated to create an overview of the test output. 
      \item It was difficult to locate the correct stakeholders to pass on the test results and mitigate the issues found.
    \end{itemize}
  \item Tools used for analyzing data
    \begin{itemize}
      \item No single tool provided an overview of all the necessary information the testers needed to perform testing and identify the root cause for the issues. 
      \item Testers kept all the tools (CANalyzer, CANape and ELEKTRA) open on different screens to get an overview of information.
      \item Navigation between tools was time-consuming, strenuous and affected their concentration.

    \end{itemize}
\end{itemize}

\hfill

\subsection{Participant Observation}
The discoveries made in the observation sessions for analyzing test output data aligned with the findings mentioned in the previous subsections. However, further details were identified during the observation.

\hfill

During the test analysis, the testers collected all the necessary files required for the process. These files included the Master Database File (MDF), Binary Logging File (BLF), and Configuration files, which are explained in Table \ref{table:testOuputDataFiles}. Their frustration during file collection was noticeable as they kept reminding themselves about where the files were stored. Once all the files were found, the testing tools were initialized and made ready to be used for test analysis. Additional time was used in setting up the tools as participants tried various configuration files in the test suite to determine if they had the correct one.

\hfill

After initializing the test suite, the test object, test requirements and test cases (DVM) were reviewed in ELEKTRA. The pre and post-conditions of the failed cases were checked in ELEKTRA, and later, the test output files were processed using either CANoe or CANalyzer, to identify failed test cases.  The testers checked each of the associated signals and ECU for the cases to find potential issues by examining the different analysis tools. They navigated between ELEKTRA and CANoe/CANalyzer to look at the test case, the requirements, and the signal output. They had all three tools open in three different monitors side-by-side and, on a few occasions, lost track of the information they were looking at on different screens. They repeated the process of finding faulty sensor signals whenever they lost their concentration. Once issues and its cause were found, the testers began searching for the stakeholder to solve the potential problem. They manually searched ELEKTRA and test object documentation to find the person involved in developing the feature for the test object. The search appeared to be frustrating for the testers as they kept opening, reading, and closing various documentation before finally locating the person responsible. They then messaged the person and notified them of the results of the test analysis.

\newpage
At the end of the observation session, a short interview was conducted. It confirmed that the most time-consuming aspects of the analysis process were gathering the necessary files from different storage platforms, identifying the cause of the problem from the unstructured data in the files, and finding the right stakeholders to be notified. The main findings from the participant observation sessions are briefly described below:

\hfill

\begin{itemize}
  \item Unstructured files and data
    \begin{itemize}
      \item Much time was spent in gathering all the files related to the test object
      \item Mental effort was spent finding the required data from the abundance of unstructured data as it involved manual searching and extracting the correct signal data.
    \end{itemize}
  
  \item Analyzing test output data
    \begin{itemize}
      \item Participants spent much time in setting up their test suite and environment to be able to visualize the necessary data in a readable and approachable manner.
      \item They had to constantly navigate between different tools to be able to make sense of the data, test cases, and its requirements, which caused loss of concentration and repetition of tasks.
    \end{itemize}

  \item Tools used for analysis
    \begin{itemize}
      \item All the tools (CANalyzer, CANape and ELEKTRA) were kept open on different screens during the testing process.
      \item Navigating between the tools frequently sometimes caused testers to lose concentration. 
      \item ELEKTRA was used to manually refer to documentation to find the responsible stakeholder for the test object, which was time-consuming for the testers.
    \end{itemize}
\end{itemize}

\newpage
\subsection{Categorizing and Organizing the Findings}
The categorization of the findings was done through the affinity diagram approach. The findings were summarized into similar problems and assigned a category. Categories were defined as data presentation, data structure, process, and communication identified from analysis of the significant problems found from the interviews, questionnaire, and observations. The result of the affinity diagram can be found in Table \ref{table:affinityDiagram}.

\hfill

\begin{table}[H]
\caption{Affinity diagram categorization and organization of the results from the initial findings.}
\label{table:affinityDiagram}
\centering
\resizebox{\textwidth}{!} {
  \begin{tabular}{ |m{8.5em}||m{8.5em}| |m{8.5em}| |m{8.5em}||m{8.5em}| }
    \hline
      \textbf{Data Presentation}&\textbf{Data Storage}&\textbf{Data Structure}&\textbf{Process}&\textbf{Communication}\\
    \hline
    \hline
      Relevant data from large files was overwhelming to find and analyze. 
      & No standard storage location for test output files and data. 
      & Test object was represented by many files.
      & Much time was spent on accumulating the files related to the test case/object.
      & Lack of contact information for responsible stakeholders for the test object. 
      \\ 
    \hline
      Lack of presentation for mapping of data. 
      & No standard storage for database, configuration, and calibration. 
      & Each file has many unstructured data according to how the testers would like to see. 
      & Much time was spent in preparing all the tools for analysis. 
      & Knowledge sharing between testers after the analysis was not always archived for historical use.
      \\ 
    \hline
      Required to switch between different interfaces of tools frequently.
      & DVM and requirements stored in another tool. 
      & The test signal output data was extensive in stature, making it difficult to search for issues.
      & Tracking historical data for test case/object was difficult as it was done manually.
      & Navigating constantly between different tools and screens caused misinterpretation and missing out on information. 
      \\ 
    \hline
      Critical information presented for the test case/object scattered across various tools.
      & No efficient way to get the required data for test analysis and diagnostics, which was time-consuming.
      & Historical data for test case/object cannot be tracked easily due to files being unstructured.
      &
      &
      \\
    \hline
    
    \hline
  \end{tabular}
}
\end{table}

\newpage
\section{Design Phase}
\label{section:designPhase}
In this section, the design phase of the artifact is described. Requirements were created from the knowledge gathered during the problem identification phase for the artifact to address problems regarding data structure and presentation. Finally, the artifact was developed into a new prototype system.

\subsection{Requirements}
\label{section:requirements}
The requirements were refined after evaluating the artifact multiple times.  Initial findings from the affinity diagram (see Table \ref{table:affinityDiagram}) were used to create the requirements. Feedback and other insight from early evaluations (see Section \ref{section:evaluationPhase}) improved existing requirements and introduced new requirements. A summarized list of the requirements can be seen in Table \ref{table:requirements}. 

\hfill

\begin{table}[H]
\caption{User requirements for the artifact.}
\label{table:requirements}
\centering
  \begin{tabular}{ |m{7em}|m{27em}| } 
    \hline
      \textbf{Requirement} & \textbf{Short Description} \\ 
    \hline
    \hline
      UR1 & View an overview of the test case. \\
    \hline
      UR2 & View and download (if available) the initiation and metadata files related to the test object and the test case. \\ 
    \hline
      UR3 & View information on a particular step in the test case. \\ 
    \hline
      UR4 & View graphs of the signals in the test case. \\
    \hline
      UR5 & Searching for sensor signals in the test output for analysis. \\
    \hline
      UR6 & View the stakeholder for the test object. \\
    \hline
      UR7 & View information of the test performers. \\
    \hline
      UR8 & View the Design Verification Method (DVM) for the test object. \\
    \hline
      UR9 & View the requirements for the test object. \\
    \hline
      UR10 & View the ECUs and its associated Bus of the test object used in the test case. \\
    \hline
      UR11 & View the details of the software and hardware read-outs. \\
    \hline
      UR12 & View the the calibration and configuration of the test object for the test case. \\
    \hline
      UR13 & View historical data for a test object. \\
    \hline
      UR14 & View and add favorite signals. \\
    \hline
      UR15 & View and add comments on signals. \\
    \hline
  \end{tabular}
\end{table}

\newpage
The requirements UR1-UR10 were created from the affinity diagram to address the problems the stakeholders faced in their current method of test analysis. The requirements UR11-UR15 were created from the feedback on the low-fidelity prototypes. These requirements are features that would be useful for test analysis. In general, the goal with the requirements was to collect all the related data related to the test case and its test object into one system and efficiently present the data for test analysis. The requirements from the initial findings (i.e., UR1-UR10) would correspond to categories from Table \ref{table:affinityDiagram} as follows: 

\hfill

\begin{itemize}
  \item \textbf{Data Presentation}: UR1, UR3, UR4, UR5, UR10
  \item \textbf{Data Structure}: UR2, UR8, UR9,
  \item \textbf{Process}: UR1-UR10
  \item \textbf{Communication}: UR6, UR7
\end{itemize}

\hfill

The user requirement specifications, including description, rationale, dependencies, and use cases, can be seen in the Appendix \ref{appendix:requirements}.  The motivation for the requirements and how it would be useful for test analysis is described in the rationale. The use case describes how the user could use the feature described by the requirement in a web-based system.

% RQ2: How should the design of a test analysis framework for automotive testing reflect the decision making models followed by testers?
% RQ3: How should data be structured, presented, and documented in this test analysis framework to improve the quality of information available during test analysis?
% RQ4: How can this test analysis framework be designed to support ease of use and accessibility of test related data during test analysis?
\subsection{Requirements and Decision Making Models}
Requirements were mapped with the decision making models from Section \ref{section:applyingTheExistingWork} to outline the factors that would be an improvement in efficiency and effort of decision making during test analysis. Mapping of the decision making models and the requirements can be seen in Table \ref{table:requirementCriteria}. The decision making models highlighted the decision making approach people tend to take to complete a task. The decision making models and their mapping with the requirements and criteria follow the thought process and approach of testers to fulfil the tasks that are described by the requirements. Hence, creating this mapping helped design a better prototype for the new system. The mapping of the decision making models with the requirements helped answer the second research question, but at the same time provided inspiration to solve the problems explained in research question three and four. 
\hfill

The mappings were done during the evaluation sessions, where the testers were asked how they would make decisions on performing each requirement. They were explained about the decision making models, and from that knowledge, they were asked which type of decision making model strategy best reflected the way the testers did the tasks to achieve the requirements. For example, if the requirements were mapped to the elimination-by-aspects model or conjunctive model, it meant that those requirements consisted of practices that would make the testers simplify and eliminate less relevant methods of achieving a task to save time and effort. The new prototype should then be designed to ensure that the testers would not spend time with unnecessary methods (like searching for files from different locations), and preferably have all files at a single place instead. If requirements were mapped with the additive/linear model, it implied that the tasks required to achieve the requirement usually involved additively forming an overall conclusion (like understanding an overview of the test cases and test diagnostic result) before performing analysis. 

\hfill

After the mapping, the testers were consulted with the prototype design in regards to the decision making model mapping. The feedback from testers suggested that data and information for requirements that focused on elimination by aspects model would be better placed in the new system where the testers would find it with the least amount of elimination possible. For requirements mapped with additive/linear model and conjunctive model, they preferred the information and data to be categorized in various views in the prototype. However, they wanted it to be logically categorized so that the views(web pages representing the data and information) could be easily accessible without needing to investigate about its whereabouts. Also, because they would require to repeatedly view such information, having a logical separation for each type of information would make it easier for them to locate it. Hence, this led to the new prototype providing information in a user-friendly interface to allow testers to make sense of multiple types of data collectively.

\hfill

\begin{table}[H]
\caption{Mapping between requirement and criteria.}
\label{table:requirementCriteria}
\centering
  \begin{tabular}{ |m{7em}|m{13em}|m{13em}| } 
    \hline
      \textbf{Requirement} & \textbf{Quality Criteria} & \textbf{Decision making Criteria} \\ 
    \hline
    \hline
      UR1 & Information Quality & Conj. Model\\
    \hline
      UR2 & System Quality & EA Model \\ 
    \hline
      UR3 & Information Quality & EA Model \\ 
    \hline
      UR4 & System Quality & EA Model \\
    \hline
      UR5 & System Quality & EA Model \\
    \hline
      UR6 & Information Quality & A/L Model \\
    \hline
      UR7 & Information Quality & A/L Model \\
    \hline
      UR8 & Information Quality & Conj. Model \\
    \hline
      UR9 & Information Quality & Conj. Model \\
    \hline
      UR10 & Information Quality & Conj. Model \\
    \hline
      UR11 & Information Quality & Conj. Model \\
    \hline
      UR12 & Information Quality & Conj. Model \\
    \hline
      UR13 & System Quality & Conj. Model \\
    \hline
      UR14 & System Quality & EA Model \\
    \hline
      UR15 & System Quality & EA Model \\
    \hline
  \end{tabular}
\end{table}

\newpage
\subsection{Design and Development of the New System}
The new system is designed as a web-based system with a dashboard-based user interface. Prototypes were created to evaluate and validate the requirements starting with a low-fidelity prototype and eventually leading to a high-fidelity prototype. 

\subsubsection{Low-fidelity Prototype}
\label{section:lowFidelityPrototype}
An overview of the low-fidelity prototype can be seen below in Figure \ref{figure:lowPrototype:test_subject} and can be seen in more detail in Appendix \ref{appendix:lowPrototype}. The prototype embodies a very abstract form of the requirements where the design principles and interface design are shown. Data and information are not concretely presented in the prototype. The low-fidelity prototype embodies the requirements UR1-UR10.

\hfill

\begin{figure}[ht!]
    \centering
    \includegraphics[width=0.6\textheight]{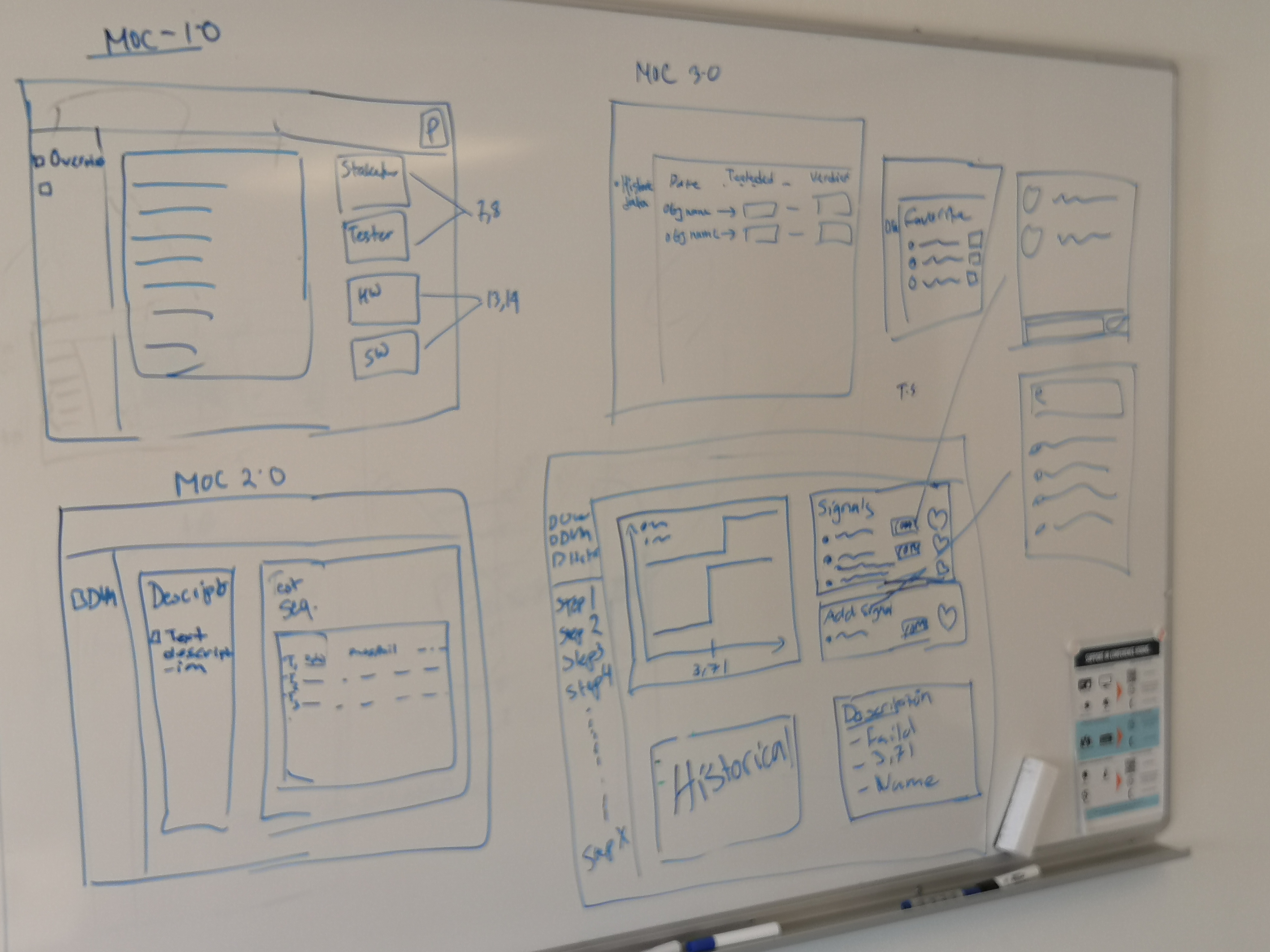}
    \caption{Low-fidelity prototype sketch drawn on whiteboard.}
    \label{figure:lowPrototype:test_subject}
\end{figure}

\newpage
\subsubsection{High-fidelity Prototype}
\label{section:high1FidelityPrototype}
An early high-level prototype can be seen in Figure \ref{figure:high1Prototype_overview}, with additional screenshots in Appendix \ref{appendix:high1Prototype}. The prototype demonstrates how data can be structured and presented. The prototype covers all the requirements (i.e., UR1-UR15) and the mapping of the requirements to the prototype can be seen in Table \ref{table:high1PrototypeValidates}.

\hfill

\begin{figure}[ht!]
    \centering
    \includegraphics[width=0.6\textheight]{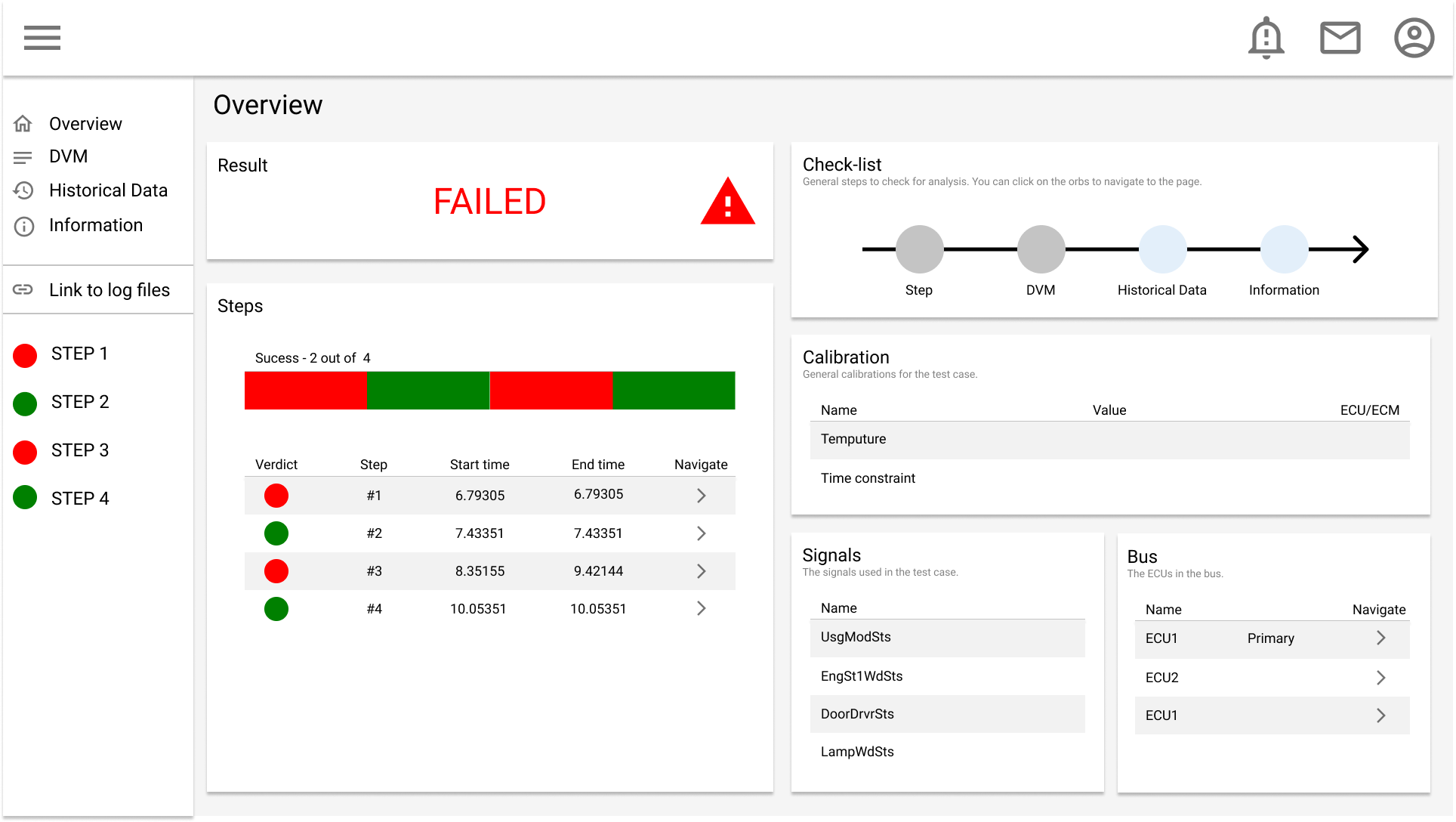}
    \caption{High-fidelity prototype of overview view.}
    \label{figure:high1Prototype_overview}
\end{figure}

\begin{table}[H]
\caption{Mapping of requirements to the High-fidelity Prototype.}
\label{table:high1PrototypeValidates}
\centering
  \begin{tabular}{ |l|l|l| } 
    \hline
      \textbf{View} & \textbf{Screenshot} & \textbf{Requirements} \\ 
    \hline
    \hline
      Overview & \ref{appendix:high1Prototype_overview} & UR1, UR10, UR12 \\
    \hline
      DVM & \ref{appendix:high1Prototype_dvm} & UR8, UR9 \\
    \hline
      Historical Data & \ref{appendix:high1Prototype_historical} & UR13 \\
    \hline
      Information & \ref{appendix:high1Prototype_information} & UR6, UR7, UR10, UR11, UR12 \\
    \hline
      Step & \ref{appendix:high1Prototype_step}, \ref{appendix:high1Prototype_stepSubview} & UR3, UR4, UR5, UR14, UR15 \\
    \hline
      Dashboard & \ref{appendix:high1Prototype} & UR2 \\
    \hline
  \end{tabular}
\end{table}

\newpage
\subsubsection{Final High-fidelity Prototype}
\label{section:high2FidelityPrototype}
A final high-fidelity prototype was created based on feedback from evaluations (see Section \ref{section:evaluationPhase}). The interface design is changed slightly as the implementation was made using React. A mapping of the requirements to this prototype can be seen in Table \ref{table:high2PrototypeValidates}. An overview of the final high-fidelity prototype can be seen in Figure \ref{figure:high2Prototype_overview} while the full set of screenshots can be seen in Appendix \ref{appendix:high2Prototype}. The design of this final version of the prototype is an approach to design a test analysis framework to support ease of use and accessibility for test related data during test analysis of automotive electronics (see research question 4), motivated from the evaluations and design process from earlier iterations.

\hfill

\begin{figure}[ht!]
    \centering
    \includegraphics[width=0.6\textheight]{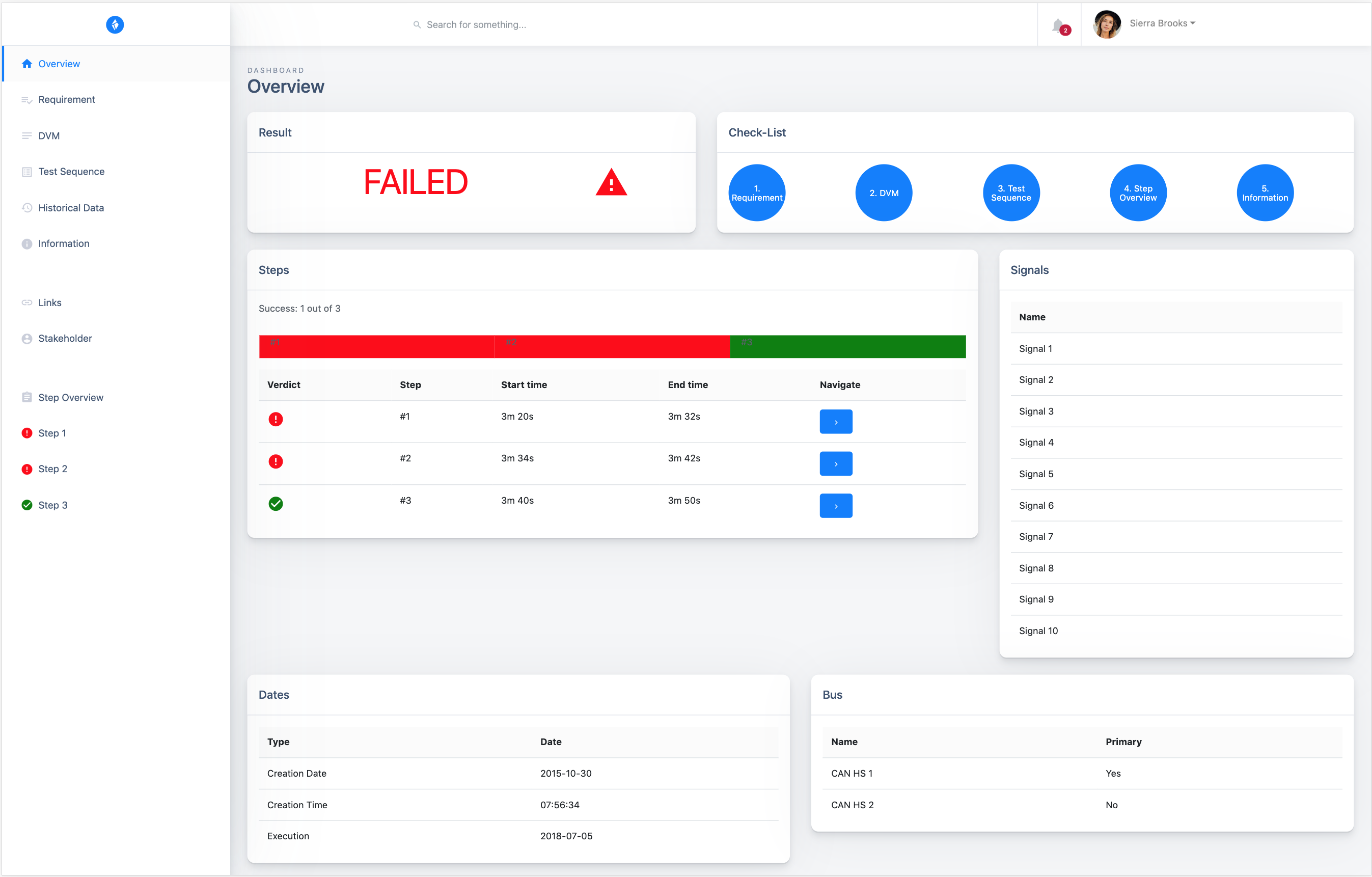}
    \caption{Final high-fidelity prototype of overview view.}
    \label{figure:high2Prototype_overview}
\end{figure}

\begin{table}[H]
\caption{Mapping of requirements to the Final High-fidelity Prototype.}
\label{table:high2PrototypeValidates}
\centering
  \begin{tabular}{ |l|l|l| } 
    \hline
      \textbf{View} & \textbf{Screenshot} & \textbf{Requirements} \\ 
    \hline
    \hline
      Overview & \ref{appendix:high2Prototype_overview} & UR1, UR10 \\
    \hline
      Requirement & \ref{appendix:high2Prototype_requirement} & UR9 \\
    \hline
      DVM & \ref{appendix:high2Prototype_dvm} & UR8 \\
    \hline
      Test Sequence & \ref{appendix:high2Prototype_testSequence} & UR8 \\
    \hline
      Historical Data & \ref{appendix:high2Prototype_historical} & UR13 \\
    \hline
      Information & \ref{appendix:high2Prototype_information} & UR10, UR11, UR12 \\
    \hline
      Links & \ref{appendix:high2Prototype_links} & UR2 \\
    \hline
      Stakeholder & \ref{appendix:high2Prototype_stakeholder}, \ref{appendix:high2Prototype_stakeholderSubview} & UR6, UR7 \\
    \hline
      Step Overview & \ref{appendix:high2Prototype_stepOverview}, \ref{appendix:high2Prototype_stepSignalCommentView}, \ref{appendix:high2Prototype_stepSignalSearchView} & UR3, UR4, UR5, UR14, UR15 \\
    \hline
      Specific Step & \ref{appendix:high2Prototype_step}, \ref{appendix:high2Prototype_stepExpanded}, \ref{appendix:high2Prototype_stepSignalCommentView}, \ref{appendix:high2Prototype_stepSignalSearchView} & UR3, UR4, UR5, UR14, UR15 \\
    \hline
  \end{tabular}
\end{table}

\newpage
\section{Evaluation Phase}
\label{section:evaluationPhase}
In this section, the findings from the evaluation phase of the low- and high-fidelity prototypes will be described. The low-fidelity prototype (see Section \ref{section:lowFidelityPrototype}) was evaluated, and the feedback was used to improve the requirements. The improved requirements were used to further develop the prototype to a high-fidelity prototype (see Section \ref{section:high1FidelityPrototype}). The high-fidelity prototype was evaluated as well, and this feedback led to improvements in data presentation and structure, resulting the final high-fidelity prototype (see Section \ref{section:high2FidelityPrototype}).

\hfill

The evaluation phase was based on cognitive walk-through and semi-structured interviews, using two testers with different experiences as participants. The methods were explained in Section \ref{section:evaluationOfTheNewSystem}.

% Task -> Result -> Analysis -> Feedback -> Improvements -> Findings
\subsection{Evaluation of Low-fidelity Prototype}
In this section, the evaluation of the low-fidelity prototype will be described.  A task analysis breakdown is listed for each task, followed by a summary of the semi-structured interview. Lastly, the feedback will be discussed together with improvements made to the requirements. 

\subsubsection{Task Analysis}
The tasks used for the low-fidelity prototype's cognitive walk-through can be found in Table \ref{table:cognitiveWalkthroughTasksForLowFiPrototypes}, in Section \ref{section:evaluationOfTheNewSystem}. The task analysis breakdown is as follow:

\hfill

\begin{itemize}
  \item \textbf{Task 1 - Overview}: The participants took the correct actions to achieve the tasks and get an overview of the information.
    \begin{itemize}
      \item \textbf{Experienced tester}: Expected more specific information on sensor signals and buses. Stakeholder, hardware. They suggested that stakeholder, hardware, and software information in Overview view could be separated.
      \item \textbf{Inexperienced tester}: Not sure what to expect in the Overview view to get a general understanding of the test case. The expectation was to see what would have failed in the test case.
    \end{itemize}
  \item \textbf{Task 2 - DVM}: Correct actions available for the participants were taken to achieve the tasks. Description of the requirements, DVM, test sequence, and expected sensor signal values were expected in this view, which was achieved. 
  \item \textbf{Task 3 - Specific step}: Correct actions available for the participants were taken to achieve the tasks. Participants noticed description, expected sensor signal values, and a graph view for sensor signals, which were expected. 
    \begin{itemize}
      \item \textbf{Experienced tester}: The graph and its feature were expected to be more customizable. Features like being able to add other sensor signals and hide sensor signals in the graph were expected.
      \item \textbf{Inexperienced tester}: The graph and its feature were expected to be more customizable. Features like being able to add additional and frequently used sensor signals to the views were desired. Knowledge sharing tools were expected, further elaborated as a hub where testers would discuss and write tips and notes for other testers' problematic signals. 
    \end{itemize}
  \item \textbf{Task 4 - Historical}: Correct actions available for the participants were taken to achieve the tasks. Being able to view a list of older test cases that were conducted, which were similar to the current test case, was expected.
    \begin{itemize}
      \item \textbf{Experienced tester}: Expected a detailed description of the older test cases including verdict, date, ECU, buses, and tester information. 
      \item \textbf{Inexperienced tester}: Expected a reference or link to the old test case to be able to open it in the tool used currently to retrieve test cases. 
    \end{itemize}
\end{itemize}

\subsubsection{Semi-structured Interview}
The interview yielded interesting information on how the prototype, could help the testers with their test analysis and diagnostics. The testers admitted that the new system provided the participants with a tool where a lot of information and data needed for test analysis was in one place. They thought it was an exciting solution to their current challenges where the data were scattered around in different tools, repositories, and storage locations. 
\hfill

The participants believed that, if the new system could provide the correct data, it would reduce the scenarios where misinterpretation or misinformation could occur during test analysis.

\subsubsection{Improvements}
The low-fidelity prototype mostly showcased the structure of the data and information in the proposed system. The feedback that was gathered from the task analysis and the semi-structured interview are as following:

\hfill

\begin{itemize}
  \item \textbf{Overview view}
    \begin{itemize}
      \item Highlight the failed steps. 
      \item Indicate which sensor signals and buses are involved in the test case. Include an overview of the calibration used in the test case.
      \item Stakeholder, software and hardware information should be split into their own separate views.
    \end{itemize}
  \item \textbf{Specific step view}
    \begin{itemize}
      \item Interactive graph features should be included like adding and hiding sensor signals.
      \item Testers should be able to see frequently used sensor signals for analysis and add them to the graph. 
      \item Testers should be able to attach a comment on test object's signals to add further information about it.
    \end{itemize}
  \item \textbf{Historical view}
    \begin{itemize}
      \item Include meta-data of the old test cases such as the verdict, date, ECU involved, bus involved and testers who conducted the test.
    \end{itemize}
  \item \textbf{Additions}
    \begin{itemize}
      \item Car information used in the test case.
      \item Configuration used in the test case, i.e., databases.
      \item References and links to log-files of the test case.
      \item A check-list, to help the tester get started with the analysis and later refer to in order to verify whether all the steps have been considered.
    \end{itemize}
\end{itemize}

\subsection{Evaluation of High-fidelity Prototype}
In this section, the evaluation of the high-fidelity prototype will be described. The task analysis is listed for each task, followed by a summary of the semi-structured interview. Lastly, the feedback and improvements of the structure and presentation of data and information will be discussed.

\subsubsection{Task Analysis}
Tasks used for the cognitive walkthrough can be found in Appendix \ref{appendix:cognitiveWalkthroughTasksHigh}. The task analysis breakdown is as follows:

\hfill

\begin{itemize}
  \item \textbf{Task 1 - Overview}: Correct actions available for the participants were taken to achieve the tasks. 
    \begin{itemize}
      \item \textbf{Experienced tester}: Expected date and time when the test case was conducted in this view. Other than that, the information in this view was satisfactory.
      \item \textbf{Inexperienced tester}: The overview of information given by this view was what was expected.
    \end{itemize}
  \item \textbf{Task 2 - DVM}: Correct actions available for the participants were taken to achieve the tasks. Participants expected information and description of the requirements as well in this view, which were missing before.
    \begin{itemize}
      \item \textbf{Experienced tester}: More detailed information was lacking in the DVM. They expected more information to be included in the DVM like pre-conditions, post-conditions, sensor signals, buses, tools, and description. The test sequence should have results and verdicts for the test cases as well. 
      \item \textbf{Inexperienced tester}: Expected the test sequence to have results and verdict for the test cases, which were missing.
    \end{itemize}
  \item \textbf{Task 3 - Historical}: Correct actions available for the participants were taken to achieve the tasks. Participants' expectations of the information and data in this view were fulfilled.
  \item \textbf{Task 4 - Information}: Correct actions available for the participants were taken to achieve the tasks. The information and data contained in this view were mostly what was expected, with minor missing information.
    \begin{itemize}
      \item \textbf{Experienced tester}: Expected software and hardware readouts that include ECU serial numbers and software part numbers, which were not included in this view. Measurements configuration was also expected for measurement tools but was missing.
      \item \textbf{Inexperienced tester}: Expected software and hardware readouts, but these were not included in this view.
    \end{itemize}
  \item \textbf{Task 5 - Links}: Actions were confusing for the participants as they expected a view to display the links to log files. 
    \begin{itemize}
      \item \textbf{Experienced tester}: Expected requirement, test case, and log file information such as name, references, and revision, which was not present.
      \item \textbf{Inexperienced tester}: Expected download links for the log files.
    \end{itemize}
  \item \textbf{Task 6 - Specific step}: Correct action available for the participants were taken to achieve the task. Expected more information in the expected results, including action description, expected result, actual result, and verdict. Apart from that, the information and data in this view were satisfactory.
\end{itemize}

\subsubsection{Semi-structured Interview}
The participants thought the prototype would be very useful for test analysis and diagnostics and would lead to efficiency improvements. The participants felt that, if the prototype could be used as a reporting tool for each test case, where testers would be required to fill the prototype's content with data and information as a requirement, it would improve the accessibility and shareability with other stakeholders involved.

\hfill

Overall, the participants said that the prototype gave a good overview of the test case if all the expected data and information was presented. By getting a good overview right from the start, the test case's context would be easier to understand, which would lead to a more confident start to the test analysis and diagnostics.

The Inexperienced participant appreciated this, as it could be challenging to know what to look during the test analysis. The more experienced participant found it easier to navigate the data and information than the current analysis method.

\hfill

According to the participants, the prototype's most useful parts were the historical view and information view. They required more effort to get this information manually in the existing test analysis process. 

\subsubsection{Improvements}
The input gathered from the task analysis, and semi-structured interview were as followed:

\hfill

\begin{itemize}
  \item \textbf{Overview view}
    \begin{itemize}
      \item The calibration should fit in the DVM view because the DVM view is responsible for this information.
      \item Date and time when the test case was conducted should be added.
    \end{itemize}
  \item \textbf{DVM view}
    \begin{itemize}
      \item DVM should include pre-conditions, post-conditions, sensor signals, buses, tools and description.
      \item Test sequence should also include the actual results and verdict.
      \item More detailed information of the requirement which the test case verifies should be shown. Only DVM is not enough.
    \end{itemize}
  \item \textbf{Historical view}
    \begin{itemize}
      \item Should also include the name and domain of the test case.
    \end{itemize}
  \item \textbf{Information view}
    \begin{itemize}
      \item Should include software and hardware readouts such as ECU serial numbers and software part numbers.
      \item Should include measurement configuration for measurements tools.
    \end{itemize}
  \item \textbf{Links}
    \begin{itemize}
      \item Have its own view which includes links and references to the requirements, DVM and log files.
    \end{itemize}
  \item \textbf{Specific step view}
    \begin{itemize}
      \item Have a summary of all steps.
      \item Favorite signals was not useful, should be removed.
      \item Historical test cases in this view were not useful, should be removed. 
    \end{itemize}
\end{itemize}

% RQ3: How should data be structured, presented, and documented in this test analysis framework to improve the quality of information available during test analysis?
% RQ4: How can this test analysis framework be designed to support ease of use and accessibility of test related data during test analysis?
\subsection{Evaluation Phase Findings}
Having the data required to perform test analysis in one place was positively perceived by participants in the evaluation phase. The prototype eases the challenges they faced in the current method of gathering the data. The risk of misinterpretation and accessing the wrong information was believed by the participants to be reduced. The efficiency in beginning test analysis was improved. Participants acknowledged that they had to manually find and locate the old system's required information and data in case another tester did not provide them. 

\hfill

It can be very overwhelming for inexperienced testers to work in many different tools to get the required data for test analysis. Most of the inexperienced testers were not fully adapted to the testing and analysis process. They admitted that it was common to get lost and ask what to look for in the current tools. The more inexperienced tester perceived the prototype as an excellent tool where the required information and data, which they needed for test analysis, were already present. The more experienced testers thought having the data in one tool made it easier to navigate between the pieces of data, making it easier to look up things they were uncertain about.

\hfill

By providing the required data that was necessary to start test analysis in a structured manner improved the overall quality of information available during test analysis by only showing relevant data and thus reducing misinterpretation of information. Having only one tool instead of multiple tools increased the efficiency to navigate between the test analysis data. Such improvements reduced the risks to get overwhelmed and lost during navigating large amount of data, by improving navigation efficiency during test analysis. 

\section{Usability Test Results}
\label{section:usabilityTestResults}
% Write a starter passage for an overall idea of this section
The usability test results comprised time for completion of tasks and calculations of the NASA Task Load Index with the Think Aloud protocol, followed by interviews to get further insights into the usability aspects of the system. This section presents the results and its analysis. Both the existing system and the new system were compared during the usability test. Testers performed four distinct tasks on both systems and their completion time of the tasks along with the NASA TLX scores were recorded.

\hfill

The usability test tasks are described in Section \ref{definingTasks} where four tasks were conducted sequentially by the participants. Each task was associated with requirements of the New System and a mapping can be seen in Table \ref{table:usabilityTestTaskRequirements}. The findings from the results of Task 1 and 4 relate to the first research question, which targeted the aspects that deteriorated decision making during test analysis and diagnostics of automotive electronics. In contrast, the findings from Task 2 and 3 relate to the second research question, which focused on data structuring, presentation and documentation to support decision making and aiding testers to communicate test analysis and diagnostics verdicts efficiently.

\begin{table}[H]
\caption{Usability test task mapping to requirements.}
\label{table:usabilityTestTaskRequirements}
\centering
  \begin{tabular}{ |l|l|l| } 
    \hline
      \textbf{Task} & \textbf{Task Name} & \textbf{Requirement} \\ 
    \hline
    \hline
      Task 1 & Understanding the issue & UR1, UR8, UR9 \\
    \hline
      Task 2 & Gathering information and data & UR2, UR6, UR10, UR11, UR12 \\
    \hline
      Task 3 & Solving the issue & UR3, UR4, UR5, UR13, UR14, UR15  \\
    \hline
      Task 4 & Reporting the issue & UR6, UR7 \\
    \hline
  \end{tabular}
\end{table}

\subsection{NASA TLX Results}
\label{section:objectiveMeasureResultWithDiscussion}
\begin{figure}[h!]
  \centering
  \begin{tikzpicture}
    \begin{axis}[
      ybar,
      ylabel=Mean score,
      symbolic x coords={Task 1, Task 2, Task 3, Task 4},
      xtick=data,
      nodes near coords,
      ymin=0,
      ymax=100,
    ]
    \addplot coordinates {(Task 1, 46.5) (Task 2, 52.8) (Task 3, 48.8) (Task 4, 49.1)};
    \addplot coordinates {(Task 1, 22.5) (Task 2, 16.8) (Task 3, 37.5) (Task 4, 22.5)};
    \legend{Old System, New System}
    \end{axis}
  \end{tikzpicture}
  \caption{NASA Task Load Index mean score of each usability test task.}
  \label{fig:usabilityTestNasaTLX}
\end{figure}
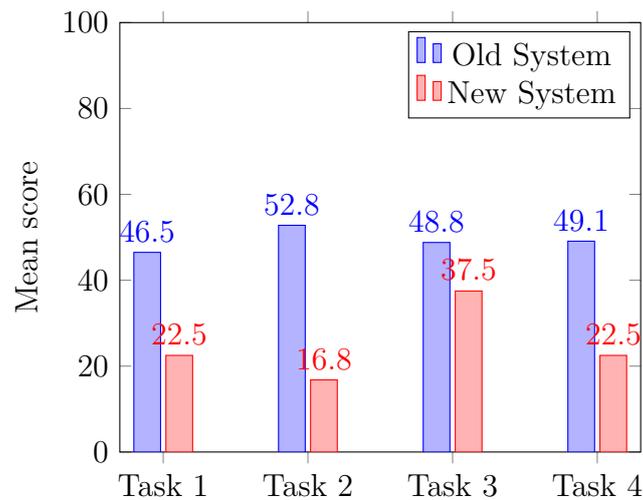

According to the NASA Task Load Index guide, each task was measured based on six subjective sub-scales, determining the workload index for each of them, as explained in Section \ref{subjectiveMeasures}. Full results of the NASA Task Load Index for the old and new systems can be found in in Appendix \ref{appendix:NasaTaskLoadIndexResultCurrentSystem} and Appendix \ref{appendix:NasaTaskLoadIndexResultNewSystem}. The mean index score for each of the tasks can be seen in Figure \ref{fig:usabilityTestNasaTLX}. The scores are in the range of 0-100 where lower scores refer to the less mental workload on test subjects while performing tasks, which is better. 

\hfill

\newpage
\subsection{Think-Aloud Protocol}
\label{section:usabilityTestThinkAloudProtocol}
The think-aloud protocol used in the usability test explained the subjects' behavior, which was posed as a supplement to the findings discussed in Section \ref{section:usabilityTestInterviews}. During the performance of tasks, the testers found it easy and intuitive to perform tasks 1, 2, and 4, especially when using the new system. Despite the new interface, they could navigate using the new system's logical information structure as a guide. The same tasks, performed in the old system, did not create an inspiring think-aloud environment as the participants were very used to those tasks. There were glimpses of relief, especially when performing Task 1, as they realized that information to complete the task was all in one place in the new system.

\hfill

For Task 3, as it required much thinking in itself, testers thought out loudly more often, regardless of the system used. The more experienced subjects talked about the need to use other measurement tools while working on the task to complement the new system on more advanced test cases. They later realized that for a test case with a more average difficulty, it wasn't required at all. 

\subsection{Interviews}
\label{section:usabilityTestInterviews}
The section explains the findings during the decoding phase of the interviews (previously discussed in Section \ref{userInterviewsUsability}), done after performing the usability test. A broad elaboration is given below along with a summary of findings in Table \ref{table:nodesForUsabilityInterview} and full data in Appendix \ref{appendix:codedUsability}. 

\hfill

The findings from the interview were generally positive for the new system. The experienced and inexperienced subjects mostly reached a consensus regarding the feedback. They favored the improved structure for storing the files for the test cases and test objects in a primary domain. All the subjects agreed that the better structure drastically improved the time required to locate information and files.

\hfill

\begin{mdframed}[style=MyFrame]
    \begin{quote}
        \begin{itshape}
        "A lot of information and data in the old system is stored in files and how we share with each other with the current tools, it can be very limiting and not everything can be shared in a optimal way. I think as the new system contains and displays all that, one can find them in faster and more efficient way."
        \end{itshape}
        - Testers' feedback.
    \end{quote}
\end{mdframed}

\hfill

The presentation of the data was improved over the old system. The subjects could relate to the efficient categorization of the information into logical pages in the user interface with faster information access. Everyone agreed that the new system gave them a quicker overview of the data the subjects required to perform test diagnosis. The subjects acknowledged that the presence of all the data in one domain was a substantial improvement as it would reduce mistakes due to misinformation and misinterpretation. Traceability of data for test objects and test cases were also better expressed due to logical structuring and presentation in a cleanly designed user interface.

\hfill

\begin{mdframed}[style=MyFrame]
    \begin{quote}
        \begin{itshape}
        "Data and files for testing done personally/ad hoc basis are not managed in a location where everyone can find it. Having such information in one place is convenient in terms of getting information fast and reliably comparing the results from the perspective of history."
        \end{itshape}
        - Testers' feedback on traceability of testing artifacts.
    \end{quote}
\end{mdframed}

\hfill

Another vital piece of feedback was that the new system drastically reduced the frequency of navigating other tools. The subjects stated that the saved time would benefit them because they could conduct a test analysis on more test cases in the same time period.

\hfill

Regarding the usability of both the systems, they noted the ease of use of the new system, which was particularly attractive for testers who were less experienced. Overall, the realization was that the new system would provide them with all the information needed to diagnose test cases problems, which were of average difficulty. However, for test cases that posed a higher complexity level, sophisticated measurement tools were still required. Nevertheless, due to test cases mostly being of an average difficulty level, they acknowledged that the new system would allow more efficient test analysis. 

\hfill
\newpage
\begin{mdframed}[style=MyFrame]
    \begin{quote}
        \begin{itshape}
        "I understand that the new system is not made to replace our existing tools but to add a layer of providing overview which is nice. I still do believe that for in-depth detailed analysis, the old system has the power of the tools to dig in, whereas the new system, is better for average cases which are indeed more in number and the new system gives a quick overview to testers."
        \end{itshape}
        - An Experienced Tester's feedback on both systems.
    \end{quote}
\end{mdframed}

\hfill

The less experienced subjects expressed that the time-consuming tasks they had to perform to begin the test analysis and diagnostics, like accumulating files and information for a test object, were tedious and repetitive. The new system, however, made such tasks simpler to perform. Hence they believed they would benefit as their primary focus would remain on understanding and finding issues rather than spending time preparing the diagnostic data for analysis.
\hfill

\begin{mdframed}[style=MyFrame]
    \begin{quote}
        \begin{itshape}
        "The testing tools we use have a learning threshold in the old system compared to this new system which is more intuitive and user-friendly. In the old system, it can be challenging to find the correct information because there are a lot of software and other tools that sometimes is used for test analysis and diagnostics. The new system on the other hand make it more appealing to the less experienced person and gives them a chance to slowly dive into our work."
        \end{itshape}
        - An Experienced Tester's feedback on both systems.
    \end{quote}
\end{mdframed}

\hfill

Lastly, the subjects agreed that the new system was simpler to understand. Learning a complicated system was not a good start to their journey as a tester. However, they suggested that, for difficult cases, they would use both the new and the old system. They revealed that features such as structuring of the files and its information could be quickly understood in the new system, while analytic precision could be gained from the old system.

\hfill

\begin{mdframed}[style=MyFrame]
    \begin{quote}
        \begin{itshape}
        "All the information needed is accessible from the left side bar menu and it is intuitive and easy to understand.  Here, I don't have to access three different systems to get an overview information of the test object and test case and these information when found faster makes it easier to do the job."
        \end{itshape}
        - An Inexperienced Tester's feedback on new system.
    \end{quote}
    
    \hfill

\end{mdframed}

\hfill

Table \ref{table:nodesForUsabilityInterview} summarizes all the findings from the usability test interviews. The table explains the categories that were made using Nvivo, which were used to collect findings in the interview. Each category has a description summarizing the findings. The pictorial representation of the table can be seen in Appendix \ref{appendix:NasaTaskLoadIndexResultNewSystem}.

\begin{table}[h!]
\caption{Major findings from the decoded interviews after usability tests with short description.}
\label{table:nodesForUsabilityInterview}
\centering
 \begin{tabular}{{ | p{4cm} | p{10cm} | }} 
    \hline
      \textbf{Findings from decoded Interviews} & \textbf{Description} \\ 
    \hline
    
    \hline
      Files structured efficiently and stored in a single domain & Testers explained that the new system created a framework that provided a much-needed structure to the files related to test object, test cases, and test diagnostics. \\ 
    \hline
      Structuring improved accessibility of information and files & The testers addressed accessibility concerning time (files and information can be accessed faster) and collectively gathered them.  \\ 
    \hline
      The old system preferred for in-depth analysis & The testers explained that, for difficult test cases, they would need to use the old system, which granted them a more sophisticated set of tools. \\ 
    \hline
      The new system preferred for overview and average test cases & For average test cases and getting an overview of the information in all cases, the new system was helpful. \\ 
    \hline
      Time for locating information reduced & Information, in general, was more comfortable to locate in the new system due to the better structure. \\ 
    \hline
      Testers got the overview of the information faster & The new system gives the testers a faster overview of the information they require to either start their evaluation of the diagnosis or come to a quick conclusion. \\ 
    \hline
    Mistakes due to interpretation or misinformation reduced & Due to all the files and information being readily accessible in the new system, lack of information or possibilities of misinformation was reduced. \\ 
    \hline
    Information and data presented efficiently for analysis & The presentation of the information and data were done efficiently, keeping it user-centered (Tester). Subjects could relate to the information presented in the user-interface. \\ 
    \hline
    Traceability of historical data improved & Testers now had the possibility of tracing testing related objects, information, and data due to the structure the new system provided. \\ 
    \hline
    Files related to testing and diagnostics were accessible faster & Testers need for requiring such files was always crucial, and the new system made this easier for them in terms of time. \\
    \hline
    The effort for navigation for data reduced & Testers did not need to navigate between various tools to make sense of the test-data and various information. \\ 
    \hline
    New system easier to use for less experienced testers & Due to the ease of use of the new system, a consensus emerged, describing it to be an attractive choice for testers new to the automotive testing field. \\ 
    \hline
  \end{tabular}
\end{table}

\newpage
\subsection{Task Measurement Results}
\label{section:usabilityTestTasks}
% Add the summary of TLX

The objective measure (i.e., time for task completion in minutes) for each task performed in both systems are tabulated in Table \ref{table:usabilityTestTime}. The table indicates the time taken for the test subjects to complete each task (explained in section \ref{definingTasks}). In the new system, the overall time taken to complete all four tasks is significantly reduced compared to the old system. The testers' experience of the task, along with analysis using the NASA TLX will be described in this section. The scores range is from 0-100, where scores closer to zero suggest low task demands and scores closer to 100 suggest high task demands. Testers 1, 2 and 3 were inexperienced, whereas testers 4, 5 and 6 were experienced.

\hfill

\begin{table}[H]
\caption{Time of completion for the Usability Test tasks in minutes.}
\label{table:usabilityTestTime}
\centering
  \begin{tabular}{ |c|c|c|c|c|c|c| } 
    \hline
      \textbf{Tester} & \textbf{System} & \textbf{Task 1} & \textbf{Task 2} & \textbf{Task 3} & \textbf{Task 4} & \textbf{Total} \\ 
    \hline
    \hline
      Tester 1 & Old & 10 & 13 & 18 & 20 & 61 \\ 
    \hline
      Tester 1 & New & 5 & 10 & 13 & 5 & 33 \\ 
    \hline
      Tester 2 & Old  & 5 & 10 & 20 & 10 & 45 \\ 
    \hline
      Tester 2 & New & 3 & 7 & 17 & 2 & 29 \\ 
    \hline
     Tester 3 & Old & 10 & 5 & 10 & 20 & 45 \\ 
    \hline
     Tester 3 & New & 6 & 3 & 7 & 8 & 24 \\ 
    \hline
      Tester 4 & Old & 11 & 10 & 15 & 15 & 51 \\ 
    \hline
      Tester 4 & New & 5 & 10 & 15 & 8 & 38 \\ 
    \hline
      Tester 5 & Old & 15 & 10 & 18 & 15 & 58 \\ 
    \hline
      Tester 5 & New & 10 & 4 & 16 & 5 & 35 \\ 
    \hline
      Tester 6 & Old & 12 & 11 & 20 & 17 & 66 \\ 
    \hline
      Tester 6 & New & 6 & 4 & 10 & 13 & 33 \\ %previous was a mistake 39! 
    \hline
  \end{tabular}
\end{table}

\hfill

The usability test was performed with four different tasks, where every task was associated with requirements mapped in Table \ref{table:usabilityTestTaskRequirements2}. Results of the usability test show the most improvements for task one, two and four.
\hfill

\begin{table}[H]
\caption{Usability test task mapping to requirements.}
\label{table:usabilityTestTaskRequirements2}
\centering
  \begin{tabular}{ |l|l|l| } 
    \hline
      \textbf{Task} & \textbf{Task Name} & \textbf{Requirement} \\ 
    \hline
    \hline
      Task 1 & Understanding the issue & UR1, UR8, UR9 \\
    \hline
      Task 2 & Gathering information and data & UR2, UR6, UR10, UR11, UR12 \\
    \hline
      Task 3 & Solving the issue & UR3, UR4, UR5, UR13, UR14, UR15  \\
    \hline
      Task 4 & Reporting the issue & UR6, UR7 \\
    \hline
  \end{tabular}
\end{table}

% For each task talk about 
% 1. Objective measure and its analysis
% 2. TLX results and its analysis
% enrich the above points with findings from the interviews and Think aloud protocol
\newpage

% Having a standard well defined way performing and reporting performing test data diagnosis analysis
% Missing a system that could manage the scattered data (META DATA example)
% Information quality
    % Having the necessary information from the many information and data (Saving time)
    % Having the correct information required for making decisions.
% System Quality
    % Having to jump between tools to get want is needed
    % Ease of use of the tool 
% Presentation of information
    % Benifits of having a dashboard interface compared to solutions practiced presently
    % Having data categorized and presented so that navigation is within the tool rather than being constantly away from the tool.
\subsubsection{Task 1 Results}
\label{tlxResultsDiscussionTask1}
\begin{figure}[H]
  \centering
  \begin{tikzpicture}
    \begin{axis}[
      ybar,
      ylabel=Time taken in minutes,
      symbolic x coords={T1, T2, T3, T4, T5, T6},
      xtick=data,
      xlabel= Testers,
      xticklabel style = {font=\small,yshift=0.5ex},
      nodes near coords,
      ymin=0,
      ymax=30,
    ]
    \addplot coordinates {(T1, 10) (T2, 5) (T3, 10) (T4, 11) (T5, 15) (T6, 12)};
    \addplot coordinates {(T1, 5) (T2, 3) (T3, 6) (T4, 5) (T5, 10) (T6, 6)};
    \legend{Old System, New System}
    \end{axis}
  \end{tikzpicture}
  \caption{Time taken to complete Task 1 by the testers (T), in both systems.}
  \label{fig:timeCompletionForTask1Results}
\end{figure}
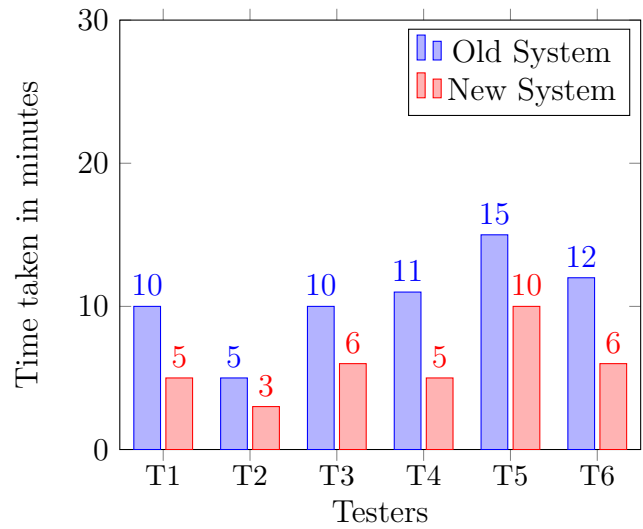

The testers, combined, spent 63 minutes on Task 1 in the old system and only 35 minutes in the new system. All the testers expressed that the data structuring had improved significantly and provided them with an overview of the information they needed to proceed with the test diagnostics and analysis. Time spent on Task 1 for each tester can be seen in Figure \ref{fig:timeCompletionForTask1Results}.

\hfill

\hfill

\begin{mdframed}[style=MyFrame]
\begin{quote}
    \begin{itshape}
    "The new tool give me a very good overview of all the basic information I need for a test case. This saves time for getting the overview of things making it faster for me to make decisions."
    \end{itshape}
    - Tester's feedback on Task 1.
\end{quote}
\end{mdframed}

\hfill

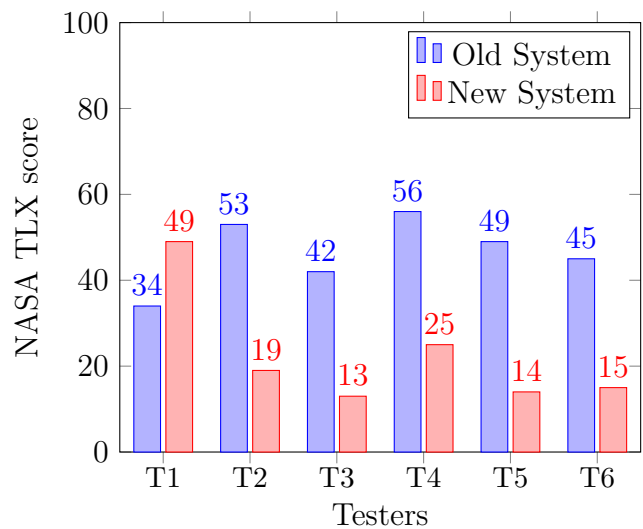
\begin{figure}[H]
  \centering
  \begin{tikzpicture}
    \begin{axis}[
      ybar,
      ylabel=NASA TLX score,
      symbolic x coords={T1, T2, T3, T4, T5, T6},
      xtick=data,
      xlabel= Testers,
      xticklabel style = {font=\small,yshift=0.5ex},
      nodes near coords,
      ymin=0,
      ymax=100,
    ]
    \addplot coordinates {(T1, 34) (T2, 53) (T3, 42) (T4, 56) (T5, 49) (T6, 45)};
    \addplot coordinates {(T1, 49) (T2, 19) (T3, 13) (T4, 25) (T5, 14) (T6, 15)};
    \legend{Old System, New System}
    \end{axis}
  \end{tikzpicture}
  \caption{NASA TLX scores of Task 1 for each tester (T), in both systems.}
  \label{fig:NASATLXTask1Score}
\end{figure}

The scores from the NASA TLX for Task 1 shows that five out of six testers felt an overall improvement on the workload. The testers' NASA TLX scores for Task 1 can be seen in Figure \ref{fig:NASATLXTask1Score}. Except for Tester 1 (T1), the workload score was improved by more than 50\% when using the new system. Tester 1's score was although not improved in the new system, but was somewhat worse. The tester was more comfortable using the old system to perform Task 1 and did not see much improvement with the new system. Tester 1 had less experience in test diagnostics in general and was more used to the old system. It was also noticed that the testers with less experience finished relatively faster than the experienced tasters. The inexperienced testers did not reassess their decisions while doing the different tasks in the test. In contrast, the experienced testers were more cautious and repeated the tasks to evaluate whether they made the correct choices.

\hfill

\begin{figure}[H]
  \centering
  \begin{tikzpicture}
    \begin{axis}[
      ybar,
      ylabel=NASA TLX mental demand score,
      symbolic x coords={T1, T2, T3, T4, T5, T6},
      xtick=data,
      xlabel= Testers,
      xticklabel style = {font=\small,yshift=0.5ex},
      nodes near coords,
      ymin=0,
      ymax=120,
    ]
    \addplot coordinates {(T1, 60) (T2, 40) (T3, 65) (T4, 65) (T5, 80) (T6, 55)};
    \addplot coordinates {(T1, 30) (T2, 20) (T3, 10) (T4, 30) (T5, 10) (T6, 15)};
    \legend{Old System, New System}
    \end{axis}
  \end{tikzpicture}
  \caption{NASA TLX mental demand scores of Task 1 for each tester (T), in both systems.}
  \label{fig:NASATLXTask1ScoreMental}
\end{figure}
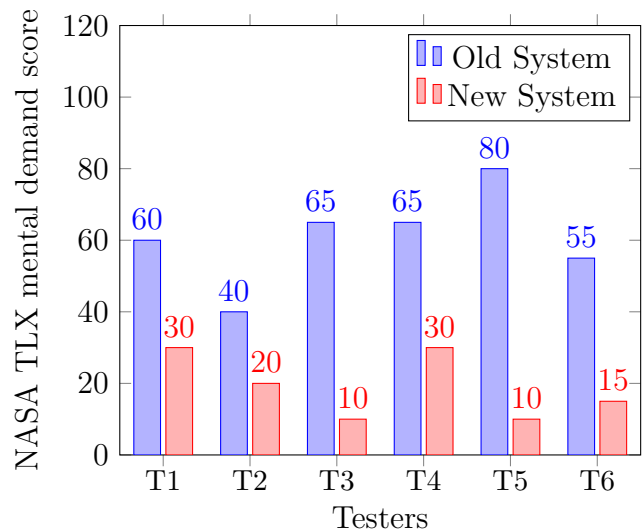

% 60, 40, 65, 65, 80, 55
% 30, 20, 10, 30, 10, 15
\textbf{Mental Demand:} The mental demand scores for Task 1 can be seen in Figure \ref{fig:NASATLXTask1ScoreMental}. No matter what experience the tester had, the mental demand was lessened when using the new system. By providing a better presentation and structure of the data and information, the task was easier to perform, and the complexity was reduced.

\hfill

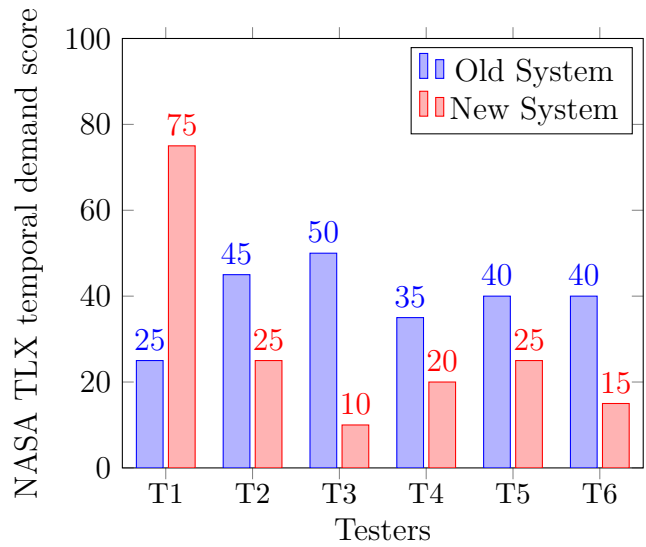
\begin{figure}[H]
  \centering
  \begin{tikzpicture}
    \begin{axis}[
      ybar,
      ylabel=NASA TLX temporal demand score,
      symbolic x coords={T1, T2, T3, T4, T5, T6},
      xtick=data,
      xlabel= Testers,
      xticklabel style = {font=\small,yshift=0.5ex},
      nodes near coords,
      ymin=0,
      ymax=100,
    ]
    \addplot coordinates {(T1, 25) (T2, 45) (T3, 50) (T4, 35) (T5, 40) (T6, 40)};
    \addplot coordinates {(T1, 75) (T2, 25) (T3, 10) (T4, 20) (T5, 25) (T6, 15)};
    \legend{Old System, New System}
    \end{axis}
  \end{tikzpicture}
  \caption{NASA TLX temporal demand scores of Task 1 for each tester (T), in both systems.}
  \label{fig:NASATLXTask1ScoreTemporal}
\end{figure}

% 25, 45, 50, 35, 40, 40
% 75, 25, 10, 20, 25, 15
\textbf{Temporal Demand:} The temporal demand scores for Task 1 can be seen in Figure \ref{fig:NASATLXTask1ScoreTemporal}. In general, the testers found the new system to be more time-efficient when understanding the test case context and getting an overview of an issue. The ease of use made the task faster to complete. For Tester 1 (T1), this was not the case because the tester was comfortable with the old system and introducing the new system introduced the need to learn a new tool.

\hfill

\begin{figure}[H]
  \centering
  \begin{tikzpicture}
    \begin{axis}[
      ybar,
      ylabel=NASA TLX effort score,
      symbolic x coords={T1, T2, T3, T4, T5, T6},
      xtick=data,
      xlabel= Testers,
      xticklabel style = {font=\small,yshift=0.5ex},
      nodes near coords,
      ymin=0,
      ymax=120,
    ]
    \addplot coordinates {(T1, 30) (T2, 70) (T3, 35) (T4, 75) (T5, 50) (T6, 65)};
    \addplot coordinates {(T1, 40) (T2, 20) (T3, 20) (T4, 40) (T5, 15) (T6, 25)};
    \legend{Old System, New System}
    \end{axis}
  \end{tikzpicture}
  \caption{NASA TLX effort scores of Task 1 for each tester (T), in both systems.}
  \label{fig:NASATLXTask1ScoreEffort}
\end{figure}
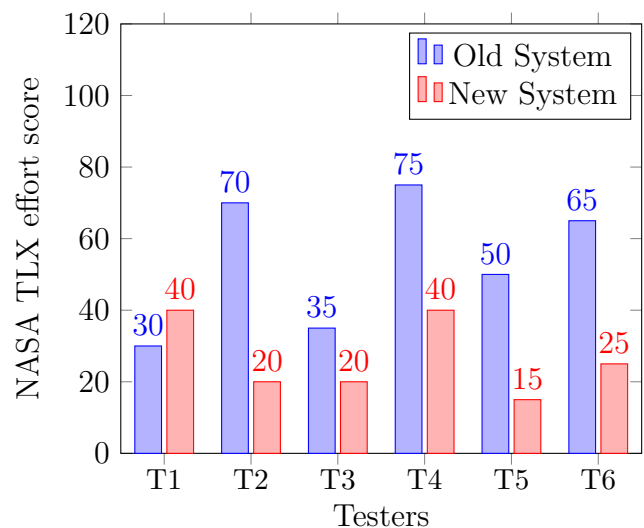

% 30, 70, 35, 75, 50, 65
% 40, 20, 20, 40, 15, 25
\textbf{Effort:} The effort scores for Task 1 can be seen in Figure \ref{fig:NASATLXTask1ScoreEffort}. The work needed for Task 1 was greatly reduced in the new system. When the data is presented in one tool, the navigation work was highly reduced. If there was some crucial information the testers forgot and wanted to look up again, it took less work for the tester to do in the new system. Once again, Tester 1 (T1) did not see any improvement in the effort score for the new system because T1 was more comfortable with the old system.

\hfill

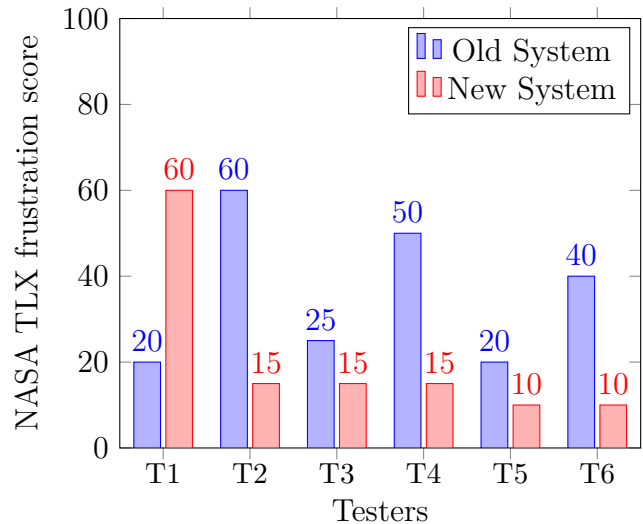
\begin{figure}[H]
  \centering
  \begin{tikzpicture}
    \begin{axis}[
      ybar,
      ylabel=NASA TLX frustration score,
      symbolic x coords={T1, T2, T3, T4, T5, T6},
      xtick=data,
      xlabel= Testers,
      xticklabel style = {font=\small,yshift=0.5ex},
      nodes near coords,
      ymin=0,
      ymax=100,
    ]
    \addplot coordinates {(T1, 20) (T2, 60) (T3, 25) (T4, 50) (T5, 20) (T6, 40)};
    \addplot coordinates {(T1, 60) (T2, 15) (T3, 15) (T4, 15) (T5, 10) (T6, 10)};
    \legend{Old System, New System}
    \end{axis}
  \end{tikzpicture}
  \caption{NASA TLX frustration scores of Task 1 for each tester (T), in both systems.}
  \label{fig:NASATLXTask1ScoreFrustration}
\end{figure}

% 20, 60, 25, 50, 20, 40
% 60, 15, 15, 15, 10, 10
\textbf{Frustration:} The frustration scores for Task 1 can be seen in Figure \ref{fig:NASATLXTask1ScoreFrustration}. The stress and irritation level during the task was reduced greatly in the new system. As the data was all in one tool, the tester was allowed to put all focus on understanding the test case. The high effort score for Tester 1 (T1) was due to general inexperience in test diagnostics and the requirement to learn a new tool.

\hfill

In conclusion, having all the necessary and correct information in one place reduced navigation work. When everything was categorized and presented in the same tool, the effort needed was lessened. More effort could then be used to understand the test case. The likelihood of finding incorrect information was reduced in the new system as the most relevant information was presented directly.

\hfill

\begin{mdframed}[style=MyFrame]
\begin{quote}
    \begin{itshape}
    "I also want to add that instead of having a lot of different tools and software windows up, but everything in one window where you can navigate around to find the required information can be good in terms of correctness as you can go back to the information easier and not get lost in all the other windows."
    \end{itshape}
    - Tester's feedback on Task 1.
\end{quote}
\end{mdframed}

\newpage
\begin{mdframed}[style=MyFrame]
\begin{quote}
    \begin{itshape}
    "Misinterpreting the requirements, using wrong databases will give faults, sometimes when you don't put the signals in the correct configuration, then the solution may look okay, but it is wrong. Having the wrong information regarding the database or configurations is a major source of errors. There are ways to get things in the old system, but you need to know a lot of details to find things. But in the new system, all this information is here and easily accessible."
    \end{itshape}
    - Tester's feedback on Task 1.
\end{quote}
\end{mdframed}

\subsubsection{Task 2 Results}
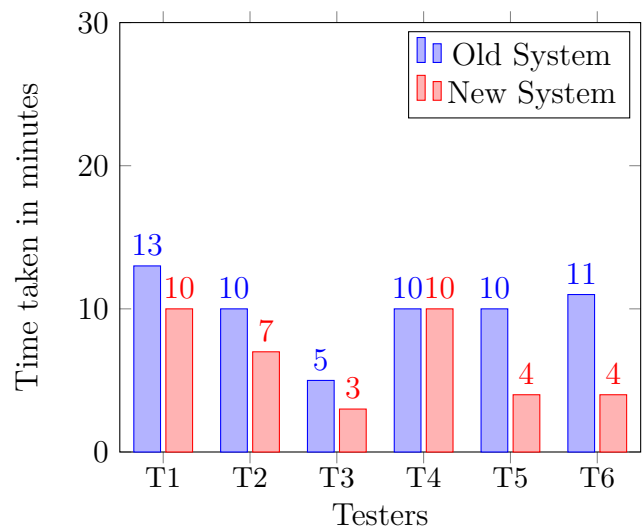
\begin{figure}[H]
  \centering
  \begin{tikzpicture}
    \begin{axis}[
      ybar,
      ylabel=Time taken in minutes,
      symbolic x coords={T1, T2, T3, T4, T5, T6},
      xtick=data,
      xlabel= Testers,
      xticklabel style = {font=\small,yshift=0.5ex},
      nodes near coords,
      ymin=0,
      ymax=30,
    ]
    \addplot coordinates {(T1, 13) (T2, 10) (T3, 5) (T4, 10) (T5, 10) (T6, 11)};
     \addplot coordinates {(T1, 10) (T2, 7) (T3, 3) (T4, 10) (T5, 4) (T6, 4)};
    
    \legend{Old System, New System}
    \end{axis}
  \end{tikzpicture}
  \caption{Time taken to complete Task 2 by the testers (T), in both systems.}
  \label{fig:timeCompletionForTask2Results}
\end{figure}

Task 2  involved collecting all the necessary files and data for the test object. The time taken summed to 59 minutes for all six testers in the old system. The testers' revealed that the unnecessary time was spent searching for files manually. The testers completed Task 2 in 38 minutes in the new system, a decrease of 21 minutes, as can be seen in Figure \ref{fig:timeCompletionForTask2Results}.

\hfill

\begin{mdframed}[style=MyFrame]
\begin{quote}
    \begin{itshape}
    "It feels like the new system has done a lot of the groundwork that I would have done manually in the old system. The data and information being in one place and is easy to follow. It is definitely more of a smoother experience when doing test analysis."
    \end{itshape}
    - Tester's feedback on Task 2.
\end{quote}
\end{mdframed}

\hfill

The testers' explained that the files related to the test object were simple to locate and download into their systems from the categorized menus in the new system (see Figure \ref{appendix:high2Prototype_links}). They also reflected on how they could find the information needed to proceed with testing in the new system using the Information and DVM menu (see Figure \ref{appendix:high2Prototype_information} and Figure \ref{appendix:high2Prototype_dvm}), without having to spend a lot of time. One of the most crucial reasons why retrieval of data was an issue was because the partnering company practised the storing of test output data in a decentralized way. One key aspect regarding decentralized systems is that they require a decentralized meta-data management system which would make the location of required data transparent to the users\cite{risse2005self}. However, in the case of the partnering company, the coordination system was not optimally implemented or used effectively, hence making it extremely difficult for the stakeholders to access all the necessary data for a test object in an efficient way. As the data was scattered across different locations, it took more time and effort to collect all the required data that represented a test object. Hence, showing the data in logical categories and web pages in the new system place allowed data and information retrieval to become faster.

\hfill

\begin{figure}[H]
  \centering
  \begin{tikzpicture}
    \begin{axis}[
      ybar,
      ylabel=NASA TLX score,
      symbolic x coords={T1, T2, T3, T4, T5, T6},
      xtick=data,
      xlabel= Testers,
      xticklabel style = {font=\small,yshift=0.5ex},
      nodes near coords,
      ymin=0,
      ymax=100,
    ]
    \addplot coordinates {(T1, 74) (T2, 74) (T3, 55) (T4, 30) (T5, 38) (T6, 46)};
    \addplot coordinates {(T1, 13) (T2, 13) (T3, 14) (T4, 22) (T5, 21) (T6, 18)};
    \legend{Old System, New System}
    \end{axis}
  \end{tikzpicture}
  \caption{NASA TLX scores of Task 2 for each tester (T), in both systems.}
  \label{fig:NASATLXTask2Score}
\end{figure}
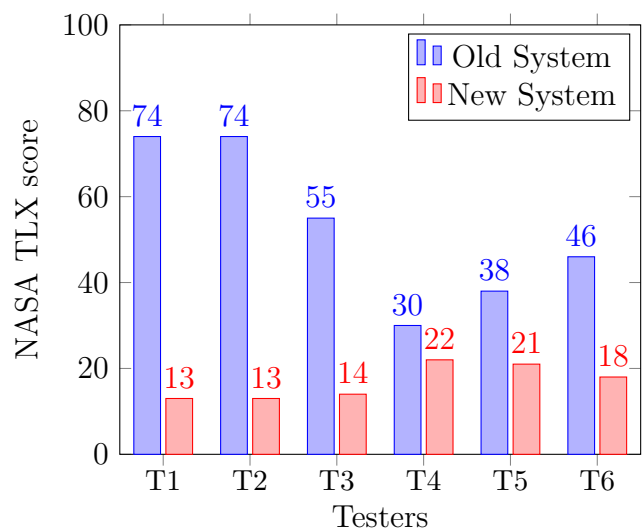

Regarding the results of the NASA TLX scores for Task 2 in the old system (see Figure \ref{fig:NASATLXTask2Score}) the overall score for the three inexperienced testers was 74, 74 and 55 respectively and 30, 38 and 46 respectively for the experience testers. The less experienced testers seemed to struggle more than their experienced counterparts. The interviews and observation uncovered the following findings.

\hfill

\newpage
\textbf{Mental Demand:} The mental demand scores for Task 2 can be seen in Figure \ref{fig:NASATLXTask2ScoreMentalDemand}. For the inexperienced testers, the mental demand was consistently high, resulting in 75, 70 and 70 for the old system. For the experienced testers, the scores were 20, 60 and 60 for the old system. Despite using the old system for at least a year, five out of the six testers believed that the task was high in mental demand. The testers collectively blamed the constant context switching between different applications and storage systems to collect the necessary data while also maintaining correctness in terms of data collection. The scores were relatively lower in the new system. The inexperienced testers scored 5, 10, and 15, while their counterparts scored 20, 25, and 20 for the new system. The testers explained that the menu categories in the new system made it simpler to interact with and gain the needed information without context switching between different applications. Here, it can be seen that the mental demand score for the new system was low relative to the old system. However, the scores were still slightly higher in the new system for the experienced testers compared with inexperienced testers. The experienced testers revealed that as they had a habit of making sure they were working with the correct information, and they reevaluated their choices often. The inexperienced testers, however, relied on the system to be correct with the data and did not reassess their choices. 

\begin{figure}[H]
  \centering
  \begin{tikzpicture}
    \begin{axis}[
      ybar,
      ylabel=NASA TLX mental demand score,
      symbolic x coords={T1, T2, T3, T4, T5, T6},
      xtick=data,
      xlabel= Testers,
      xticklabel style = {font=\small,yshift=0.5ex},
      nodes near coords,
      ymin=0,
      ymax=100,
    ]
    \addplot coordinates {(T1, 75) (T2, 70) (T3, 70) (T4, 20) (T5, 60) (T6, 60)};
    \addplot coordinates {(T1, 5) (T2, 10) (T3, 15) (T4, 20) (T5, 25) (T6, 20)};
    \legend{Old System, New System}
    \end{axis}
  \end{tikzpicture}
  \caption{NASA TLX metal demand scores of Task 2 for each tester (T), in both systems.}
  \label{fig:NASATLXTask2ScoreMentalDemand}
\end{figure}

\hfill

\begin{figure}[H]
  \centering
  \begin{tikzpicture}
    \begin{axis}[
      ybar,
      ylabel=NASA TLX temporal demand score,
      symbolic x coords={T1, T2, T3, T4, T5, T6},
      xtick=data,
      xlabel= Testers,
      xticklabel style = {font=\small,yshift=0.5ex},
      nodes near coords,
      ymin=0,
      ymax=100,
    ]
    \addplot coordinates {(T1, 85) (T2, 60) (T3, 50) (T4, 20) (T5, 20) (T6, 35)};
    \addplot coordinates {(T1, 15) (T2, 15) (T3, 15) (T4, 20) (T5, 20) (T6, 15)};
    \legend{Old System, New System}
    \end{axis}
  \end{tikzpicture}
  \caption{NASA TLX temporal demand scores of Task 2 for each tester (T), in both systems.}
  \label{fig:NASATLXTask2ScoreTemporalDemand}
\end{figure}
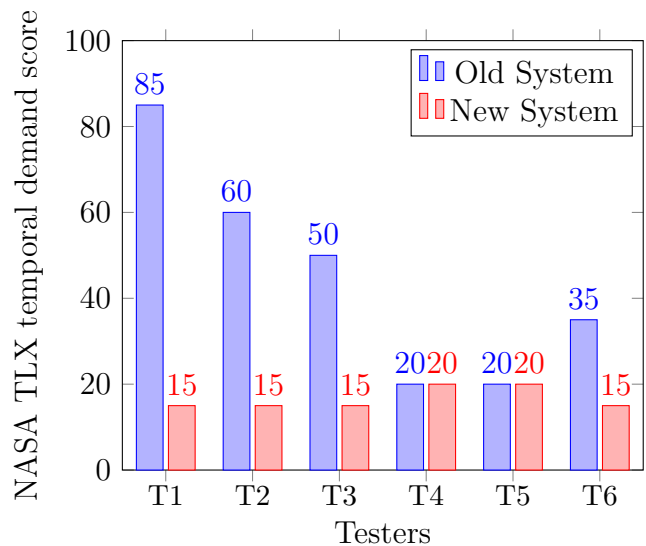

\textbf{Temporal Demand:} The temporal demand scores for Task 2 can be seen in Figure \ref{fig:NASATLXTask2ScoreTemporalDemand}. The inexperienced testers found it overwhelming to perform this task in the old system as it required constantly moving between various applications and looking for the required data. On top of that, due to their low experience level, they had to frequently confirm whether the outcome of the task resulted in desired data, hence making it more time-consuming. For the experienced testers, the task appeared to be slightly trivial, as they were used to it in their day to day work. However, they acknowledged struggling with such type of task during their early stages of being a tester. The inexperienced testers had a significantly lower score with the new system, accounting for a score of 15 for each. They explained the lack of context switching between testing tools provided an immense advantage because all the necessary data were categorized accordingly in the new system. As the data was categorized, presented, and labelled in the new system's interfaces, navigation and searching for information felt easier.
The testers' explained they no longer had to apply guesswork or rely on intuition to collect data for the test cases as they were well structured and traceable in the new system. Hence, this reduced performance time, giving them more time to conduct tests analysis. As for the experienced testers, all of them scored 20 each in the new system, stating the same reasoning as the Inexperienced testers.

\hfill

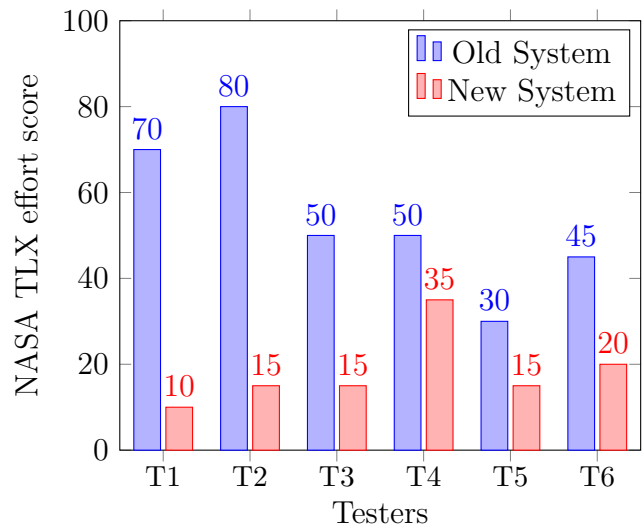
\begin{figure}[H]
  \centering
  \begin{tikzpicture}
    \begin{axis}[
      ybar,
      ylabel=NASA TLX effort score,
      symbolic x coords={T1, T2, T3, T4, T5, T6},
      xtick=data,
      xlabel= Testers,
      xticklabel style = {font=\small,yshift=0.5ex},
      nodes near coords,
      ymin=0,
      ymax=100,
    ]
    \addplot coordinates {(T1, 70) (T2, 80) (T3, 50) (T4, 50) (T5, 30) (T6, 45)};
    \addplot coordinates {(T1, 10) (T2, 15) (T3, 15) (T4, 35) (T5, 15) (T6, 20)};
    \legend{Old System, New System}
    \end{axis}
  \end{tikzpicture}
  \caption{NASA TLX effort scores of Task 2 for each tester (T), in both systems.}
  \label{fig:NASATLXTask2ScoreEffort}
\end{figure}

\textbf{Effort:} The effort scores for Task 2 can be seen in Figure \ref{fig:NASATLXTask2ScoreEffort}. The inexperienced testers scored 70, 80 and 50 respectively in the old system. They blamed the manual process of the task. The testers all agreed that the old system was not tailored for the inexperienced testers in terms of usability. The consensus amongst the testers suggested that they had to spend additional efforts adapting to the flaws of the old system, along with the task, hence increasing the time and labour. As for the experienced testers, they had slightly better scores ranging from 30 to 50 in the old system. Despite their knowledge and practice with the existing system, they still believed that it was not a pleasant experience. Intuition played an important role when they performed the task because they could anticipate where and what to find. Despite that, they found certain aspects of the task like the data and file collection for the test cases a nuisance rather than a challenge. For the inexperienced testers, the scores were much better in the new system. They explained that certain aspects of the new system made the overall task and its experience intuitive and simple. Amongst those were the neatly categorized menus which held all the necessary information and data for the test case logically and concisely. Hence, the inexperienced testers felt that the interface brought a sense of predictability in regards to what they wanted to see, making the task consume less effort. As for the experienced testers, they too scored better here, accounting for 35, 15 and 20 in the new system.  They discussed the implementation of the interface and how it saved them time on collecting data, files, and information which should have always been at the testers' disposal.

\begin{mdframed}[style=MyFrame]
    \begin{quote}
        \begin{itshape}
        "In the current way of doing this (old system), when there is new project or unfamiliar test cases, I have to pinpoint a lot of the things such as version numbers, databases etc myself, so it can be kinda tedious as I have to go around and find it on documents and web pages. The new system provides a lot of this stuff, so I think this can be very valuable
."
        \end{itshape}
        \hfill
        - Tester's feedback on Task 2.
    \end{quote}
\end{mdframed}

\begin{figure}[H]
  \centering
  \begin{tikzpicture}
    \begin{axis}[
      ybar,
      ylabel=NASA TLX frustration score,
      symbolic x coords={T1, T2, T3, T4, T5, T6},
      xtick=data,
      xlabel= Testers,
      xticklabel style = {font=\small,yshift=0.5ex},
      nodes near coords,
      ymin=0,
      ymax=120,
    ]
    \addplot coordinates {(T1, 70) (T2, 85) (T3, 65) (T4, 30) (T5, 30) (T6, 65)};
    \addplot coordinates {(T1, 20) (T2, 15) (T3, 10) (T4, 15) (T5, 20) (T6, 15)};
    \legend{Old System, New System}
    \end{axis}
  \end{tikzpicture}
  \caption{NASA TLX frustration scores of Task 2 for each tester (T), in both systems.}
  \label{fig:NASATLXTask2ScoreFrustration}
\end{figure}
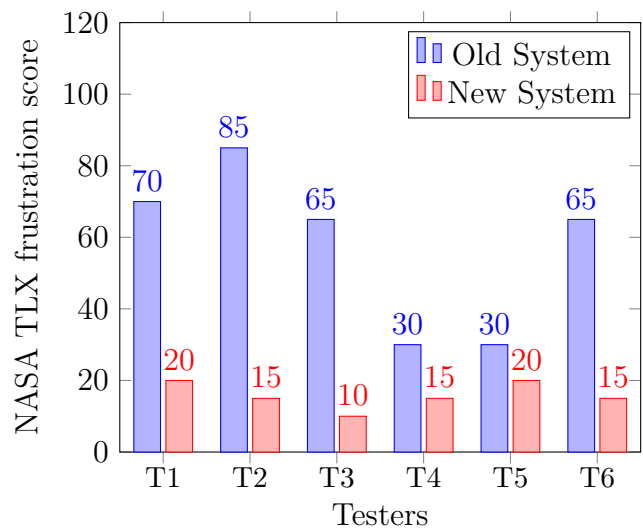

\textbf{Frustration:} The frustration scores for Task 2 can be seen in Figure \ref{fig:NASATLXTask2ScoreFrustration}. The inexperienced testers altogether found this task extremely frustrating with the old system. Their TLX scores were 70, 85 and 65 individually, suggesting a rather frustrating experience overall due to the redundant routine of manually accumulating test-related data. The experienced testers scored less, accounting to 30, 30 and 65 individually in the old system. As they had more experience of gathering such data for test cases, they could intuitively realize whether the data they collected was correct or not. Both groups of testers explained that they almost always had to go through this routine each time they faced a new test case or during debugging/re-running test cases which were performed a long time in the past. Hence, finding the correct data was always a challenge because data could be missing or not archived for more than a certain period, making the whole process very frustrating.

\hfill

The TLX scores for the inexperienced testers were better in the new system. They scored 20, 15 and 10 for frustration. The experienced testers group had relatively less deviation, apart from one who scored 15 (65 in the old system). They claimed to have exerted significantly less effort due to the information and data being appropriately structured. Even though the proposed system was new for them, it proved to be intuitive in terms of navigation. They said that the interface was mostly self-explanatory, which made the task not frustrating. The need for data to be in a common platform was crucial from their perspective. Locating the necessary data to make decisions on a test case would always require time and unnecessary work.

\hfill

\begin{mdframed}[style=MyFrame]
    \begin{quote}
        \begin{itshape}
        "The experience was smooth and quick overall. Not much time needed to look for stuff. You know what I mean? We as testers need to get some information on a fast basis and the tool gives that overview quickly."
        \end{itshape}
        \hfill
        - Tester's feedback on Task 2.
    \end{quote}

\end{mdframed}

\newpage
\subsubsection{Task 3 Results}
\begin{figure}[h!]
  \centering
  \begin{tikzpicture}
    \begin{axis}[
      ybar,
      ylabel=Time taken in minutes,
      symbolic x coords={T1, T2, T3, T4, T5, T6},
      xtick=data,
      xlabel= Testers,
      xticklabel style = {font=\small,yshift=0.5ex},
      nodes near coords,
      ymin=0,
      ymax=30,
    ]
    \addplot coordinates {(T1, 18) (T2, 20) (T3, 10) (T4, 15) (T5, 18) (T6, 20)};
     \addplot coordinates {(T1, 13) (T2, 17) (T3, 7) (T4, 15) (T5, 16) (T6, 10)};
    
    \legend{Old System, New System}
    \end{axis}
  \end{tikzpicture}
  \caption{Time taken to complete Task 3 by the testers (T), in both systems.}
  \label{fig:timeCompletionForTask3Results}
\end{figure}
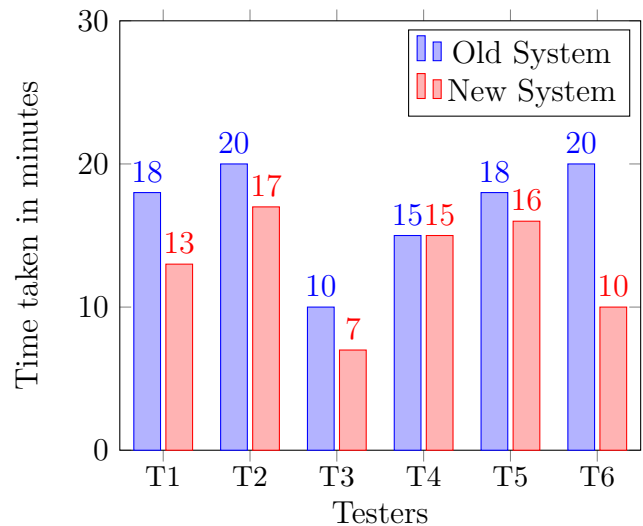

Task 3 involved readying the testing tools using the data/files from Task 2 and creating an overview of information about the test to find the root cause of the failure. The total time spent on Task 3 (see Figure \ref{fig:timeCompletionForTask3Results}) on the old system summed to 101 minutes and 78 minutes for the new.

According to the testers, the typical issues while performing this task included the time and effort spent in preparing the tools for analysis, navigating between the multiple tools and files and lastly, working with large and unstructured data for test cases. In the new system, the efforts needed to prepare tools was no longer an issue as testers(mostly Inexperienced) agreed that the files and data required to prepare the tools were present in the new system on the dashboard. All of them agreed that the time for locating information was also reduced as they did not have to switch between tools to find what they needed. 

\hfill

\begin{mdframed}[style=MyFrame]
    \begin{quote}
        \begin{itshape}
        "Faster overview(in the new system), helps give insights to the problem, and saves time from the eventual analysis in the analysis tools."
        \end{itshape}
        - Tester's feedback on getting a good overview.
    \end{quote}
\end{mdframed}

\hfill

Furthermore, the information was presented efficiently and intuitively based on the requirements of the testers, giving them more time to focus on analyzing the test case rather than doing unnecessary work. The testers finally concluded that finding issues depended on the experience and the two systems did little to help there apart from aiding to set up the analysis. 

\hfill

\begin{mdframed}[style=MyFrame]
    \begin{quote}
        \begin{itshape}
        "You can make a decision on easier test cases I think directly in the new system. When you see the expected signal values in the new interface, you can see from the pattern directly in the graph and find what went wrong which could save time. For more complicated analysis, this can also help you get a head start but for in-depth analysis the old system provides more tools."
        \end{itshape}
        - Tester's feedback on finding issues in tests.
    \end{quote}

\end{mdframed}

\hfill 

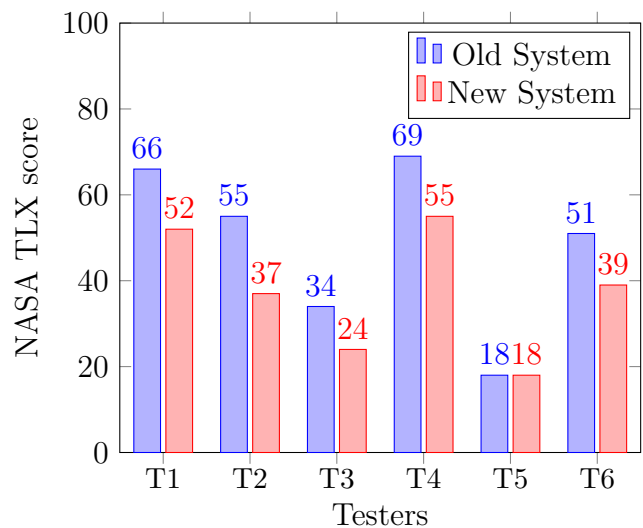
\begin{figure}[H]
  \centering
  \begin{tikzpicture}
    \begin{axis}[
      ybar,
      ylabel=NASA TLX score,
      symbolic x coords={T1, T2, T3, T4, T5, T6},
      xtick=data,
      xlabel= Testers,
      xticklabel style = {font=\small,yshift=0.5ex},
      nodes near coords,
      ymin=0,
      ymax=100,
    ]
    \addplot coordinates {(T1, 66) (T2, 55) (T3, 34) (T4, 69) (T5, 18) (T6, 51)};
    \addplot coordinates {(T1, 52) (T2, 37) (T3, 24) (T4, 55) (T5, 18) (T6, 39)};
    \legend{Old System, New System}
    \end{axis}
  \end{tikzpicture}
  \caption{NASA TLX scores of Task 3 for each tester (T), in both systems.}
  \label{fig:NASATLXTask3Score}
\end{figure}

The NASA TLX scores for Task 3 can be seen in Figure \ref{fig:NASATLXTask3Score}. The NASA TLX scores in the old system for the experienced testers were 66, 55 and 34 and 69, 18 and 51 for the inexperienced testers. However, in the new system, the experienced testers dropped to 52, 37 and 24 while the inexperienced fell to 55, 18 and 39, respectively. The new system did reduce the workload compared to the old system, but only slightly. This is because the actual sensor signal analysis in the current tools offers a lot of advanced features, and if the testers know how to utilize them, they made it easier to complete the task. For simpler tasks where those advanced features were not needed, the testers could use the new system for faster judgment and decisions.

\begin{figure}[H]
  \centering
  \begin{tikzpicture}
    \begin{axis}[
      ybar,
      ylabel=NASA TLX mental demand score,
      symbolic x coords={T1, T2, T3, T4, T5, T6},
      xtick=data,
      xlabel= Testers,
      xticklabel style = {font=\small,yshift=0.5ex},
      nodes near coords,
      ymin=0,
      ymax=120,
    ]
    \addplot coordinates {(T1, 85) (T2, 60) (T3, 40) (T4, 80) (T5, 20) (T6, 55)};
    \addplot coordinates {(T1, 65) (T2, 30) (T3, 30) (T4, 70) (T5, 20) (T6, 40)};
    \legend{Old System, New System}
    \end{axis}
  \end{tikzpicture}
  \caption{NASA TLX mental demand scores of Task 3 for each tester (T), in both systems.}
  \label{fig:NASATLXTask3ScoreMentalDemand}
\end{figure}
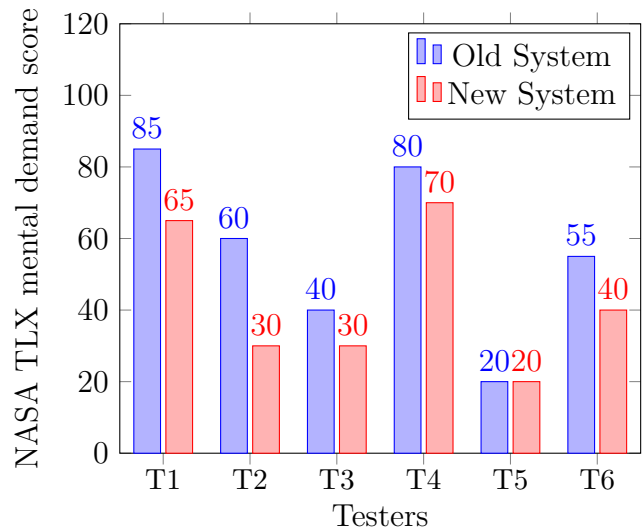

\textbf{Mental Demand:} The mental demand scores for Task 3 can be seen in Figure \ref{fig:NASATLXTask3ScoreMentalDemand}. As the task required to finding issues in test cases, their ability to solve this task was not the major drawback for the testers. It was rather the inconvenience of needing to find things they needed in various systems. Perceptual activities like constantly needing to move between windows of different tools made the task cumbersome and mentally taxing. In the old system, the inexperienced testers gave a score 85, 60 and 70 while the experienced testers gave 80, 20 and 55, respectively. The testers at the initial stages of the task did the most context switching between tools using the old system, which was cumbersome, mentally frustrating and annoying. In the new system, however, the testers spent less time gathering an overview of the task due to the information being present already in the overview section. The new system did not explicitly help them in identifying the root-cause as this depends on the experience and knowledge of the tester. Still, it reduced the mental effort by decreasing perceptual activities and context switching.

%How much time pressure did you feel due to the pace at which the tasks or task elements occurred? Was the pace slow or rapid 
%SUPER IMPORTANT, we have to make sure we mention that each tas was TIME-BOXED to 20 minutes to avoid testers to have infinite amount of time to complete the tasks.

\hfill

\begin{mdframed}[style=MyFrame]
    \begin{quote}
        \begin{itshape}
        "The new tool gives me a very good overview of all the basic information I need for a test case. This saves time for getting the overview of things making it faster for me to make decisions."
        \end{itshape}
        \hfill
        - Tester's feedback for Task 3.
    \end{quote}
\end{mdframed}

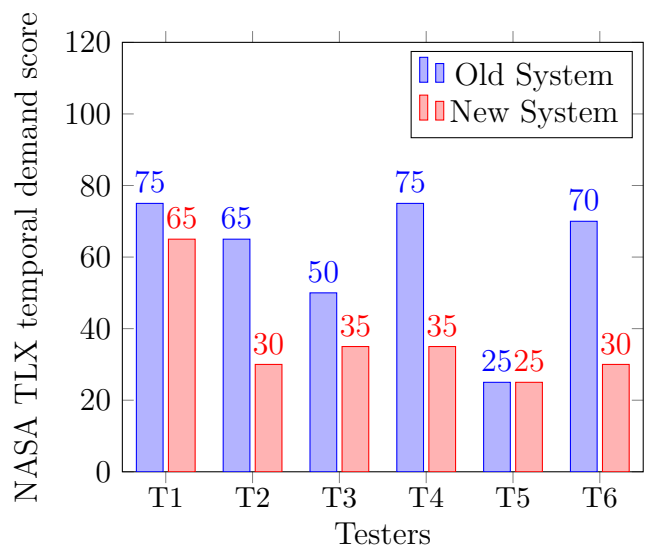
\begin{figure}[H]
  \centering
  \begin{tikzpicture}
    \begin{axis}[
      ybar,
      ylabel=NASA TLX temporal demand score,
      symbolic x coords={T1, T2, T3, T4, T5, T6},
      xtick=data,
      xlabel= Testers,
      xticklabel style = {font=\small,yshift=0.5ex},
      nodes near coords,
      ymin=0,
      ymax=120,
    ]
    \addplot coordinates {(T1, 75) (T2, 65) (T3, 50) (T4, 75) (T5, 25) (T6, 70)};
    \addplot coordinates {(T1, 65) (T2, 30) (T3, 35) (T4, 35) (T5, 25) (T6, 30)};
    \legend{Old System, New System}
    \end{axis}
  \end{tikzpicture}
  \caption{NASA TLX temporal demand scores of Task 3 for each tester (T), in both systems.}
  \label{fig:NASATLXTask3ScoreTemporalDemand}
\end{figure}

\newpage
\textbf{Temporal Demand:} The temporal demand scores for Task 3 can be seen in Figure \ref{fig:NASATLXTask3ScoreTemporalDemand}. The testers noticed that, if they could not create an overall understanding of the task rather quickly, it would hurt their chances of finding the root-cause. The old system had an overall higher score for temporal demand due to that. As the task was time-boxed, the testers felt a sense of urgency while accomplishing the task in the old system. In the new system, however, the time required to develop and comprehend an overview of the artifacts was less. The testers also felt fewer time constraints in the new system. The experienced testers noted a reduction of time needed to locate information and gather an overview, leaving more time for root-cause analysis.

\hfill

\begin{figure}[H]
  \centering
  \begin{tikzpicture}
    \begin{axis}[
      ybar,
      ylabel=NASA TLX effort score,
      symbolic x coords={T1, T2, T3, T4, T5, T6},
      xtick=data,
      xlabel= Testers,
      xticklabel style = {font=\small,yshift=0.5ex},
      nodes near coords,
      ymin=0,
      ymax=120,
    ]
    \addplot coordinates {(T1, 75) (T2, 65) (T3, 25) (T4, 75) (T5, 20) (T6, 50)};
    \addplot coordinates {(T1, 70) (T2, 50) (T3, 25) (T4, 75) (T5, 15) (T6, 50)};
    \legend{Old System, New System}
    \end{axis}
  \end{tikzpicture}
  \caption{NASA TLX effort scores of Task 3 for each tester (T), in both systems.}
  \label{fig:NASATLXTask3ScoreEffort}
\end{figure}
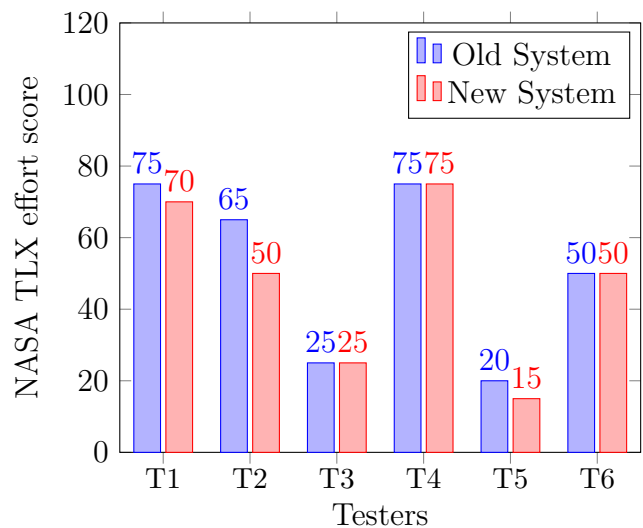

\newpage
\textbf{Effort:} The effort scores for Task 3 can be seen in Figure \ref{fig:NASATLXTask3ScoreEffort}. For the effort, the inexperienced testers scored 75, 65 and 25 in the old system. The first two testers suggested that the effort to find the required information from within the different tools presented itself as a hurdle in the task. For the experienced testers, the numbers were similarly high (70, 25 and 75) in the old system with the same feedback as the inexperienced testers. Another reason for such high effort was due to the time spent in not misinterpreting the information they found from the different tools and files. All the testers admitted that the new system was good for getting an overview and a better choice in attempting to solve easy-medium difficulty test cases, however, they would still prefer the old system for in-depth analysis. Even though the new system gave them easier access to information for the test artifacts, finding the root cause of the issue was not made easier with this change. The new system helped the testers to locate and utilize data quickly and efficiently but did not improve the testers ability to think. The mental efforts needed to find the root-cause would always remain high as that is based on testers experience. 

\hfill

\hfill

\begin{figure}[H]
  \centering
  \begin{tikzpicture}
    \begin{axis}[
      ybar,
      ylabel=NASA TLX frustration score,
      symbolic x coords={T1, T2, T3, T4, T5, T6},
      xtick=data,
      xlabel= Testers,
      xticklabel style = {font=\small,yshift=0.5ex},
      nodes near coords,
      ymin=0,
      ymax=120,
    ]
    \addplot coordinates {(T1, 60) (T2, 35) (T3, 30) (T4, 60) (T5, 15) (T6, 45)};
    \addplot coordinates {(T1, 30) (T2, 40) (T3, 15) (T4, 50) (T5, 15) (T6, 45)};
    \legend{Old System, New System}
    \end{axis}
  \end{tikzpicture}
  \caption{NASA TLX frustration scores of Task 3 for each tester (T), in both systems.}
  \label{fig:NASATLXTask3ScoreFrustration}
\end{figure}
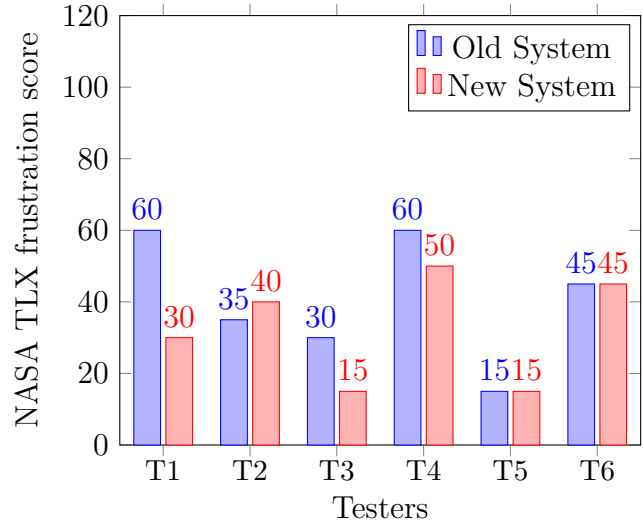

 %IE/E ==> 60,35,30 --- 60, 15, 45 OLD SYSTEM
%IE/E ==> 30, 40, 15 --- 50, 15, 45 New SYSTEM
\textbf{Frustration:} The frustration scores for Task 3 can be seen in Figure \ref{fig:NASATLXTask3ScoreFrustration}. The most frustrating aspects of the task for the testers were finding the needed data, artifacts and their information from the files, before attempting to find the root cause in the old system. Everyone agreed that they were habituated to the old system to the point that the once frustrating aspects that caused them trouble during the starting of their career eased eventually. All three inexperienced testers thought that the new system reduced efforts for navigation of data, reducing time for locating information. One tester opted to give a slightly higher value (from 35 to 40)  because the new system was something that the person needed to get used to. For the inexperienced testers, the score was practically the same. The new system did not help them deal with the frustration encountered when attempting to identify the issue in the test case. A common consensus here was that the required knowledge for finding root-causes was still unique to the individual testers, which is something that the new system does not improve. They did agree that it aided them during the task.  

\subsubsection{Task 4 Results}
\label{tlxResultsDiscussionTask4}
\begin{figure}[H]
  \centering
  \begin{tikzpicture}
    \begin{axis}[
      ybar,
      ylabel=Time taken in minutes,
      symbolic x coords={T1, T2, T3, T4, T5, T6},
      xtick=data,
      xlabel= Testers,
      xticklabel style = {font=\small,yshift=0.5ex},
      nodes near coords,
      ymin=0,
      ymax=30,
    ]
    \addplot coordinates {(T1, 20) (T2, 10) (T3, 20) (T4, 15) (T5, 15) (T6, 17)};
    \addplot coordinates {(T1, 5) (T2, 2) (T3, 8) (T4, 8) (T5, 5) (T6, 13)};
    \legend{Old System, New System}
    \end{axis}
  \end{tikzpicture}
  \caption{Time taken to complete Task 4 by the testers (T), in both systems.}
  \label{fig:timeCompletionForTask4Results}
\end{figure}
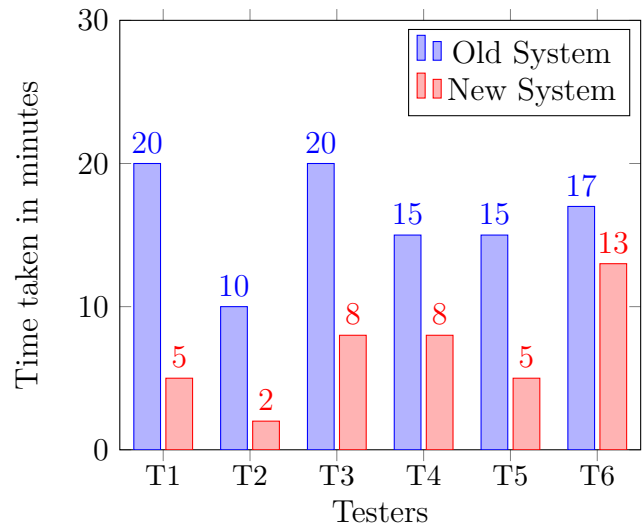

Time spent on Task 4 was greatly decreased in the new system compared to the old system. Introducing a feature where where all the stakeholders involved or responsible are listed in the new system was appreciated by the testers as they now had an option to narrow down which stakeholder they should contact to perform this task. The time taken for each testers for Task 4 can be seen in Figure \ref{fig:timeCompletionForTask4Results}. The total time for the testers in the old system was 97 minutes and 41 minutes in the new system, making it more than 50\% more efficient. 

\begin{mdframed}[style=MyFrame]
\begin{quote}
    \begin{itshape}
    "The stakeholder view did stick out (in the New System) and I think it could be very useful as it is lacking in the old system."
    \end{itshape}
    - Tester's feedback on Task 4.
\end{quote}
\end{mdframed}

\hfill

\begin{figure}[H]
  \centering
  \begin{tikzpicture}
    \begin{axis}[
      ybar,
      ylabel=NASA TLX score,
      symbolic x coords={T1, T2, T3, T4, T5, T6},
      xtick=data,
      xlabel= Testers,
      xticklabel style = {font=\small,yshift=0.5ex},
      nodes near coords,
      ymin=0,
      ymax=100,
    ]
    \addplot coordinates {(T1, 35) (T2, 54) (T3, 49) (T4, 68) (T5, 38) (T6, 51)};
    \addplot coordinates {(T1, 19) (T2, 33) (T3, 20) (T4, 23) (T5, 22) (T6, 18)};
    \legend{Old System, New System}
    \end{axis}
  \end{tikzpicture}
  \caption{NASA TLX scores of Task 4 for each tester (T), in both systems.}
  \label{fig:NASATLXTask4Score}
\end{figure}
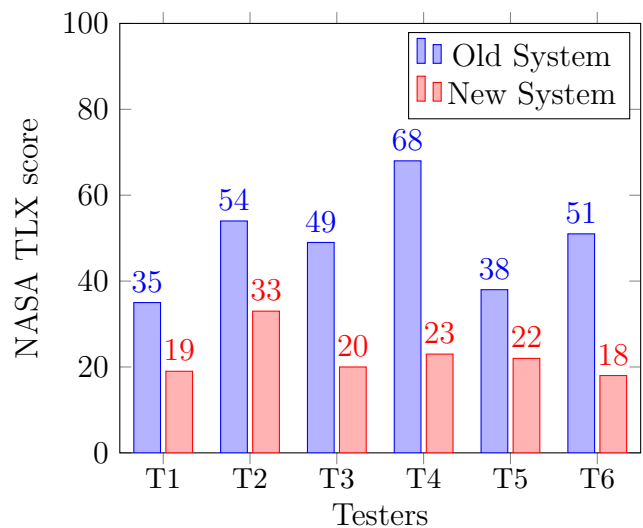

The NASA TLX scores in Task 4 can be seen in Figure \ref{fig:NASATLXTask4Score}. The scores were greatly improved in the new system compared to the old system for all the testers. In the old system, the information about stakeholders was not easy to find because it required backtracking to retrieve the information, and the experience and knowledge of the tester were needed to know where to find the information about stakeholders within the company or department. By presenting a list of stakeholders in the new system, the work effort was significantly reduced for the testers as the work was done by the tool itself. From the stakeholder list, the testers could make decisions on who was the most suitable stakeholder to contact regarding the test case. Even if an incorrect stakeholder was chosen, it was straightforward to choose another more suitable stakeholder from the stakeholder list in the new system.

\hfill

The stakeholder list provided a good additional feature to support testers decision making. By narrowing down and eliminating stakeholders until only the most suitable stakeholders were left, testers could complete this task without much work effort.

\hfill

\begin{mdframed}[style=MyFrame]
\begin{quote}
    \begin{itshape}
    "In the old system it’s actually hard to really know who is responsible and I usually just go to my product owner and ask him who to contact. By providing a list of the stakeholders, it makes it easier to narrow it down."
    \end{itshape}
    - Tester's feedback on Task 4.
\end{quote}
\end{mdframed}

\subsubsection{Summarizing Task Improvement Results}
Tasks 1, 2 and 4 showed more improvement in terms of task load index and time consumption scores than task 3. The reason was that task 1, 2 and 4 consisted of a lot of repetitive manual tasks which were difficult to perform in the old system. Task 3 required the testers to investigate root-cause issues in the test case, which depended a lot on how the testers evaluated the test results and situation. Aspects of Task 3 related to inspecting information were reduced in effort and time consumption. However, the thought process of each tester was unique and did not change when using either system. As a result, there was less improvement for Task 3 from the use of the new system.

%% file: include/Discussion.tex
% Creating a standardized way to structure data for everyone
% Structuring data saves time to do other productive tasks(Perspective of testers)
% Performance (Completion of test cases and tasks ) increase due to having all data structured
% Wrong cfg/db/measurement files, leading to wrong output, correctness is hampered. This improves in the new system 
% Models:
    %1. Elimination by aspects can be done faster, when they look through signals
    %2. Additive model increases, by having more usable data in a structured (prototype) / faster access to information / information is present, thus making correctness a key
    %3. Conjunction Mode, helps in the new system, by having information structured (Stakeholders)
% Requirement revision is hard to track, and people might use the old req instead of the new, such things need to be checked manually but in the new system, these things are shown to the testers as an overview (Saving time and improving performance)
% Using the new tool would help the new testers (Requires less exp in the testing of automotive domain) 
% Frequency of average test cases is more, and test cases that are that, can be quickly covered using the new system.
% System Quality, Information Quality and Information Presentation discussion. 

\chapter{Discussion}
\label{chapter:discussion}
This chapter starts with a discussion on the research questions of this study based on the results from Chapter \ref{chapter:results}. A framework will also be presented to improve automotive testing in the context of data presentation. Lastly, future work to extend this study will be discussed along with threats to the validity of this study.

%RQ1 What aspects deteriorate decision making during test analysis and diagnostics of automotive electronics?
% Having a standard well defined way performing and reporting performing test data diagnosis analysis
% Missing a system that could manage the scattered data (META DATA example)
% Information quality
    % Having the necessary information from the many information and data (Saving time)
    % Having the correct information required for making decisions.
% System Quality
    % Having to jump between tools to get want is needed
    % Ease of use of the tool 
% Presentation of information
    % Benifits of having a dashboard interface compared to solutions practiced presently
    % Having data categorized and presented so that navigation is within the tool rather than being constantly away from the tool.
\section{Deteriorating Factors for Decision Making}
\label{section:deterioratingFactorsDecisionMaking}
Recall RQ1: \textit{What aspects deteriorate decision making during test analysis and diagnostics of automotive electronics?}

\hfill

Various aspects lead to deterioration of decision making which were gradually uncovered throughout this study. Looking at the current challenges in test analysis and diagnostics for automotive electronics at the partner company, unstructured data and superfluous information were the primary challenges. Working with multiple tools, storage, and systems made finding and using the necessary information to conduct test analysis difficult. The causes of such problems were also related to excessive navigation between different tools and working with an abundance of information. All the findings that helped answer the RQ1 were categorized and can be seen in Table \ref{table:affinityDiagram}. Factors that made decision making more complicated than necessary for the testers were thoroughly investigated in the problem identification phase through interviews, questionnaires, observations and in the evaluation phases. The deteriorating factors were as follow: 

\hfill

\begin{itemize}
  \item Spending unnecessary time doing redundant things(such as manually collecting related files) in the testing process.
  \item Lacking a good presentation of test data to create an overview for the testers. 
  \item Not storing test-related data in a standard way.
  \item Needing to traverse and navigate large unstructured data and information using the different tools all at the same time.
  \item Needing to locate stakeholders to aid with fault mitigation of test cases.
\end{itemize}

\hfill

The testers workload and focus were used primarily to retrieve the relevant data and information to conduct test analysis and diagnostics. The correctness of the data was affected when it was not easily accessible. If the testers wanted to double-check the information to confirm its correctness, their workflow was slowed because of the difficulty in accessing data. By providing accessibility to the data, the workload and focus were instead shifted to the actual test analysis and diagnostic, making it easier for the tester to put in more effort on during analysis.

\section{Design of Test Analysis Framework with Decision Making Models}
\label{section:designOfTestAnalysisFrameworkWithDecisionMakingModels}
Recall RQ2: \textit{How should the design of a test analysis framework for automotive testing reflect the decision making models followed by testers?}

\hfill

% REFACTOR THIS TO FIT THE RQ2

The testers mostly had a typical process of thinking and making decisions in the old test system. For completion of each test task, the testers relied on the related data to decide the outcome. However, as data was not efficiently available to the testers, they could not make informed decisions effectively. For example, to begin the test analysis process, testers had to read and understand the requirements and collect all the files for starting the test analysis. The files were not easy to find because the testers had to search for them manually from an abundance of other files and eliminate the unnecessary ones. The approach and design of this process were not efficient as it increased effort to complete the tasks. To improve the design of the new system, we opted to understand and map how the testers perform the requirements during the test analysis process and what kind of decision making model best described their decision approach. Data for requirements that were mapped with the elimination by aspects model were placed in the new system where the testers would find it with the least amount of searching and elimination of unnecessary information possible. 

\hfill

The data for the requirements that were mapped with the additive/linear model and the conjunctive model were logically categorized in various views(web pages) in the new system. Each view represented a specific type of data. For example, the data related to requirements, test case steps, test output files (log, signal) and stakeholders were separated into separate web pages. Through evaluations of the design, the testers agreed that the design choice allowed them to better understand the data, which aided decision making because the information structure and presentation did not overwhelm or confuse them. By logically categorizing the data, the accessibility of information and decision making satisfaction increased. 

\hfill

Due to the logical categorization of information into their separate views, the navigation of information improved in the new system. The decision making models helped us to design the features of the requirement to the new system in an efficient manner to aid decision making by understanding the tester's decision making process for each feature and design the data structure accordingly.

\section{Structuring and Presenting Data}
\label{section:structuringPresentingData}
Recall RQ3: \textit{How should data be structured, presented, and documented in this test analysis framework to improve the quality of information available during test analysis?}

\hfill

Through evaluation and interviews with the testers, the issues that affected the old system regarding structuring test data and files, presenting test-related information and data efficiently and documenting stakeholders were all identified. Improvements were implemented by building prototypes of a new system. The low fidelity prototypes were improved upon by evaluating them with expert testers, finally coming to a consensus on the user interfaces, experience and data presentation. To ensure a better structure for data, we opted to take inspiration from web-based dashboards because they are intuitive in design and easy to understand. In the dashboard, we chose to present similar types of data as logical groups.

\hfill

As testers needed to use a lot of test-related data, the data was neatly organized in the left sidebar as can be seen in Figure \ref{appendix:high1Prototype_overview}. Each category in the sidebar revealed a new interface with additional information and data. The requirements for the test cases can be seen in Figure \ref{appendix:high1Prototype_dvm} and the signals from the test object for test data analysis and finding the root cause can be seen in Figure \ref{appendix:high1Prototype_step} and Figure \ref{appendix:high1Prototype_stepSubview}. The prototype made using Figma was later was converted to React.js to improve usability, and it became the final high fidelity prototype which was later used in the usability tests. The reason for opting to create the final high fidelity prototype using React.js was to make sure that, during the usability tests, the testers felt as if they were interacting with a web-based application. Throughout the development of the prototype, the discussions we had during the evaluation phases helped identify improvements. Efficiently structuring data and presenting it had a positive impact on the testers ability to solve issues in test cases. They acknowledged that the implemented improvements helped aid them in analyzing test cases by providing testers with a good overview of information. Table \ref{table:nodesForUsabilityInterview} describes the findings from the usability tests.

\hfill

The requirements that were developed and the design of the new system can be used as an inspiration to structure and efficiently present data for test analysis. This study showed that the new system was more time-efficient and required less effort for the testers to perform testing related tasks. General guidelines on how to structure and present data to support decision making and communicate verdicts efficient for test analysis of automotive electronics are as follows:

\hfill

\begin{itemize}
    \item Use a familiar design for the tool used by the testers such as a web-based system or dashboard. 
    \item Structure and present data of different categories in different views to not overwhelm the tester. For example, the test requirement should be in one view, while test object data should be in another view. 
    \item Give an overview of the test case as the landing page to help testers understand the context more efficient. Include the result of the overall test case and each step, test case dates and other essential data about the test case. 
    \item Offer relevant data that can be used for test analysis easily accessible, such as functional and test requirements, test object information and measurement configuration.
    \item Include revisions numbers on the data and information. Revision numbers are the different versions or states of the objects, data or information. If a requirement gets updated for instance, a new revision number is assigned which is usually an increment of the previous revision number.
    \item Include a list of stakeholders with contact information that are involved or relevant for the test object, requirements and functionality.
    \item Show the different test sequences or steps in the test case with an overview of the output. The output can be a summarized graph with sensor signal values and information.
\end{itemize}

\section{Accessibility of Test Analysis}
\label{section:accessibilityOfTestAnalysis}
Recall RQ4: \textit{How can this test analysis framework be designed to support ease of use and accessibility of test related data during test analysis?}

\hfill

The data used for test analysis should be placed in one tool that is easy for the testers to access. Having one tool instead of multiple tools increased the navigation efficiency between pieces of analysis data. By giving the testers the data related to requirements, test object, test cases, test diagnostics and stakeholders in one single tool, we increased the overall accessibility and efficiency of data collection. The categorization of data and presenting them into separate views in the same tool according to its association also increased accessibility because the testers could easily find and access the data they needed for the test analysis.

\hfill

Because we used a common web-based design for the system, the testers felt more comfortable with using the new tool. The motivation of using a web-based design was because it was commonly used for decision making systems \cite{bhargava2001decision} and is well known for all internet users. The design was intended to be user-centered and focus on the tester's needs for test analysis and diagnostics. The testers' responses from the evaluation phases and usability test indicated that the new system's design was easy to work with and made it more efficient to access the data needed for test analysis. 

\hfill

An interesting finding in this study was how much the testers appreciated the new system with its improved navigation. By providing more efficient navigation while working with large unstructured data, it made it easier for the tester to confirm their findings before making a decision, which reduced potential misinterpretation of information. In the study performed by Bharati and Chaudhury \cite{bharati2004empirical}, information presentation does not have a positive effect on decision making satisfaction where navigation efficiency is a part of information presentation. Their research was conducted for customers decision making during shopping. Given the different context between their work and this study, navigation efficiency was a useful factor when working with a large amount of unstructured data as it can sometimes be overwhelming for the user to remember everything. Information presentation also includes graphics, color and presentation style, but these factors were not something that was noticed by the testers in this study.

\section{Summarizing the Findings}
% write about the following
% talking points
    % pn ositives of the new system
       
        % Accessibility of information increased due to Struncturing of files
        % Time for locating information reduced
        
        % information presentation done efficiently
            % files needed to solve test cases are accessible faster due to 'requirment oriented' user interface reducing traversal efforts.
             % Mistakes due to interpretation or mis-information reduced
        
        % Suitable for solving easy-medium difficulty level test cases, as they don't require precise digging in the data
        
        % Highly recommended tool for testers who are in the early stages of their career or are simply less experienced, as it is easier to learn.
    
    % negative points on the new system
        % Granular level analysis
            % Not suitable in root-cause analysis for testcases that are difficult to solve as they need more precision in terms of data visualization(graphs) and a much more sophisticated data analysis tool which the old system provides (CanAnalyzer, Canoe, MatLab)
        
\begin{figure}[H]
  \centering
  \includegraphics[width=\textwidth]{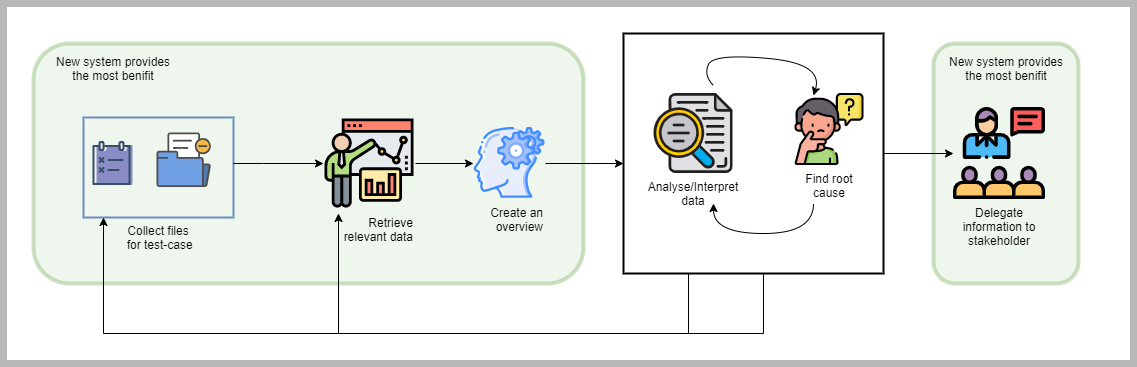}
  \centering
  \caption{Summarizing for the findings from the interview after usability test.}
  \label{fig:usabilityDesign}
\end{figure}

        % Does not enhance the ability to solve test cases but aids the process by making the preliminary aspects as stated in the (Positives) faster.
The findings revealed various problematic aspects in both the old and the new system. All testers agreed that information accessibility increased due to better structuring of the test-case related files and data. As a result, the time required to locate data reduced significantly. This can be seen in Figure \ref{fig:timeCompletionForTask2Results} that depicts the time taken for the Task 2 specifically. The presentation of information and data increased the perceived value of the new system. Improved presentation of data resulted in a decrease in time and effort for Tasks 1, 2 and 4. Information was accessible faster with little to no context switching, which helped in decreasing misinterpretation of information. All of this was achieved due to fulfilling user requirements 1-10. The benefits of the new system can be summarized in Figure \ref{fig:usabilityDesign}.

\hfill

\hfill

The testers agreed that the new system would be a better choice to analyze test cases with easy-to-medium difficulty. The new system would also be a better option for testers who were reasonably new to testing. It would be a stepping stone for them, which would gradually train them to move into more sophisticated tools and systems eventually.

\newpage
\begin{mdframed}[style=MyFrame]
    \begin{quote}
        \begin{itshape}
        "It’s a higher learning threshold in the old system compared to this new system which is more intuitive and user-friendly. In the old system, it can be challenging to find the correct information because there is a lot of software and other tools that sometimes is not that useful for test analysis and diagnostics. The new system on the other hand make it more appealing to the less experienced person and gives them a chance to slowly dive into our work."
        \end{itshape}
        - Tester's feedback on the new system.
    \end{quote}
\end{mdframed}

\hfill

The old system, however, had the advantage when dealing with difficulty test cases. The testers, regardless of their experience, considered the old system to be better at a low-level analysis of test cases. The need for further analysis requires working with tools that offer precision. If they needed to work their way around a difficult test-case, they could use the new system to retrieve and visualize data and use the old system for analyzing issues and finding root-causes. Analyzing test cases also requires experience and knowledge on the domain, which neither of the systems can provide as they are merely tools at a tester's disposal. However, the new system aided the overall process of testing. The testers felt that the four individual tasks felt more relaxed compared to the old system because they did not have to perform several repetitive and trivial tasks in the new system.

\hfill

%ionormat COMM anENTED FOR LATER USE
\newcommand{\commentout}[1]{
\begin{figure}[h!]
  \centering
  \begin{tikzpicture}
    \begin{axis}[
      ybar,
      ylabel= Mean scores for task load,
      symbolic x coords={Mental, Temporal, Effort, Frustration},
      xtick=data,
      nodes near coords,
      ymin=0,
      ymax=100,
    ]
    \addplot coordinates {(Mental, 46.5) (Temporal, 52.8) (Effort, 48.8) (Frustration, 49.1)};
    \addplot coordinates {(Mental, 22.5) (Temporal, 16.8) (Effort, 37.5) (Frustration, 22.5)};
    \legend{Current System, New System}
    \end{axis}
  \end{tikzpicture}
  \caption{NASA Task Load Index mean score of each usability test task.}
  \label{fig:usabilityTestNasaTLX}
\end{figure}

\hfill
}

\section{Benefits and Implication On Practice}
\label{section:discussionFramework}

\begin{figure}[ht!]
  \centering
  \includegraphics[scale=0.3]{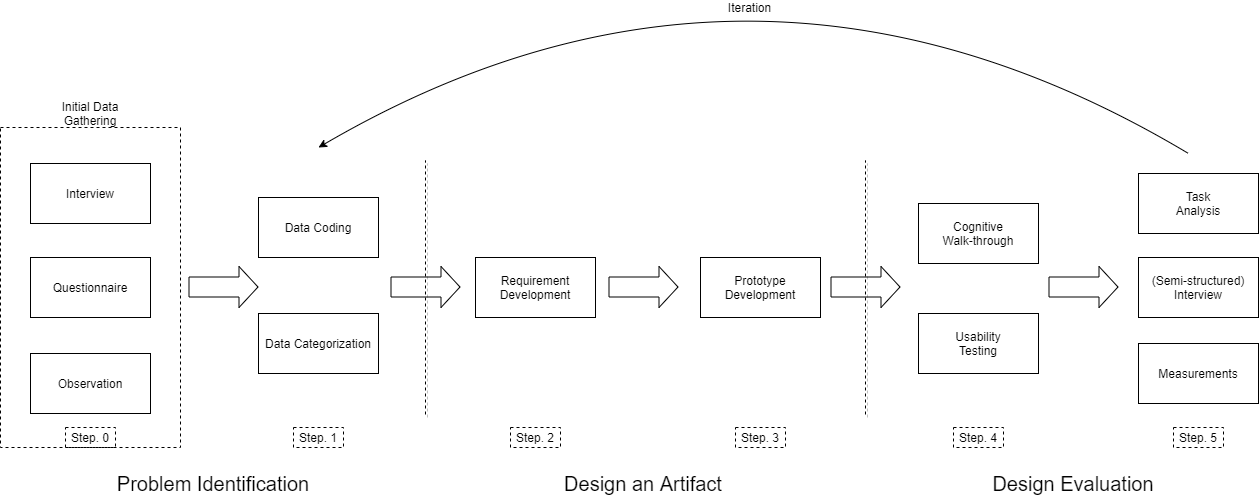}
  \centering
  \caption{The framework based on design science methodologies used in the study.}
  \label{fig:frameworkUserInTheStudy}
\end{figure}

% Tangible and In-Tangible benefits, recap it based on both the research questions targetting improvements in terms of business and technicality (Testing diagnostics improvement)
Many problems and challenges could be investigated in the domain of testing in the automotive industry. Little work has been done regarding how individuals perform tasks and how specific task workload affects their work in the automotive industry. In this study, an approach to improve task performance and the workload was conducted by improving the data structure and presentation. The methodologies used in this study to find improvements can be presented in a generic framework (see Figure \ref{fig:frameworkUserInTheStudy}) that companies can use to identify areas to improve in regards of data structure and presentation. Companies can use this framework as a guide to investigate problems and challenges within the testing domain in their existing business process by carrying out internal research. Due to the abstract nature of the framework, we believe that it can be adapted and used in similar domains to improve and overcome similar challenges investigated by this study.  

\hfill

The requirements we developed along with the testers can also help create awareness of the problems testers face and how those problems can be rectified by implementing solutions that help elevate the stakeholders. The findings from this study also show areas where performance in automotive testing can be improved by creating awareness of unstructured data, inefficient test analysis practices, presentation of test data and advantages/disadvantages of the testing tool within the testing domain. By starting on one area to improve the unstructured data, such as testing analysis, the awareness can be spread to other areas with challenges of unstructured data within the company. General guidelines based on the findings of this study to improve test analysis challenges are as follows:

\hfill

\begin{itemize}
    \item Introduce a standard method for stakeholders to store and manage data. When there is no standard way to store data, it will require more time and effort to find the relevant data.
    \item Use the right tools for the task and minimize the usage of unnecessary tools to decrease navigation and context switching during test analysis as this can be overwhelming for testers and can lead to loss of concentration. 
    \item Communication between stakeholders can be improved by mapping responsible stakeholders to their relevant object, artifact or domain to ask for aid efficiently. 
    \item Requirements revisions can be tracked to reduce misinterpretation of information during test analysis. 
\end{itemize}

\hfill

The unstructured nature of the data is not an exclusive challenge in the testing domain as testing analysis. This study looks into how the data structure and presentation can be used as a solution to improve task performance and workload when interacting with a system. The design and evaluation methods used in this study can be applied to data visualization research to identify further challenges regarding the relationship between data presentation and workload performance on analytical tasks.

\hfill

A previous study by H. Wang et al. \cite{wang2012mavis} showed how better visibility of the results and outputs of tests benefits testers with a quicker general perspective. With this study, the same results were achieved by providing a better overview of the results, outputs and information of the test case. We also used NASA TLX as a tool to evaluate the performance to justify the findings. We believe that the NASA TLX is a powerful tool to assess the performance on users when conducting research for visualization tools to improve effectiveness and efficiency of test analysis as it is essential to understand the interaction of testers with the visualization tool as well.

% time wasting in preparing for tests 
% unstructured data awarness
\section{Threats to Validity}
\label{section:threatsToValidity}

\subsection{Internal Validity}
Controlling external factors such as participants fatigue levels when performing usability tests is important, and this was done by giving them a fixed recovery time between tasks. Still, different participants may need different amounts of time to recover, and that could influence the performance of the participants in the usability tests. We assumed that 15 minutes of rest time would be sufficient for all participants. Another possible threat to internal validity was that the participants in the usability tests, might have had discussions with the testers involved in evaluating the prototypes. There was no way for us to completely isolate the testers used for a usability test and the ones used for evaluating the prototypes. However, we explained the importance of such isolation to the testers beforehand. We also made sure that learning effects were kept at a minimum by not having the testers repeat the same specific task in both systems during the usability tests.

\hfill

Qualitative coding of the results may be biased by our existing hypotheses regarding the research questions, and gathered qualitative data could be interpreted subjectively. We tried to minimize this effect by doing the coding together without being biased towards our personal beliefs and perspectives. We had discussions with each other so that the findings from the coding sessions reflected the data with limited personal interpretation.

\subsection{External Validity}
The sample size for the interview, questionnaire, evaluation, and usability test subjects could have been larger. Due to time constraints of the study, we could not perform the usability tests with testers from other departments in the company. We focused on one department (Power and Energy dept), which could hamper external validity. To partially deal with this problem, we tried to broaden our data collection to incorporate the knowledge from testers working in various departments so that the new system would be created from a broader perspective. Since this study was conducted with one specific industrial partner, the generalization might be limited due to the particular setting of the company. Due to time constraints, empirical evidence from usability tests, the study of the existing testing process for testing and its impacts on the stakeholders could not always be studied to a greater extent. However, to mitigate this, we had multiple interviews with the same participants on different days to evaluate consistency in their answers while taking the extra time and effort to validate our understanding of their replies through follow-up discussion. We believe the results of this analysis are applicable, at minimum, at other companies in the automotive domain.

\subsection{Construct Validity}
As our research relies on using subjective measurements to answer the research questions, there could have been times when we might have unwillingly and unknowingly influenced the stakeholder actions. During interviews, we tried not to affect the stakeholders' answers, but we might have designed the questions in ways that would help influence the resulting responses. Hence, this could be a threat to construct validity because research biases may have been reflected in the questions in the interviews. During the participant observation, we tried to isolate ourselves from the participants completely. Despite our best efforts, the feeling in the participants was noticeable regarding, being watched and could have affected their way of working during the observation session. Finally, there could have been additional test cases for usability testing for both the systems because the participants could have had better/worse scores in the TLX their feelings when analyzing a particular test. The participants were given time to rest to minimize such effects.

\section{Future Work}
\label{section:futureWork}
% Can provide otehr studies with inspriation and data as this area is not that %  researched about in automotive industry

% investigate more in the domain of cognitve workload and the automotive domain specially testing done by humans.

%Applying the framewok in another setting in another organization
Cognitive research, especially regarding workload on individuals in the testing domain within the automotive industry, has not been given the focus it needs. Hence, further investigation in cognitive workload within the automotive testing domain could be done to broaden the problems, and aspects of the testing process addressed. The current research for testing of automotive electronics and ECU testing has focused largely on generating and optimizing the output data. With this study, we also hope there will be more interest in examining other aspects, such as data presentation and structure. 

\hfill

Data structuring and presentation are aspects which can be further investigated as this study showed improvements in terms of NASA TLX. The focus should also be put upon how the data (files, sensor signal data, documentation) is being used in an organization rather than within different domains. Furthermore, the study can inspire researchers to take into consideration the aspects of mental workload and how it affects decision making.

%% file: include/Conclusion.tex
% CREATED BY DAVID FRISK, 2016
%You may consider to instead divide this chapter into discussion of the results and a summary. 

% Clearly state the answer to the main research question
    % This is mostly done in Discussion Chapter already
% Summarize and reflect on the research
    % While X limits the generalizability of the results, this approach provides new insight into Y.
    % This research clearly illustrates X, but it also raises the question of Y.
% Make recommendations for future work on the topic
    % Based on these conclusions, practitioners should consider…
    % To better understand the implications of these results, future studies could address…
    % Further research is needed to determine the causes of/effects of/relationship between…
% Show what new knowledge you have contributed
    % Returning to your problem statement to explain how your research helps solve the problem.
    % Referring back to the literature review and showing how you have addressed a gap in knowledge.
    % Discussing how your findings confirm or challenge an existing theory or assumption.

\chapter{Conclusion}
\label{section:conclusion}

In this study, we explored the testing domain in the automotive industry, mainly focusing on testing of automotive electronics and its challenges in regards to data structure and presentation. An exploratory case study was performed with Volvo Cars, where we conducted multiple elicitation and evaluation sessions to investigate the current challenges in data structure and presentation within the testing of automotive electronics. Thirteen stakeholders and six testers in the partner company were involved in this study.

\hfill

A prototype was created for this study to illustrate better data structure and presentation. The design of the prototype was based on good design practices and stakeholders' feedback, knowledge, experience and evaluation results. The final evaluation was conducted in the form of a usability test on the prototype to gather data for analysis to answer the study's research question. The prototype's results show that providing better data structure significantly increases the efficiency and reduces the workload for testers when conducting test analysis and diagnostics. The correctness of the data and information was also improved by increasing the accessibility of the information and aided with decision making during the test analysis. 

\hfill

In conclusion, managing data is a challenging task when there is an overwhelming amount of data present. The management of overwhelming data will always be a challenge for a company, and it can be not easy to have perfect data management, structure and presentation. Still, challenges can be minimized with the right practices. Good data structure and presentation is essential to increase the efficiency and correctness of the testing process for the stakeholders involved. 

%% file: include/backmatter/Appendix_1.tex
% CREATED BY DAVID FRISK, 2016
\chapter{Interview Guides}

\section{Interview guide for product owner (Initial round)}
\label{appendix:interviewProductOwner}

\textbf{Goals for the interview}
\begin{itemize}
    \item Understanding the existing testing process overview.
        \begin{itemize}
            \item Protocols for testing.
            \item Testing process as a whole.
            \item Involved stakeholders in the testing process.
        \end{itemize}
    \item Understanding the problems in the process from the perspective of the interviewee
\end{itemize}

\hfill

\textbf{Questions}
\begin{enumerate}
    \item What is your name?
    \item Which department do you work in?
    \item Which domain does your department expertise in?
    \item What can you tell about the testing process?
    \item What can you tell about the testing process?
    \item Who performs the tests in your department?
        \begin{enumerate}
            \item What other responsibilities do they have apart from testing?
        \end{enumerate}
    \item We heard that the department is migrating from manual to automated testing can you tell something about this?
    \item Would you describe the process of the existing testing as a whole which is practiced in the department?
        \begin{enumerate}
            \item Which stakeholder groups are involved here?
            \item Are there any typical problems with the current testing process? If yes, what are they?
            \item Which stakeholder group suffers the most due to the problems in the testing process?
        \end{enumerate}
    \item What is the role of a tester in the test process?
    \item Which tester groups should we talk to in order to know more about the work of a tester?
\end{enumerate}

\newpage
\section{Interview guide for testers (Initial round)}
\label{appendix:interviewTester}

\textbf{Goals for the interview}
\begin{itemize}
    \item Further investigate the role of a tester in the testing process.
    \item Find out the practices testers do during and after the test process is done
    \item Understanding the problems in the process from the perspective of the interviewee
\end{itemize}

\hfill

\textbf{Questions}
\begin{enumerate}
    \item Would you kindly introduce yourselves?
    \item Are you all performing testing in your teams?
    \item Is the testing process common in this department?
    \item How is the test process in your perspective, as a tester?
        \begin{enumerate}
            \item What are the objects that you test on?
            \item Do you write test-cases for the tests you do on test-objects?
            \item Can you show us examples of test-cases for test-objects?
            \item Which tools do you use for testing and why?
        \end{enumerate}
    \item What are the problems in the existing testing process?
    \item What do you do after the testing is done?
    \item How do you perform analysis on the test output data?
    \item Do you use any tools to perform your analysis?
\end{enumerate}

\newpage
\section{Interview guide for Evaluation Phase}
\label{appendix:interviewEvaluationPhase}

\textbf{Goals for the interview}
\begin{itemize}
    \item Evaluate and validate the requirements of the artifact (prototype).
    \item Find improvements in current requirements or new additional requirements.
    \item Investigate the data/information structure and presentation of the prototype.
\end{itemize}

\hfill

\textbf{Questions}
\begin{enumerate}
    \item Did the prototype provide useful for test analysis and diagnostics?
    \item What parts in the prototype were the most useful?
    \item What part in the prototype were not useful?
    \item Is there some parts in the prototype that could be improved?
    \item Is there something that could be added to help with the test analysis and diagnostics?
    \item Compared to the current system and process you are using for test analysis and diagnostics, are there something missing in the prototype?
    \item How was the data structure of the data in the prototype? Described in each specific part.
    \item Did the new data structure in the prototype help to draw conclusions/make decisions in term of efficiency and correctness? 
    \item How was the data and information presentation in the prototype? Described in each specific part.
    \item Did the new data and information presentation in the prototype help to draw conclusions/make decisions in term of efficiency and correctness?
    \item Is there something else you want to add?
\end{enumerate}

\newpage

\section{Interview guide for Usability Test}
\label{appendix:interviewUsabilityTest}

\textbf{Goals for the interview}
\begin{itemize}
    \item Compare the two systems in regards to Effort, Efficiency, and Correctness.
    \item Investigate the research questions:
        \begin{itemize}
            \item How the data structure and presentation helped in the new system.
            \item What aspects and factors the new system provided was an improvement.
            \item What in the current system was causing it to be difficult.
        \end{itemize}
    \item Target research questions:
        \begin{itemize}
            \item How to structure and present data to improve decision making from unstructured data in the context of test analysis and diagnostics of automotive electronics?
            \item How to structure and present data in order to find the relevant information needed to make decisions more efficiently and with reduced effort?
            \item Which aspects in the presentation of data deteriorate decision making during test analysis and diagnostics of automotive electronics?
        \end{itemize}
\end{itemize}

\subsection{Comparison of the two systems}
\begin{enumerate}
    \item Comparing the old and the new system, how do you think the new system fares in terms of usefulness?
    \item How was the accessibility of the required information to solve tasks in the new system different from the old? Was it an improvement, if yes then why?
    \item While performing sub-task involving collecting the necessary files and information for the test diagnostics using both systems, can you describe your experience in terms of:
        \begin{enumerate}
            \item How was the effort spent?
            \item How efficient was it?
            \item How prone to making incorrect decisions?
        \end{enumerate}
\end{enumerate}

\subsection{Questions targeting unstructured data and information presentation [RQ1]}
\begin{enumerate}
    \item How is the data presented that can be challenging for the test diagnostics in the current system? In the new system?
    \item Does the information for performing test diagnostic in the current system, in regards to DVM, log files, information about the test case, software, hardware and stakeholder, difficult to accumulate/gather?
    \item Do you believe that having all the information easily accessible would have made your diagnosis better? If your answer is yes in regards to:
        \begin{enumerate}
            \item Effort needed
            \item Efficiency
            \item Correctness
        \end{enumerate}
    \item How did the new system do compared to the existing? If there is an improvement can you elaborate in terms of effort and efficiency?
    \item How does the data structure in the new system compare to the old system?
        \begin{enumerate}
            \item Does it bring any improvement? If yes, then elaborate.
            \item Does it bring any improvement in terms of effort and efficiency? If yes, which aspects do you believe improves the effort spent and makes it more efficient?
        \end{enumerate}
\end{enumerate}

\subsection{Questions targeting decision making in terms of data presentation and structure [RQ2]}
\begin{enumerate}
    \item What are the factors in the current system that could make you take the wrong conclusions or decisions? In regards to:
        \begin{enumerate}
            \item Unstructured files and information
            \item Amount of information viewed at once
        \end{enumerate}
    \item Would categorization of relevant information be visually appealing while dealing with test diagnosis?
    \item Do you think the new system improves decision making? If so, how?
        \begin{enumerate}
            \item Do you think the presentation of the data has an influence on the improvement?
            \item Do you think the unified structure of allocating all the necessary files and data in a central place would improve the time and effort required to take or make decisions?
            \item Do you think that the factors you referred to are dealt with in the new system?
        \end{enumerate}
\end{enumerate}

\subsection{Questions targeting the structural representation done in the new system [RQ2]}
\begin{enumerate}
    \item What is your thought on the structural representation of the data in the new system?
    \item Was having an overview of the test case helpful?
    \item Was having information about car information or measurement configuration helpful?
    \item Was having information about hardware/software information helpful?
    \item How important was the requirement and DVM section? Is it useful for test diagnosis? Did it aid in decision making?
    \item How important was the historical test-case section? Is it useful for test diagnosis? Did it aid in decision making?
    \item Was the step overview section useful for the diagnosis? Did it aid in decision making?
    \item The steps were specifically shown in the new system. How useful was this? Did it aid in decision making?
    \item The signals represented using graphs, how useful were they in terms of understanding the failed signals and making decisions off it?
    \item Was the signals with comment fields to share knowledge between each other useful?
    \item Will it be useful to add other signals to the steps directly from the new system? Will it provide useful?
\end{enumerate}

%% file: include/backmatter/Appendix_2.tex
\chapter{Questionnaire Guide}

\section{Questionnaire guide for Problem Identification}
\label{appendix:questionnaireProblemIdentification}

\begin{enumerate}
    \item How much are you involved within the testing process?
        \begin{enumerate}
            \item Tester: Prepare and conducting tests. Also, analyzing the results of the test.
            \item Analyzing: Solving the errors from the test results. Backtrack to the test results if necessary.
            \item Both: Doing both of the above.
        \end{enumerate}
    \item Does your team or the other teams see the testing as an important part of the development life-cycle? Rate them within a scale of 1-5. 
    \item Questions about DVMs. Rate them within a scale of 1-5. 
        \begin{enumerate}
            \item How much is the effort of creating a DVM?
            \item Is it time consuming to create a DVM?
            \item Is the current way of creating DVM sufficient and optimal?
            \item Do you believe that having DVMs adds value to the test process?
            \item Do you think that the test sequences effectively target the outcome of the test?
        \end{enumerate}
    \item Are there some challenges or limitation when DVM? (multiple choice options)
        \begin{enumerate}
            \item Cannot express what I want to test.
            \item The process of writing DVM feels limited.
            \item The knowledge of the function is low when writing a DVM in the early phases.
            \item No, not what I can think of.
            \item Other.
        \end{enumerate}
    \item If have concerns about DVMs, what are the issues with them?
    \item How do you ensure the quality of the test sequence?
        \begin{enumerate}
            \item I verify it against the DVMs Guidance.
            \item I have enough prior experience in writing test sequences.
            \item I map out requirements and verify whether the test sequence fulfill them. I use my own methods to perform this.
            \item I use guidelines provided by the team or department.
            \item Other.
        \end{enumerate}
    \item Questions about the effective time in preparation of the tests. Rate them from 1-5: 1-2 hours, 3-4 hours, 5-6 hours, 7-8 hours, 9+ hours.
        \begin{enumerate}
            \item How much effective time does it take to prepare the testing environment? i.e software, car
            \item How much effective time does it take to conduct the actual test?
        \end{enumerate}
    \item What is the time consuming task during the preparation of the testing environment? (multiple choice options)
        \begin{enumerate}
            \item Setting up the testing software.
            \item Setting up the environment cars.
            \item Configurations (e.g software, hardware).
            \item Other.
        \end{enumerate}
    \item What is the time consuming task while conducting the actual test? (multiple choice options)
        \begin{enumerate}
            \item Waiting for it to finish.
            \item Faults related to software or hardware.
            \item Badly written DVM.
            \item Badly written requirement.
            \item Other.
        \end{enumerate}
    \item Questions about the post testing process. Will focus on the process after the tests are done and when an error occurs or a fault is found in the test results. Rate them within a scale of 1-5.
        \begin{enumerate}
            \item How often is it because of the test sequence?
            \item How often is it because of the DVM?
            \item Is the results given from the tests sufficient to analyse and draw conclusion?
            \item Does it take a lot of time work out and analyse error or fault in the results given from the tests?
            \item How would you rate the effort when analysing the error or fault in the results given from the tests?
            \item When you found the source of the problem, how often is the correct  stakeholder that is contacted the right one?
        \end{enumerate}
    \item How much time do you usually spend working with the results given by the test before making a conclusion?
        \begin{enumerate}
            \item <1 hour
            \item 1-2 hours
            \item 3-4 hours
            \item 5-6 hours
            \item Other.
        \end{enumerate}
    \item What are the issues or challenges when moving from manual testing to automated testing? (multiple choice options)
        \begin{enumerate}
            \item The functions or steps I want to test cannot be automated at the moment.
            \item It takes time to learn automated testing.
            \item It takes time to prepare necessary preparation for automated testing.
            \item I cannot backtrack the results provided from the automated testing effectively.
            \item I prefer manual testing. 
            \item Other.
        \end{enumerate}
    \item What kind of issues do the test data, log files and results from the automated testing have? (multiple choice options)
        \begin{enumerate}
            \item The large amount of data.
            \item The large amount of irrelevant data.
            \item It is hard to find the relevant data.
            \item Missing information or data.
            \item The overall structure of how the test data and log files are stored.
            \item The data/log files are not in the same place. \item Lacking better visualization.
            \item Difficult to map the test case(DVM) with the output of the data.
            \item The files related to the test-object are stored in various places, making it difficult to collectively analyze them.
            \item Other.
        \end{enumerate}
    \item Which software do you use for analyzing the results of the tests? (multiple choice options)
        \begin{enumerate}
            \item CANalyzer
            \item CANape
            \item Other.
        \end{enumerate}
    \item Is there some other problems or challenges you want to add?
    \item How long have you been in your current position in the department?
        \begin{enumerate}
            \item 0-1 years
            \item 1-2 years
            \item 3-4 years
            \item 5+ years
        \end{enumerate}
    \item Demographic: Age
        \begin{enumerate}
            \item <20
            \item 21-30
            \item 31-40
            \item 41-50
            \item 50+
        \end{enumerate}
    \item Demography: Nationality
\end{enumerate}

\newpage

\section{Questionnaire guide for gathering subjects for Usability Test}
\label{appendix:questionnaireSubjectUsabilityTest}

\begin{enumerate}
    \item Name
    \item Email
    \item Department and Team
    \item Experience in the field of testing (in years)
\end{enumerate}

%% file: include/backmatter/Appendix_3.tex
\chapter{Cognitive Walk-through Tasks}
\label{appendix:cognitiveWalkthroughTasks}

Tasks for the participants to perform for the cognitive walk-through. The tasks are made for the participants to find the necessary information and data in the prototype that could be helpful during test analysis and diagnostics. 

\section{Tasks for Low-Fidelity Prototype}
\label{appendix:cognitiveWalkthroughTasksLow}

\begin{enumerate}
    \item Get information about the test case.
        \begin{enumerate}
            \item Open the system.
            \item Go to the Overview.
            \item Get general information.
            \item Get Stakeholder and Tester information.
            \item Get hardware and software information.
        \end{enumerate}
    \item Get information about the requirement and DVM of the test case.
        \begin{enumerate}
            \item Open the system.
            \item Go to DVM.
            \item Get description.
            \item Get test sequence.
        \end{enumerate}
    \item Get information about a specific step.
        \begin{enumerate}
            \item Open the system.
            \item Go to a step number.
            \item Get the sensor signals involved in that specific step.
            \item Read the description of the specific step.
            \item Compare the sensor signals outputs in the graph.
        \end{enumerate}
    \item Get information about test cases which are similar with the current test case.
        \begin{enumerate}
            \item Open the system.
            \item Go to Historical Data.
            \item Get the list of similar test cases.
        \end{enumerate}
\end{enumerate}

\newpage

\section{Tasks for High-Fidelity Prototype}
\label{appendix:cognitiveWalkthroughTasksHigh}

\begin{enumerate}
    \item Get an overview of the test case.
        \begin{enumerate}
            \item Open the system.
            \item Go to Overview.
            \item Get the results of the test case.
            \item Get an overview of the steps in the test case.
            \item Get the sensor signals used in the test case.
            \item Get the buses involved in the test case.
        \end{enumerate}
    \item Get information about the requirement and DVM of the test case.
        \begin{enumerate}
            \item Open the system.
            \item Go to the DVM.
            \item Get description.
            \item Get test sequence.
        \end{enumerate}
    \item Get information about test cases which are similar with the current test case.
        \begin{enumerate}
            \item Open the system.
            \item Go to Historical Data.
            \item Get the list of similar test cases.
        \end{enumerate}
    \item Get the information related to the test case.
        \begin{enumerate}
            \item Open the system.
            \item Get the software and hardware information.
            \item Get the car information, such as database and configurations.
            \item Get the tester information.
            \item Get the stakeholder information.
        \end{enumerate}
    \item Get the links to the file-repository of the log-files used in the test case.
        \begin{enumerate}
            \item Open the system.
            \item Go to link to log files.
        \end{enumerate}
    \item Get information about a specific step.
        \begin{enumerate}
            \item Open the system.
            \item Go to a step number.
            \item Get the sensor signals involved in that specific step.
            \item Read the description of the specific step.
            \item Compare the sensor signals outputs in the graph.
        \end{enumerate}
\end{enumerate}

%% file: include/backmatter/Appendix_10.tex
\chapter{Coded Interviews}
\label{appendix: codedInterviews}

\begin{sidewaysfigure}
    \centering
    \includegraphics[width=.9\textheight]{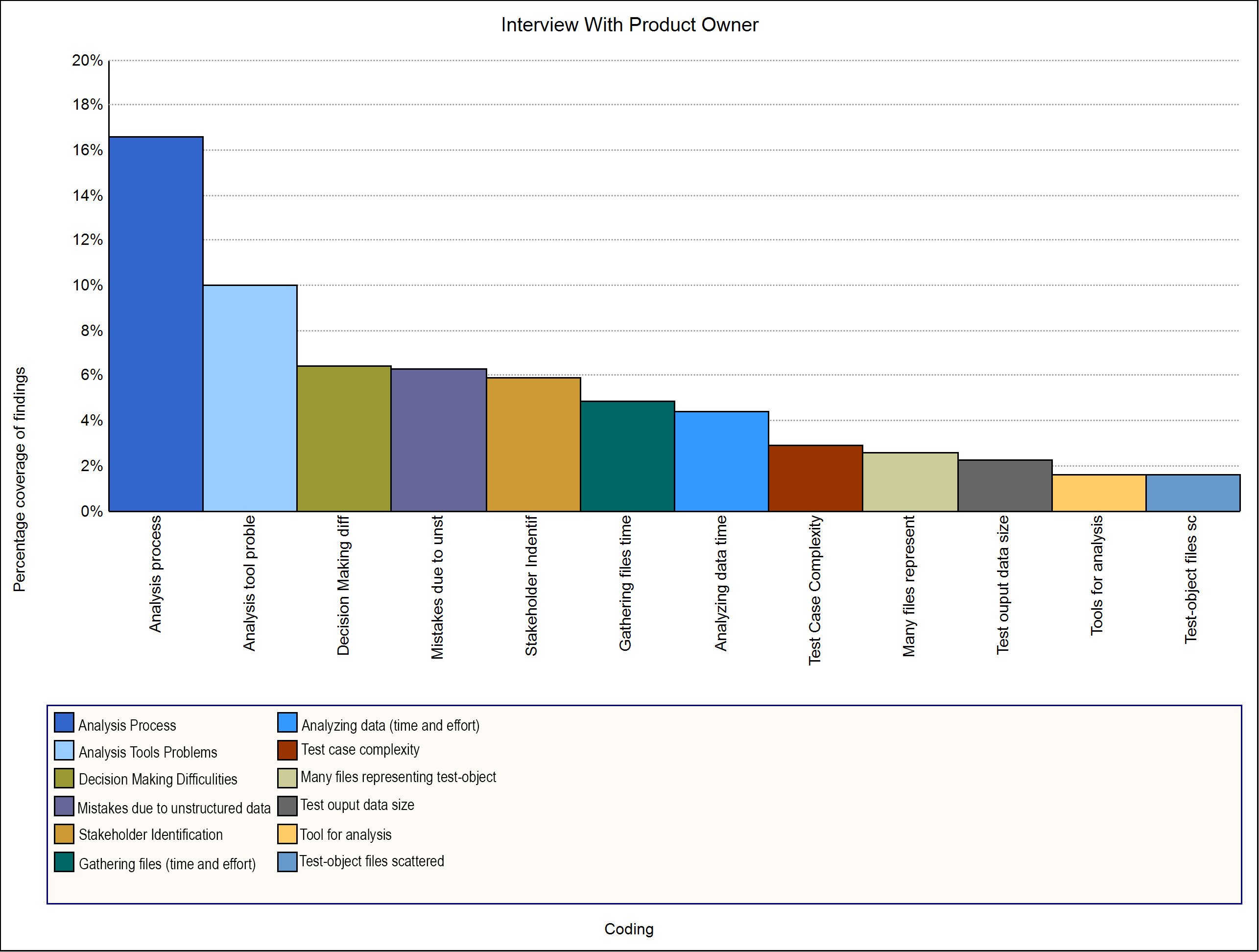}
    \caption{Findings from the interview with product-owner.}
    \label{appendix:codedInterviews:productOwner}
\end{sidewaysfigure}

\begin{sidewaysfigure}
    \centering
    \includegraphics[width=.9\textheight]{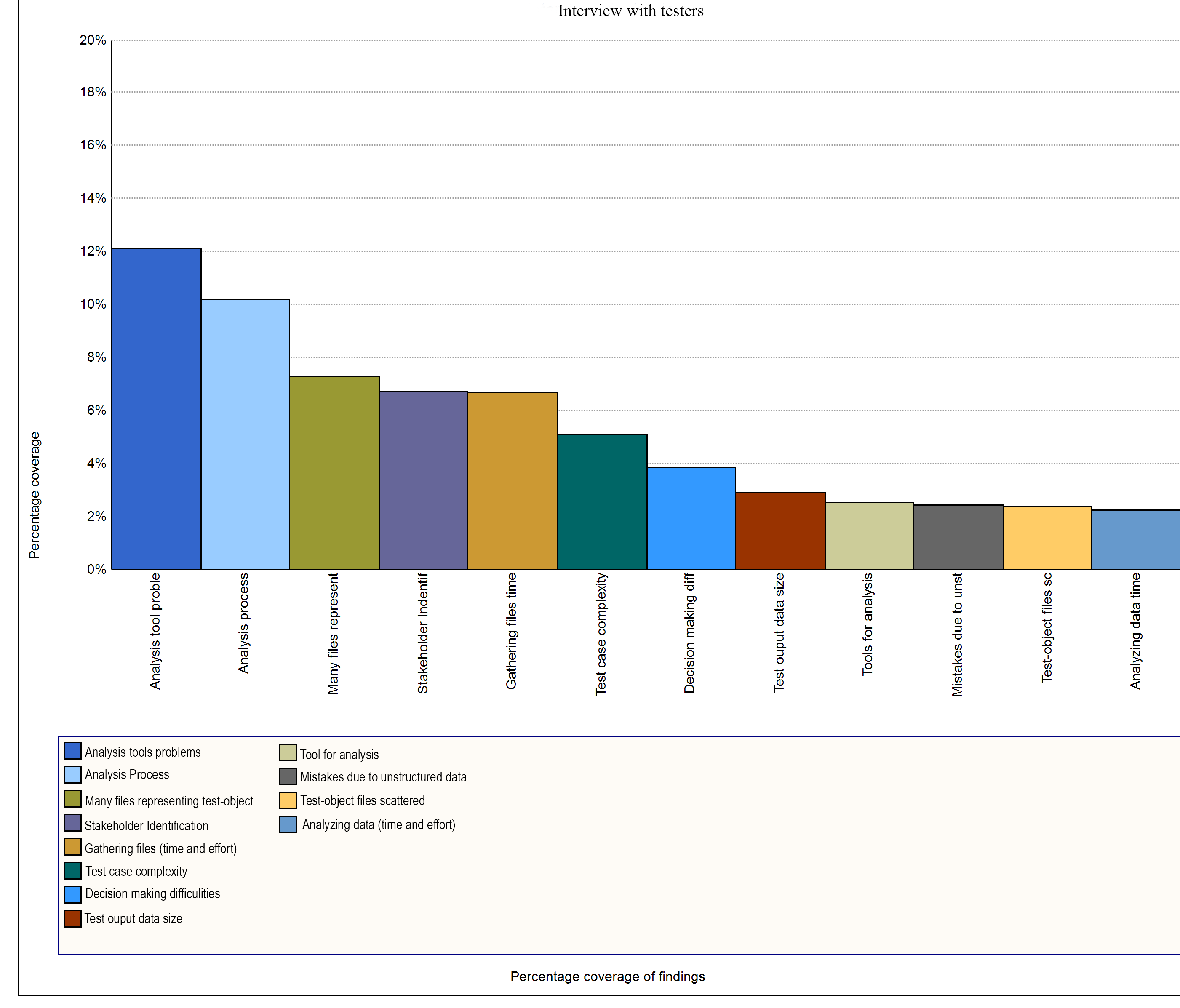}
    \caption{Findings from the interview with testers group 1.}
    \label{appendix:codedInterviews:testerGroup1Interview}
\end{sidewaysfigure}

\begin{sidewaysfigure}
    \centering
    \includegraphics[width=.9\textheight]{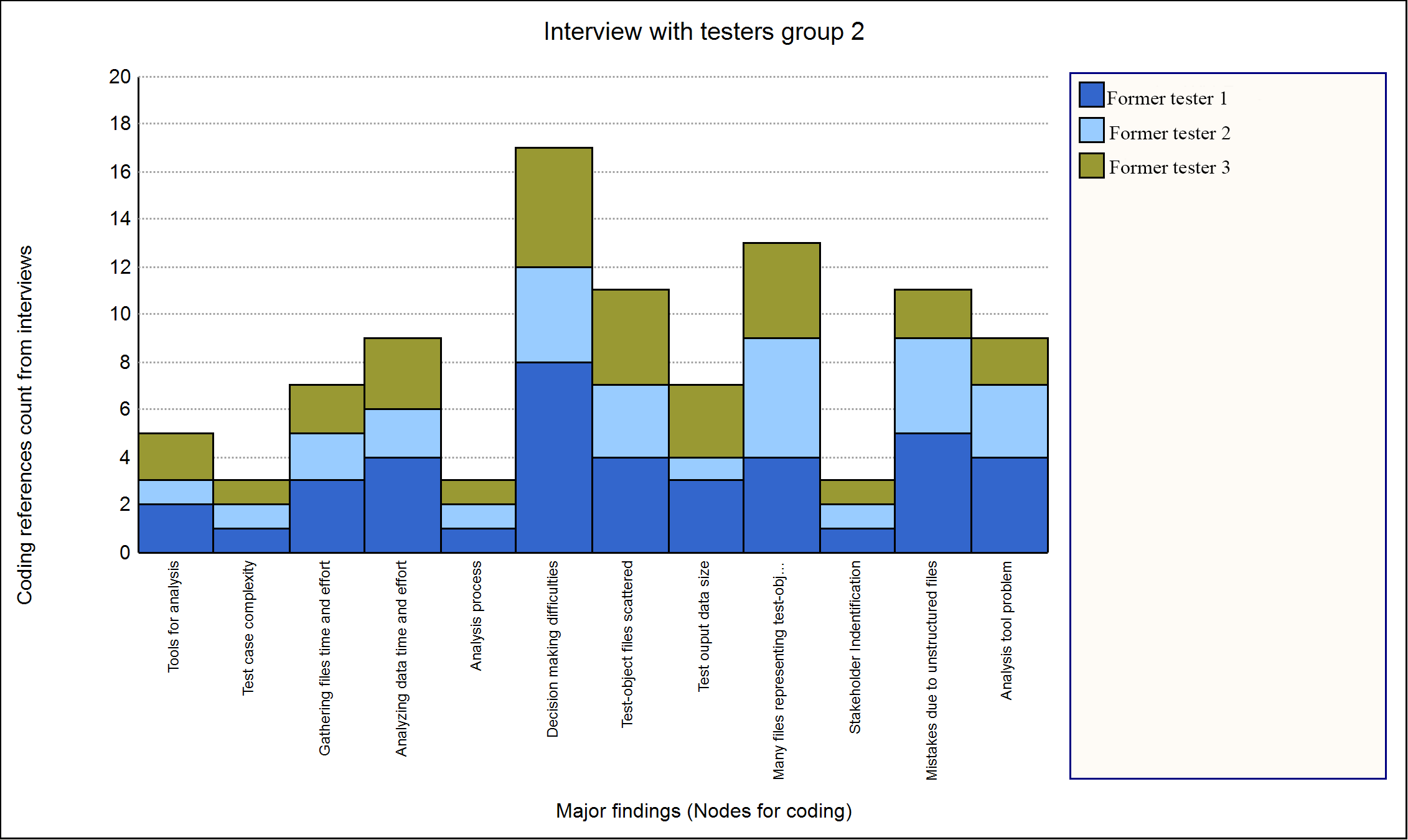}
    \caption{Findings from the interview with former testers.}
    \label{appendix:codedInterviews:formerTestersInterview}
\end{sidewaysfigure}

%% file: include/backmatter/Appendix_4.tex
\chapter{Requirements}
\label{appendix:requirements}

The user requirement specifications, including description, rationale, dependencies and use case can be seen below.  The motivation for the requirements and how it would be useful for test analysis were described in the rationale. Use case described how the user could use the feature of the requirement in the context of a web-based system (Dashboard) which was the artifact.

\begin{table}
\centering
\begin{tabular}{ |p{.2\textwidth}|p{.7\textwidth}| }
    \hline
        \textbf{UR1} & View an overview of the test case. \\ 
    \hline
        \textbf{Description} & User can view an overview of the test case. The overview page of the test case was the landing page of Dashboard and that included: 
        \begin{itemize}
            \item The results of the test (i.e if it failed or passed) 
            \item Overview of each step in the test case
            \item Overview of signals involved in the test case
            \item Overview of the buses involved in the test case
            \item Time and dates for the test case (Time-stamp)
        \end{itemize} \\ 
    \hline
        \textbf{Rationale} & This feature lets the user understand and get brief information on the test case and its test object. By getting this overview, it will help the user to quickly navigate to more interesting and relevant aspects/information during the analysis of the test object. \\ 
    \hline
        \textbf{Dependencies} & UR3, UR10  \\ 
    \hline
        \textbf{Use Case} & 
        \begin{enumerate}
            \item Open Dashboard.
            \item View the overview of the test object:
            \begin{itemize}
                \item The overview page was the landing page of the Dashboard, no more action needed.
                \item If already browsing the Dashboard, click on the “Overview” tab on the sidebar.
            \end{itemize}
        \end{enumerate} \\ 
    \hline
\end{tabular}
\end{table}

%\vspace*{1 cm}

\begin{table}
\label{appendix:getAllFilesForTestCase}
\centering
\begin{tabular}{ |p{.2\textwidth}|p{.7\textwidth}| }
    \hline
        \textbf{UR2} & View and download (if available) the initiation and metadata files related to the test object and the test case.  \\ 
    \hline
        \textbf{Description} & 
        User can see the location of initiation and metadata files containing the data and information related to the test object and the test case. It also includes the directory, identification, or name of the files making it easier to locate them if required. 
        \newline\newline
        The data and information presented in the Dashboard used these files as a reference to present to the most relevant context for the test case. Not everything presented in the Dashboard had access to those files, but a reference link to where the information was stored was indicated regardless. \\ 
    \hline
        \textbf{Rationale} & 
        This feature lets the user access all the initiation and metadata files that are associated with the test-object. Hence it also allowed the users to get the files quickly for initializing their test suites. They were required to perform the test again or open log files in their preferred analysis tools for a more in-depth analysis, which was necessary for more advanced test cases. 
        \newline\newline
        Providing faster and direct access to the user with the initiation and metadata files would increase the correctness, reduce effort, and increase the test process efficiency and analysis compared to finding the files manually. \\ 
    \hline
        \textbf{Dependencies} & - \\ 
    \hline
        \textbf{Use Case} & 
        \begin{enumerate}
            \item Open Dashboard.
            \item Click the “Links” on the sidebar.
        \end{enumerate} \\ 
    \hline
\end{tabular}
\end{table}

%\vspace*{1 cm}

\begin{table}
\centering
\begin{tabular}{ |p{.2\textwidth}|p{.7\textwidth}| }
    \hline
        \textbf{UR3} & View information on a particular step in the test case. \\ 
    \hline
        \textbf{Description} & 
        System should map test sequence with sensor signal at particular timestamp and see the information of what happened at that timestamp. The information that will be included are as:
        \begin{itemize}
            \item The expected values of the signals and result in the step.
            \item The signals that are being tested in the step.
            \item Graphs for the signals and the timestamp for the step.
            \item Being able to comment on the signals, for knowledge sharing and archiving.
            \item Marking the frequently analyzed/investigated signal as "Favorite", 
        \end{itemize} \\ 
    \hline
        \textbf{Rationale} & 
        This feature will give the user a brief overview of what happen during each step of the test sequence during the testing. The signals showed here will be analyzed here by comparing the result with the expected outputs, as well as between other signals.
        \newline\newline
        The goal of this feature is to provide a brief overview of a particular step, to allow the user to visualize the relevant and correct information regarding the sensor signals for that step. Information gathered here, could be used for quickly performing test analysis for preferably medium difficulty level tests. For more in-depth analysis, other testing tools could be used instead. \\ 
    \hline
        \textbf{Dependencies} & UR4, UR5, UR14, UR15 \\ 
    \hline
        \textbf{Use Case} & 
        \begin{enumerate}
            \item Open Dashboard.
            \item Click on:
                \begin{itemize}
                    \item “Step Overview” on the sidebar to get all the steps.
                    \item “Step [number]” on the sidebar to get a specific step.
                \end{itemize}
        \end{enumerate} \\ 
    \hline
\end{tabular}
\end{table}

%\vspace*{1 cm}

\begin{table}
\centering
\begin{tabular}{ |p{.2\textwidth}|p{.7\textwidth}| }
    \hline
        \textbf{UR4} & View graphs of the sensor signals in the test case. \\ 
    \hline
        \textbf{Description} & User can visualize the graphs of the sensor signals, generated from test output. The graph included the sensor signals values/outputs in the y-axis and timestamp in the x-axis. \\ 
    \hline
        \textbf{Rationale} & 
        This feature will provide the user with better visualization of the signals and the steps, which would help with analysis by getting a better overview of each step. By displaying the sensor signals associated with the test case, this would increase the efficiency and correctness. Furthermore, the users will not be overwhelmed by witnessing many sensor signal data at once due to how they were displayed according to steps.

        Overall, this feature provided the users with a simple graph that could be useful for simple analysis by comparing signals output values, tailoring towards analyzing test cases with medium difficulty level. More in-depth analysis could be done on other analysis programs if required. \\ 
    \hline
        \textbf{Dependencies} & - \\ 
    \hline
        \textbf{Use Case} & 
        \begin{enumerate}
            \item Open Dashboard.
            \item Click on:
                \begin{itemize}
                    \item “Step Overview” on the sidebar to get all the steps.
                    \item “Step [number]” on the sidebar to get a specific step.
                \end{itemize}
        \end{enumerate} \\ 
    \hline
\end{tabular}
\end{table}

%\vspace*{1 cm}

\begin{table}
\centering
\begin{tabular}{ |p{.2\textwidth}|p{.7\textwidth}| }
    \hline
        \textbf{UR5} & Search for signals. \\ 
    \hline
        \textbf{Description} & User could search for the signals in the test output for analysis. \\ 
    \hline
        \textbf{Rationale} & This feature would let users attach and analyze additional sensor signal outputs that were not the default ones provided by for each step. This would help the users see correlations and perform analysis directly in the Dashboard. \\ 
    \hline
        \textbf{Dependencies} & - \\ 
    \hline
        \textbf{Use Case} & 
        \begin{enumerate}
            \item Open Dashboard.
            \item Click on:
                \begin{itemize}
                    \item “Step Overview” on the sidebar to get all the steps.
                    \item “Step [number]” on the sidebar to get a specific step.
                \end{itemize}
        \end{enumerate} \\ 
    \hline
\end{tabular}
\end{table}

%\vspace*{1 cm}

\begin{table}
\centering
\begin{tabular}{ |p{.2\textwidth}|p{.7\textwidth}| }
    \hline
        \textbf{UR6} & View the stakeholder for the test object. \\ 
    \hline
        \textbf{Description} & User could see the responsible stakeholder for a test object. \\ 
    \hline
        \textbf{Rationale} & 
        This feature would help the user to see the stakeholders who were responsible for test object, which would make it easier to contact them if fixing an error/issue found in the test was their responsibility or ask domain specific question about the test case/object.
        \newline\newline
        Providing a list of the stakeholders and their roles, would  help the users decide which stakeholder to contact efficiently and reduce the efforts compared to manually finding the correct stakeholder. \\ 
    \hline
        \textbf{Dependencies} & - \\ 
    \hline
        \textbf{Use Case} & 
        \begin{enumerate}
            \item Open Dashboard.
            \item Click on the “Stakeholder” in the sidebar:
                \begin{itemize}
                    \item More information about the stakeholder by clicking the “i” icon next to the stakeholder.
                \end{itemize}
        \end{enumerate} \\ 
    \hline
\end{tabular}
\end{table}

%\vspace*{1 cm}

\begin{table}
\centering
\begin{tabular}{ |p{.2\textwidth}|p{.7\textwidth}| }
    \hline
        \textbf{UR7} & View information of the test performers. \\ 
    \hline
        \textbf{Description} & User could see the information of the test performers. \\ 
    \hline
        \textbf{Rationale} & This feature would help the users to see the testers who conducted the tests, which would make it easier to contact them for more information or questions about test case or object. \\ 
    \hline
        \textbf{Dependencies} & - \\ 
    \hline
        \textbf{Use Case} & 
        \begin{enumerate}
            \item Open Dashboard.
            \item Click on the “Stakeholder” in the sidebar:
                \begin{itemize}
                    \item More information about the tester by clicking the “i” icon next to the testers.
                \end{itemize}
        \end{enumerate} \\ 
    \hline
\end{tabular}
\end{table}

%\vspace*{1 cm}

\begin{table}
\centering
\begin{tabular}{ |p{.2\textwidth}|p{.7\textwidth}| }
    \hline
        \textbf{UR8} & View the Design Verification Method (DVM) for the test object. \\ 
    \hline
        \textbf{Description} &
        Users will see the DVM for the test object. DVM includes information about the pre-condition, post-condition, test requirements, and test object, and test case description. The DVM name and revision version are also included for the users to avoid making mistakes, like refereeing the wrong DVM version for test analysis.
        \newline\newline
        The user will also see the test sequence, which describes how the test was supposed to be conducted step by step and its verdicts, including expected and actual results.
        \\ 
    \hline
        \textbf{Rationale} & This feature will be used to verify test-cases with the DVM by visually interpreting it. By providing this feature, users will have an easier and faster access to the correct DVM. \\ 
    \hline
        \textbf{Dependencies} & - \\ 
    \hline
        \textbf{Use Case} & 
        \begin{enumerate}
            \item Open Dashboard:
                \begin{itemize}
                    \item DVM: Click “DVM” on the sidebar.
                    \item Test Sequence: Click “Test Sequence” on the sidebar.
                \end{itemize}
        \end{enumerate} \\ 
    \hline
\end{tabular}
\end{table}

%\vspace*{1 cm}

\begin{table}
\centering
\begin{tabular}{ |p{.2\textwidth}|p{.7\textwidth}| }
    \hline
        \textbf{UR9} & View the requirements for the test object. \\ 
    \hline
        \textbf{Description} & User will be able to see the requirements for the test object. Requirements includes description of the test object’s constraints, expected outputs, limitations and objects(e.g signals/ECU) involved. The requirement and revision version is also included. \\ 
    \hline
        \textbf{Rationale} & This feature is used to verify test-cases with the feature requirement by visually interpreting it. By providing this feature, users will have an easier and faster access to the correct requirement. \\ 
    \hline
        \textbf{Dependencies} & - \\ 
    \hline
        \textbf{Use Case} & 
        \begin{enumerate}
            \item Open Dashboard.
            \item Click “Requirement” on the sidebar.
        \end{enumerate} \\ 
    \hline
\end{tabular}
\end{table}

%\vspace*{1 cm}

\begin{table}
\centering
\begin{tabular}{ |p{.2\textwidth}|p{.7\textwidth}| }
    \hline
        \textbf{UR10} & View the ECUs and its associated Bus of the test object used in the test case. \\ 
    \hline
        \textbf{Description} & User will be able to see a list of ECUs and its associated Bus of the test object used in the test case. Name of the ECUs and Buses are included to allow users to quickly identify them as opposed to doing it manually. \\ 
    \hline
        \textbf{Rationale} & This feature allows the user to get an overview of the context for the test case to be able to see what kind ECUs and buses are involved in the testing. This feature will help the user be more efficient when locating relevant data and information during analysis in regards to buses and ECUs. \\ 
    \hline
        \textbf{Dependencies} & - \\ 
    \hline
        \textbf{Use Case} & 
        \begin{enumerate}
            \item Open Dashboard.
            \item View the overview of the test object:
                \begin{itemize}
                    \item The overview page is the landing page of the Dashboard, no more action needed.
                    \item If already browsing the Dashboard, click on the “Overview” tab on the sidebar.
                \end{itemize}
        \end{enumerate} \\ 
    \hline
\end{tabular}
\end{table}

%\vspace*{1 cm}

\begin{table}
\centering
\begin{tabular}{ |p{.2\textwidth}|p{.7\textwidth}| }
    \hline
        \textbf{UR11} & View the details of the software and hardware read-outs. \\ 
    \hline
        \textbf{Description} & User will be able to see the software and hardware information of the test object. The read-outs are generated by the testing tools and provide useful information to the user for test analysis. The read-out specific information were:
        \begin{itemize}
            \item Identifiers
            \item Response outputs
            \item ECU (serial number and part number)
            \item Timestamps
            \item Request and response time
        \end{itemize} \\ 
    \hline
        \textbf{Rationale} & This feature gives the user the relevant read-outs of software and hardware information. As the read-outs usually contains a large amount of information, providing them briefly and concisely here will help to users reduce the workload when trying to understand and find read-out specific information. \\ 
    \hline
        \textbf{Dependencies} & - \\ 
    \hline
        \textbf{Use Case} & 
        \begin{enumerate}
            \item Open Dashboard.
            \item Click “Information” on the sidebar.
        \end{enumerate} \\ 
    \hline
\end{tabular}
\end{table}

%\vspace*{1 cm}

\begin{table}
\centering
\begin{tabular}{ |p{.2\textwidth}|p{.7\textwidth}| }
    \hline
        \textbf{UR12} & View the the calibration and configuration of the test object for the test case. \\ 
    \hline
        \textbf{Description} & User will be able to see the calibration and configuration information of the test when it was conducted. The calibration is needed by the test suite to let users perform/re-produce tests. The calibration information included:
        \begin{itemize}
            \item Topology: Car ID and Car Family ID
            \item ECU name and version
            \item Bus name and version
            \item Database name and version for signal sensor signals
            \item Measurement and analysis tools configuration
        \end{itemize} \\ 
    \hline
        \textbf{Rationale} & This feature let the user to set-up the calibration and configuration for more in-depth analysis, using their preferred analysis tool more efficient. Also, by providing the correct calibration and configuration data, users will not require to manually look for them in different storages. Furthermore, this feature will increase the correctness during the test analysis and diagnostics as testers usually had to do trail and error to validate correct calibration and configuration files amongst the incorrect ones. \\ 
    \hline
        \textbf{Dependencies} & - \\ 
    \hline
        \textbf{Use Case} & 
        \begin{enumerate}
            \item Open Dashboard.
            \item Click “Information” on the sidebar.
        \end{enumerate} \\ 
    \hline
\end{tabular}
\end{table}

%\vspace*{1 cm}

\begin{table}
\centering
\begin{tabular}{ |p{.2\textwidth}|p{.7\textwidth}| }
    \hline
        \textbf{UR13} & View historical data for a test object. \\ 
    \hline
        \textbf{Description} & User can view previous test related data for similar test object. The similarity of a test object is determined by the test object, test case, requirement and DVM. The metadata displayed for the user are: 
        \begin{itemize}
            \item The verdict (i.e success or failed)
            \item Name of the test case
            \item Date and time
            \item Domain
            \item The author who conducted the test
            \item Link to the directory to initiation and metadata files off the test-case
        \end{itemize} \\ 
    \hline
        \textbf{Rationale} & This feature will let the user backtrack the previous test results of similar test cases/test-objects to get references and knowledge for their analysis. \\ 
    \hline
        \textbf{Dependencies} & - \\ 
    \hline
        \textbf{Use Case} & 
        \begin{enumerate}
            \item Open Dashboard.
            \item Click “Historical Data” on the sidebar.
        \end{enumerate} \\ 
    \hline
\end{tabular}
\end{table}

%\vspace*{1 cm}

\begin{table}
\centering
\begin{tabular}{ |p{.2\textwidth}|p{.7\textwidth}| }
    \hline
        \textbf{UR14} & View and add favorite signals. \\ 
    \hline
        \textbf{Description} & User can favorite signals to store them locally to access them more easily. \\ 
    \hline
        \textbf{Rationale} & This feature lets the user get their most frequently used sensor signals faster and more efficiently. In some cases, some signals should always be checked during the analysis. \\ 
    \hline
        \textbf{Dependencies} & - \\ 
    \hline
        \textbf{Use Case} & 
        \begin{enumerate}
            \item Open Dashboard.
            \item Click on:
                \begin{itemize}
                    \item “Step Overview” on the sidebar to get all the steps.
                    \item “Step [number]” on the sidebar to get a specific step.
                \end{itemize}
            \item A heart button next to the signals:
                \begin{itemize}
                    \item Click a hollow heart to like.
                    \item Click a filled heart to unlike.
                \end{itemize}
        \end{enumerate} \\ 
    \hline
\end{tabular}
\end{table}

%\vspace*{1 cm}

\begin{table}
\centering
\begin{tabular}{ |p{.2\textwidth}|p{.7\textwidth}| }
    \hline
        \textbf{UR15} & View and add comments on signals. \\ 
    \hline
        \textbf{Description} & Users can write and read comments on signals. \\ 
    \hline
        \textbf{Rationale} & This feature allows knowledge sharing between users. It works as a forum to discuss issues. \\ 
    \hline
        \textbf{Dependencies} & - \\ 
    \hline
        \textbf{Use Case} & 
        \begin{enumerate}
            \item Open Dashboard.
            \item Click on:
                \begin{itemize}
                    \item “Step Overview” on the sidebar to get all the steps.
                    \item “Step [number]” on the sidebar to get a specific step.
                \end{itemize}
            \item Click on the comment icon next to the sensor signal, a pop-up windows of comments will be shown.
            \item Action:
                \begin{itemize}
                    \item Read comments in chronological order.
                    \item Write a comment and then post it.
                \end{itemize}
        \end{enumerate} \\ 
    \hline
\end{tabular}
\end{table}

%% file: include/backmatter/Appendix_5.tex
\chapter{Low-fidelity Prototype}
\label{appendix:lowPrototype}

\begin{sidewaysfigure}
    \centering
    \includegraphics[width=.9\textheight]{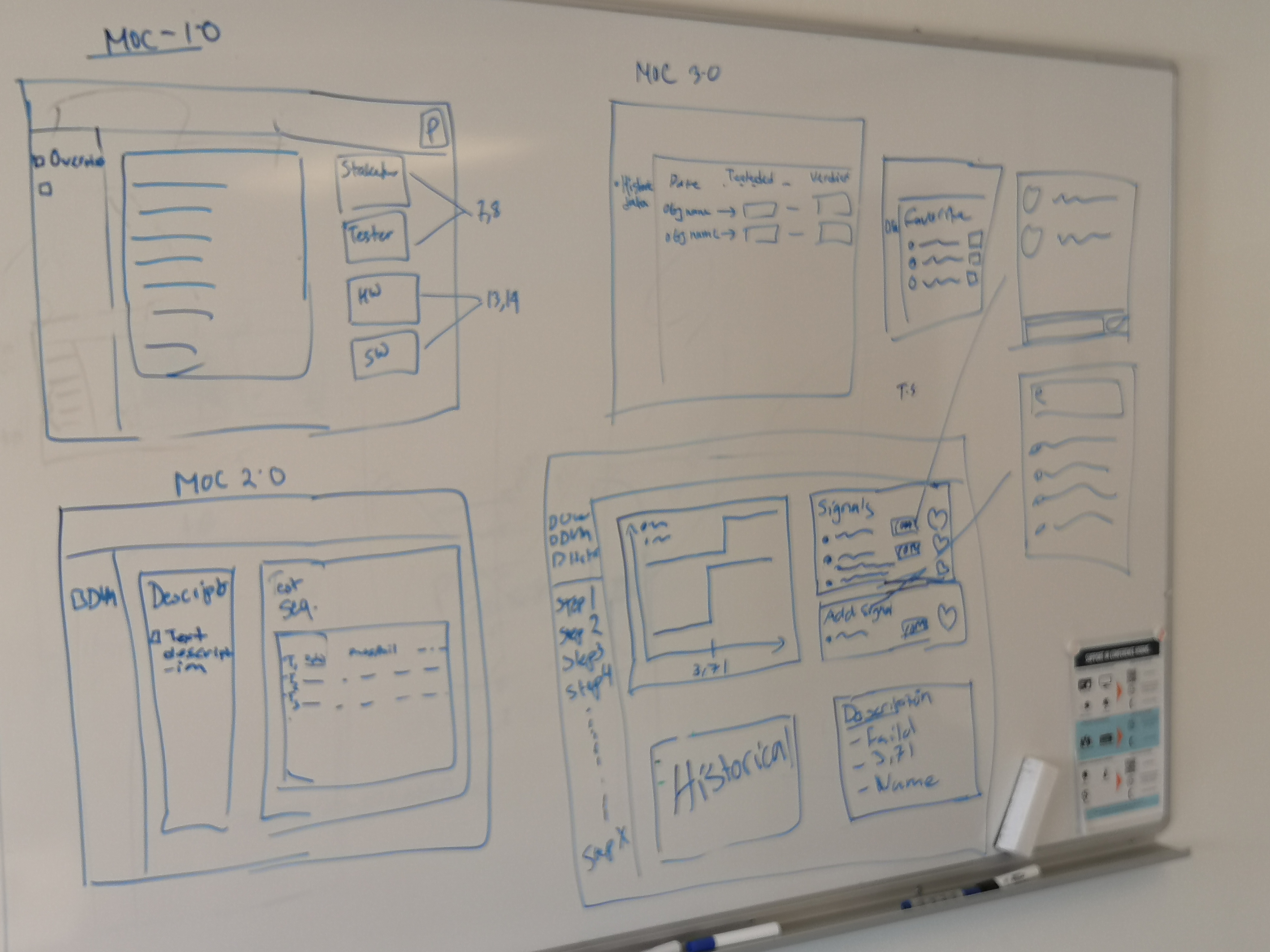}
    \caption{Low-fidelity prototype sketch drawn on whiteboard.}
    \label{appendix:lowPrototype:test_subject}
\end{sidewaysfigure}

%% file: include/backmatter/Appendix_6.tex
\chapter{High-fidelity Prototype}
\label{appendix:high1Prototype}

\begin{sidewaysfigure}
    \centering
    \includegraphics[width=.9\textheight]{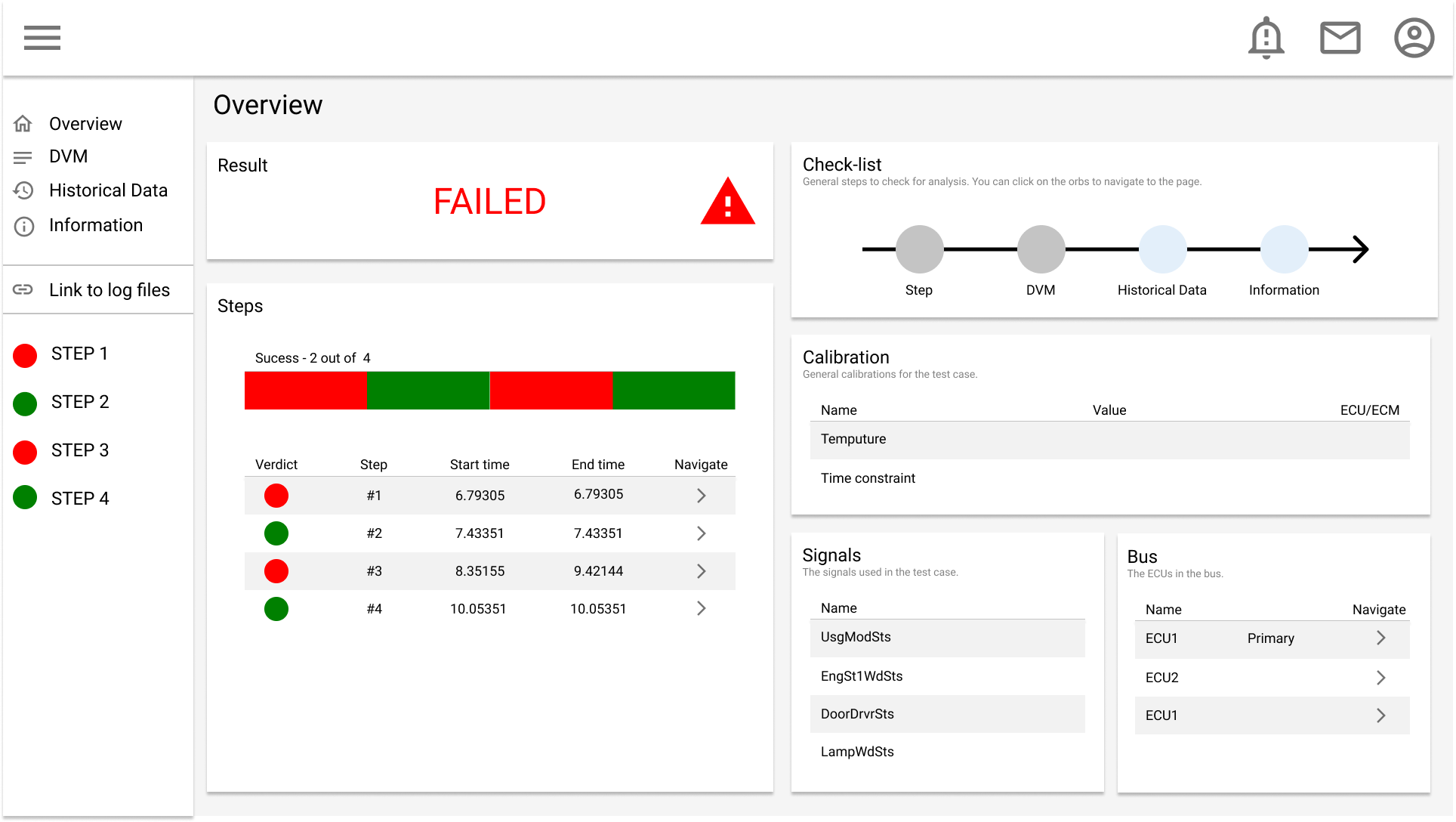}
    \caption{High-fidelity prototype of overview view.}
    \label{appendix:high1Prototype_overview}
\end{sidewaysfigure}

\begin{sidewaysfigure}
    \centering
    \includegraphics[width=.9\textheight]{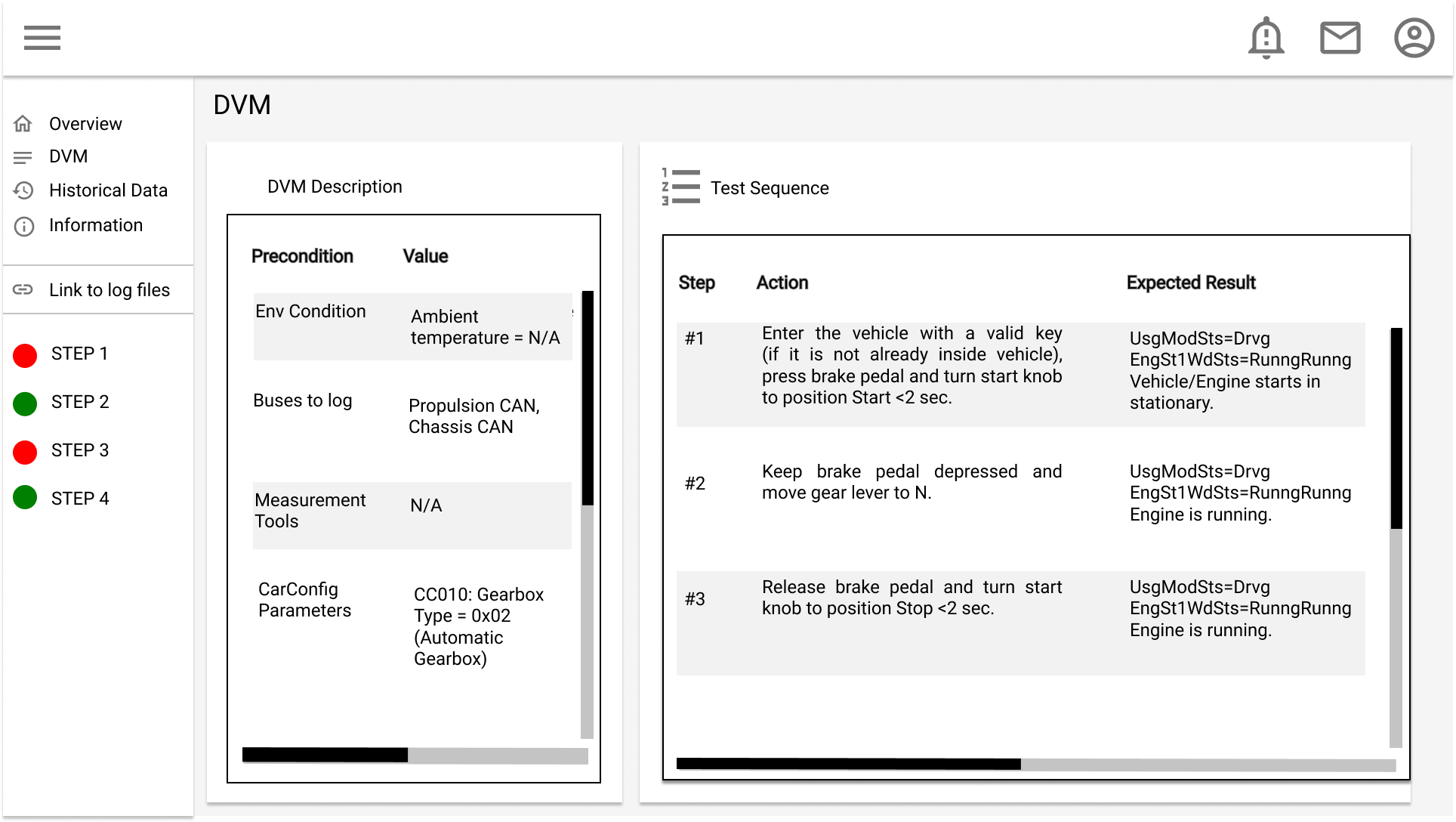}
    \caption{High-fidelity prototype of DVM view.}
    \label{appendix:high1Prototype_dvm}
\end{sidewaysfigure}

\begin{sidewaysfigure}
    \centering
    \includegraphics[width=.9\textheight]{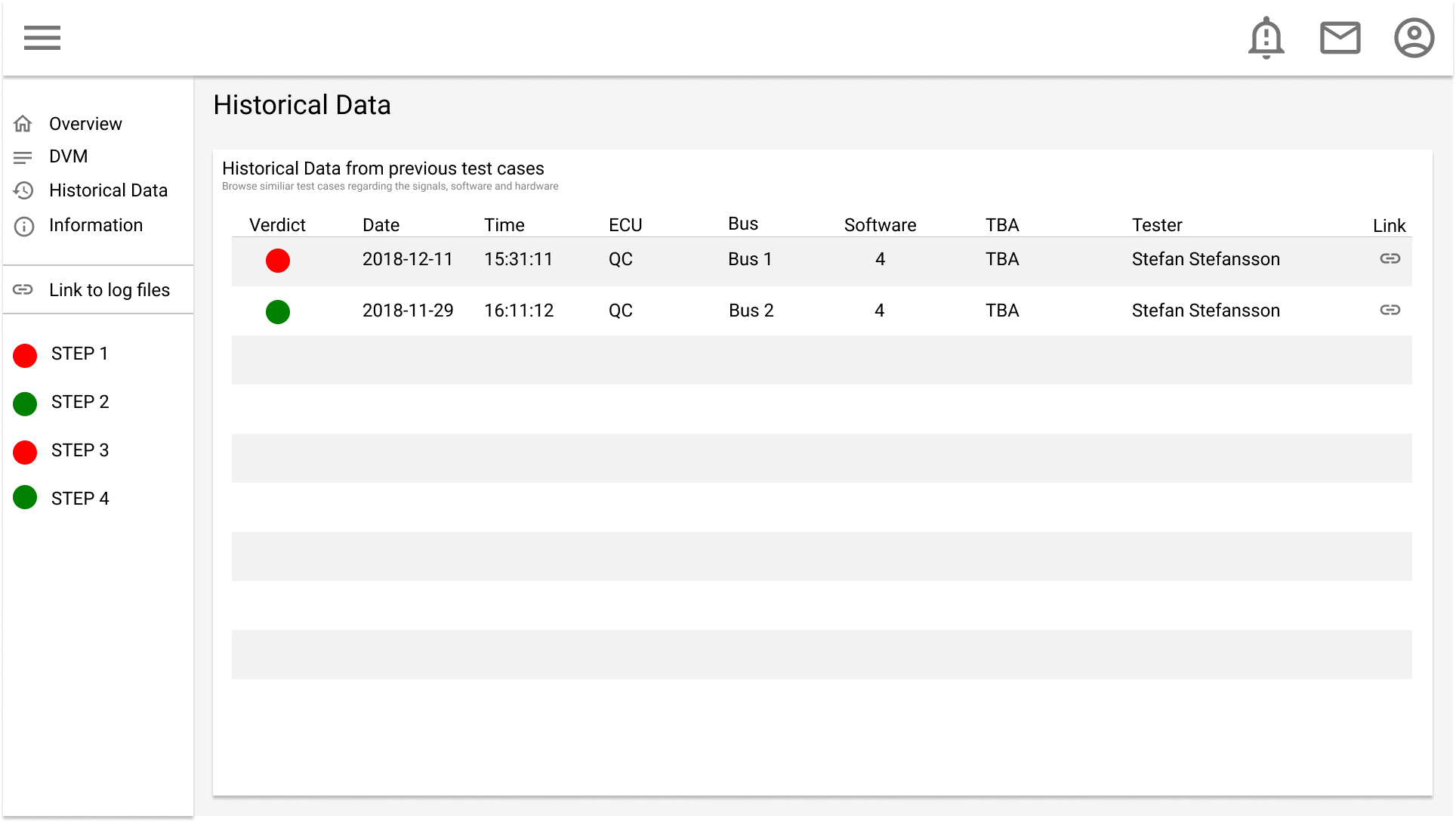}
    \caption{High-fidelity prototype of historical view.}
    \label{appendix:high1Prototype_historical}
\end{sidewaysfigure}

\begin{sidewaysfigure}
    \centering
    \includegraphics[width=.9\textheight]{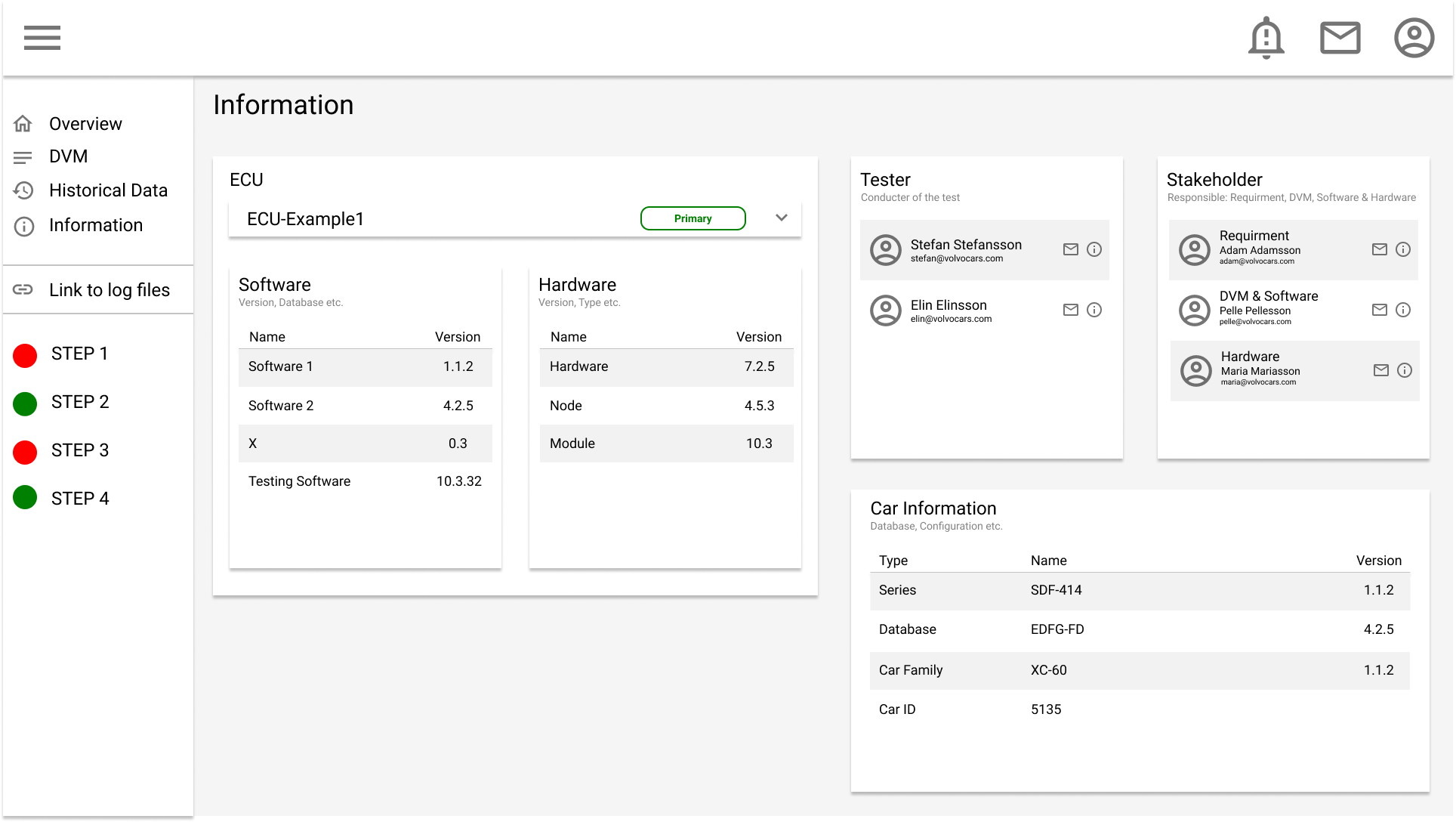}
    \caption{High-fidelity prototype of information view.}
    \label{appendix:high1Prototype_information}
\end{sidewaysfigure}

\begin{sidewaysfigure}
    \centering
    \includegraphics[width=.9\textheight]{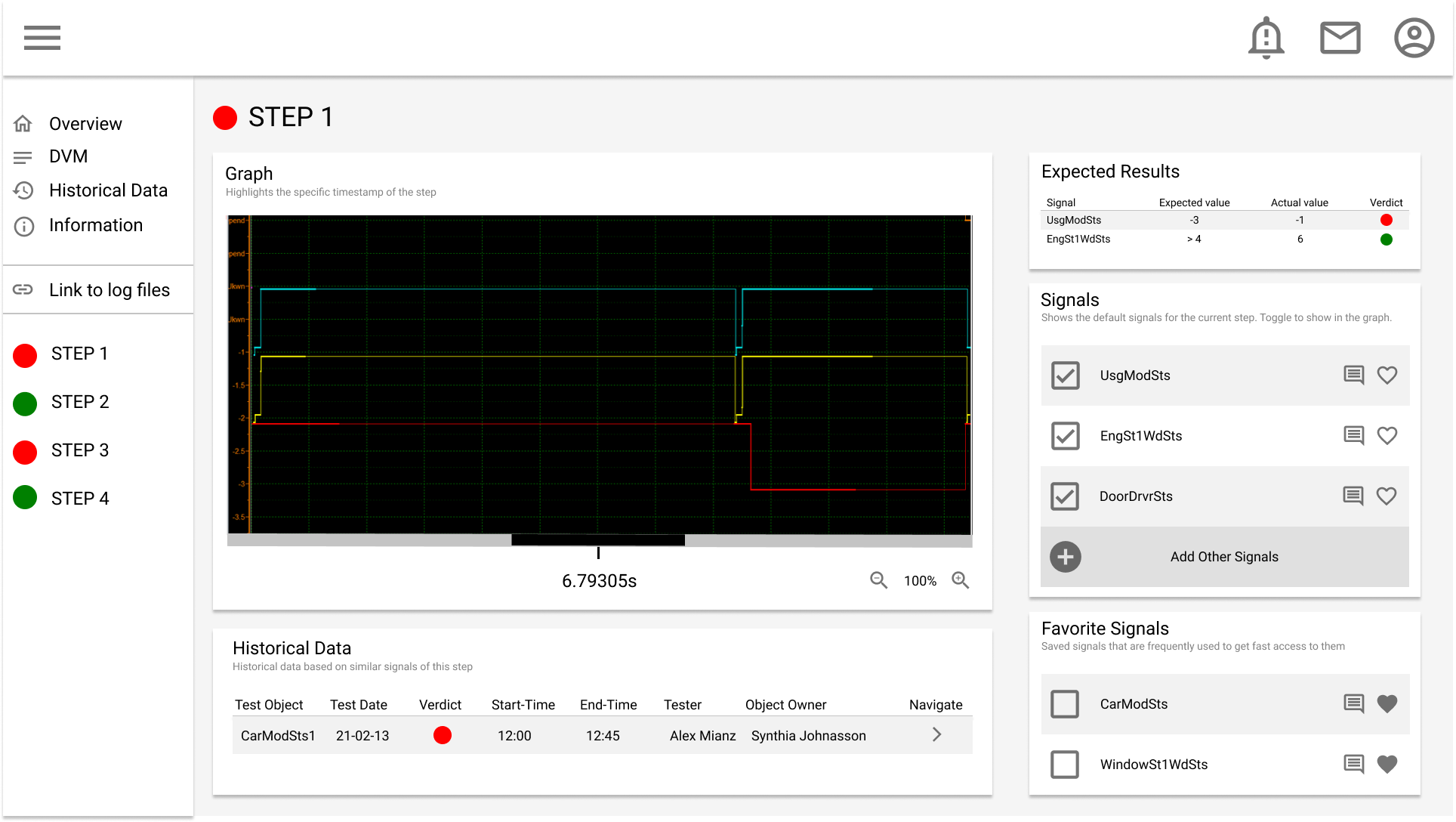}
    \caption{High-fidelity prototype of step view.}
    \label{appendix:high1Prototype_step}
\end{sidewaysfigure}

\begin{sidewaysfigure}
    \centering
    \includegraphics[width=.9\textheight]{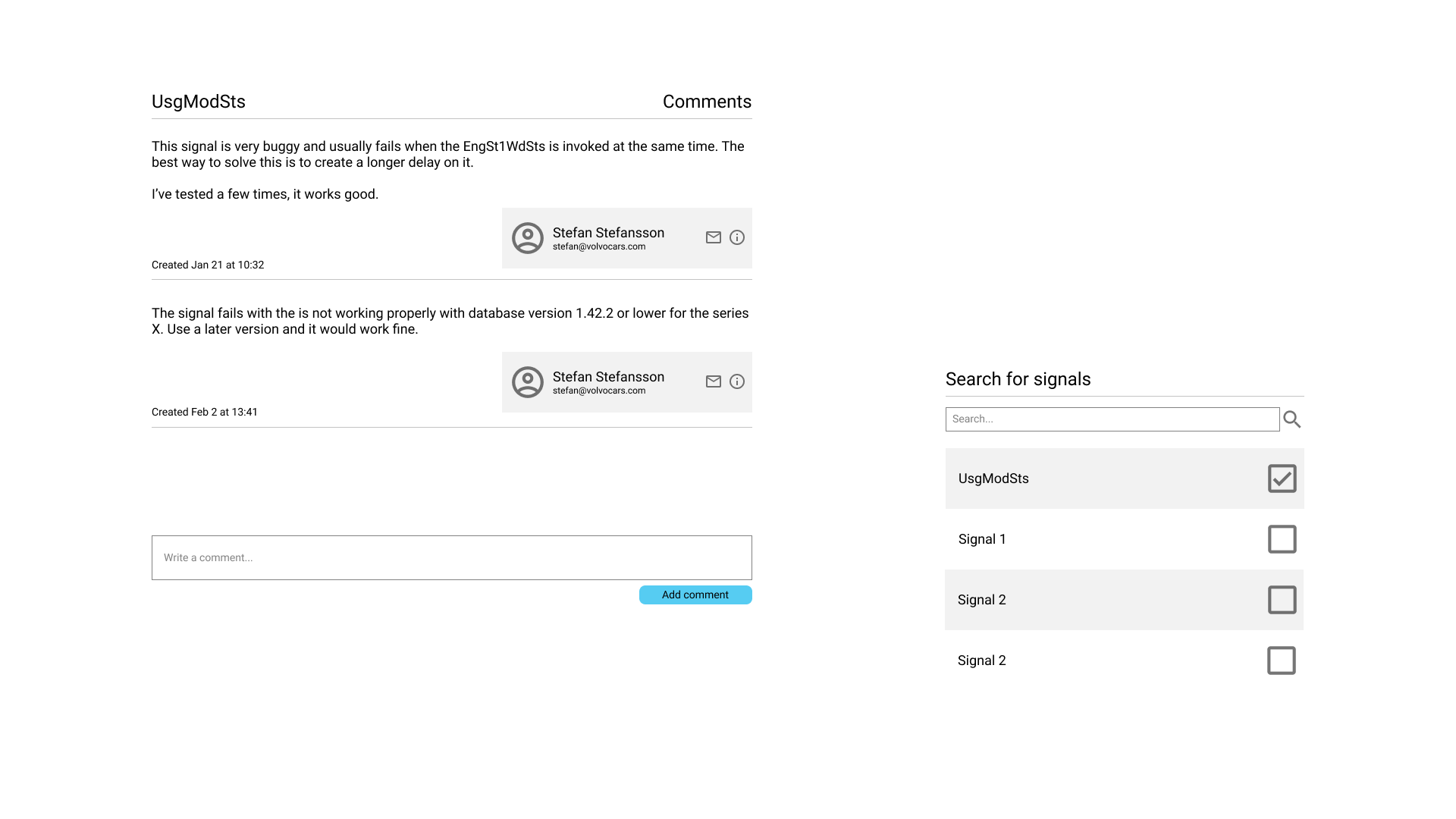}
    \caption{High-fidelity prototype of step sub-views.}
    \label{appendix:high1Prototype_stepSubview}
\end{sidewaysfigure}

%% file: include/backmatter/Appendix_7.tex
\chapter{Final High-fidelity Prototype}
\label{appendix:high2Prototype}

\begin{sidewaysfigure}
    \centering
    \includegraphics[width=.9\textheight]{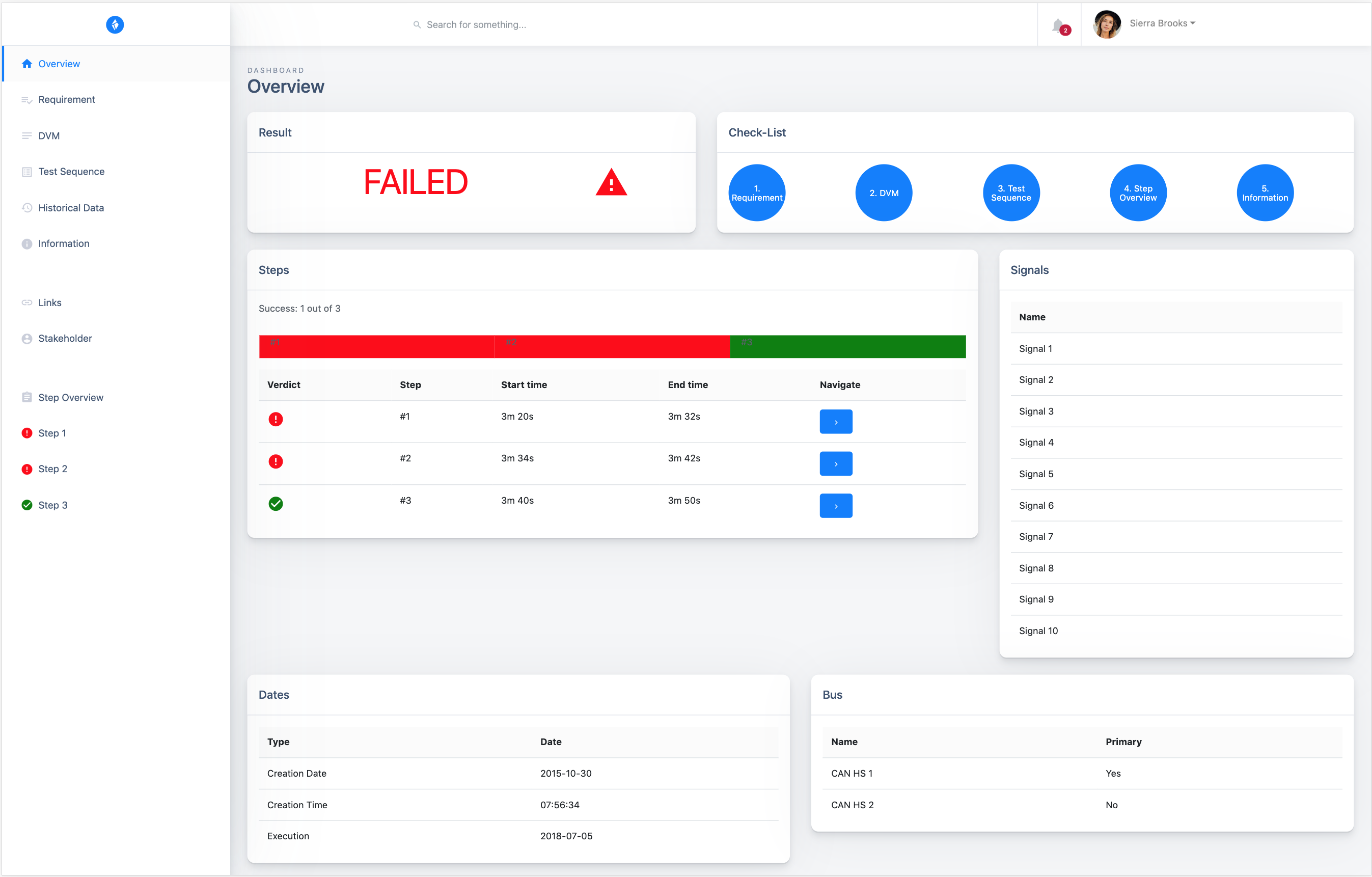}
    \caption{Final high-fidelity prototype of overview view.}
    \label{appendix:high2Prototype_overview}
\end{sidewaysfigure}

\begin{sidewaysfigure}
    \centering
    \includegraphics[width=.9\textheight]{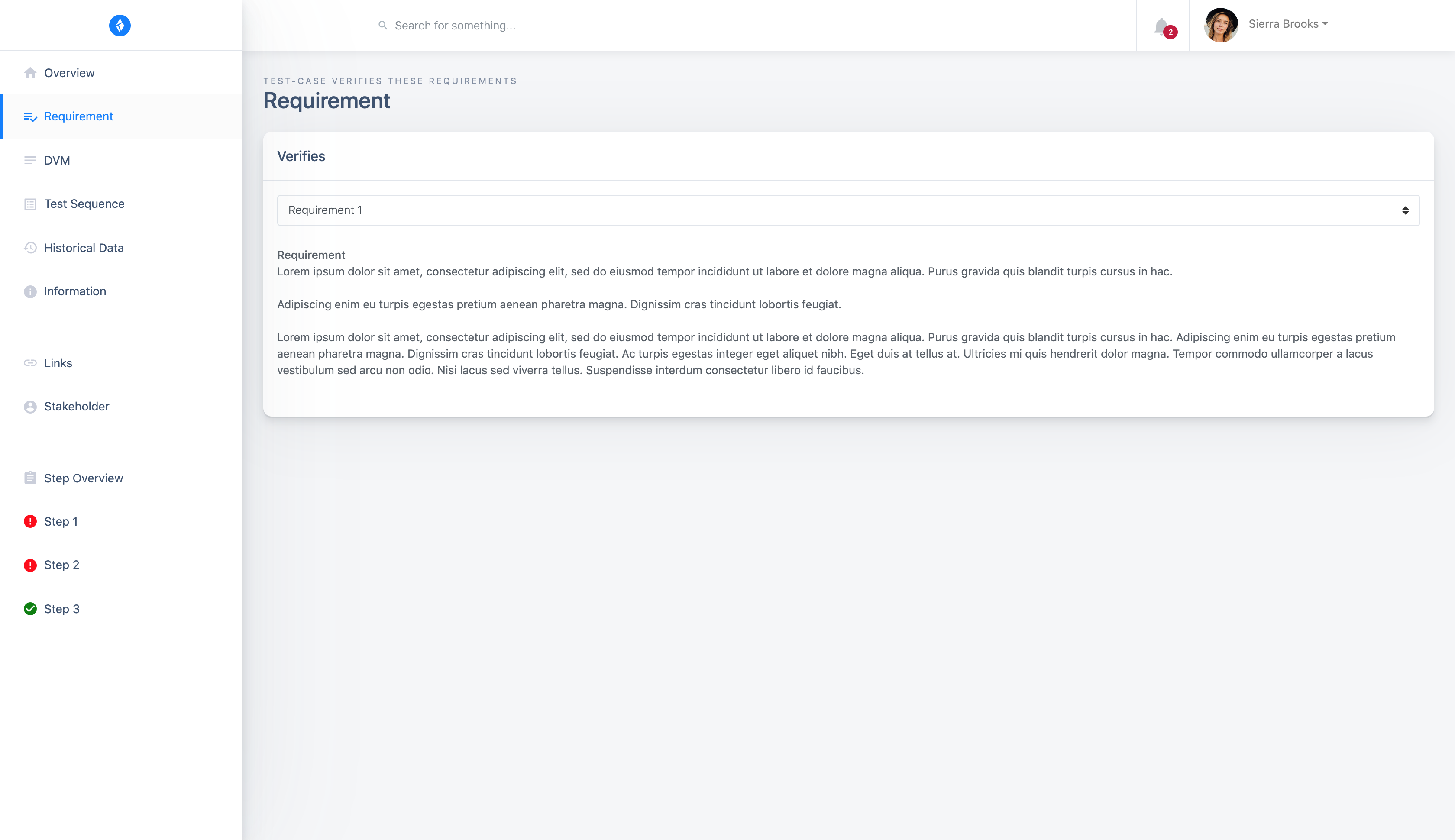}
    \caption{Final high-fidelity prototype of requirement view.}
    \label{appendix:high2Prototype_requirement}
\end{sidewaysfigure}

\begin{sidewaysfigure}
    \centering
    \includegraphics[width=.9\textheight]{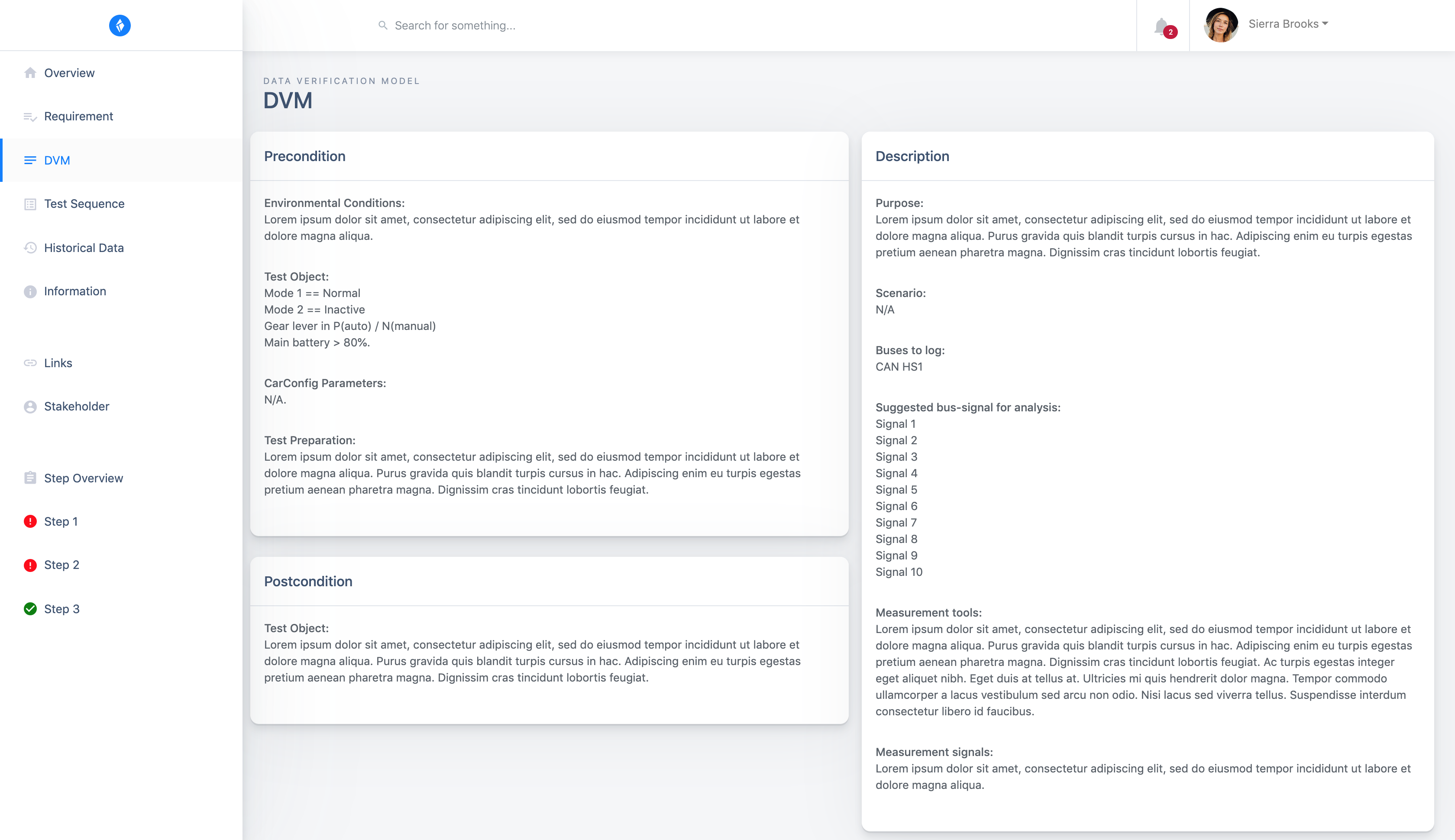}
    \caption{Final high-fidelity prototype of DVM view.}
    \label{appendix:high2Prototype_dvm}
\end{sidewaysfigure}

\begin{sidewaysfigure}
    \centering
    \includegraphics[width=.9\textheight]{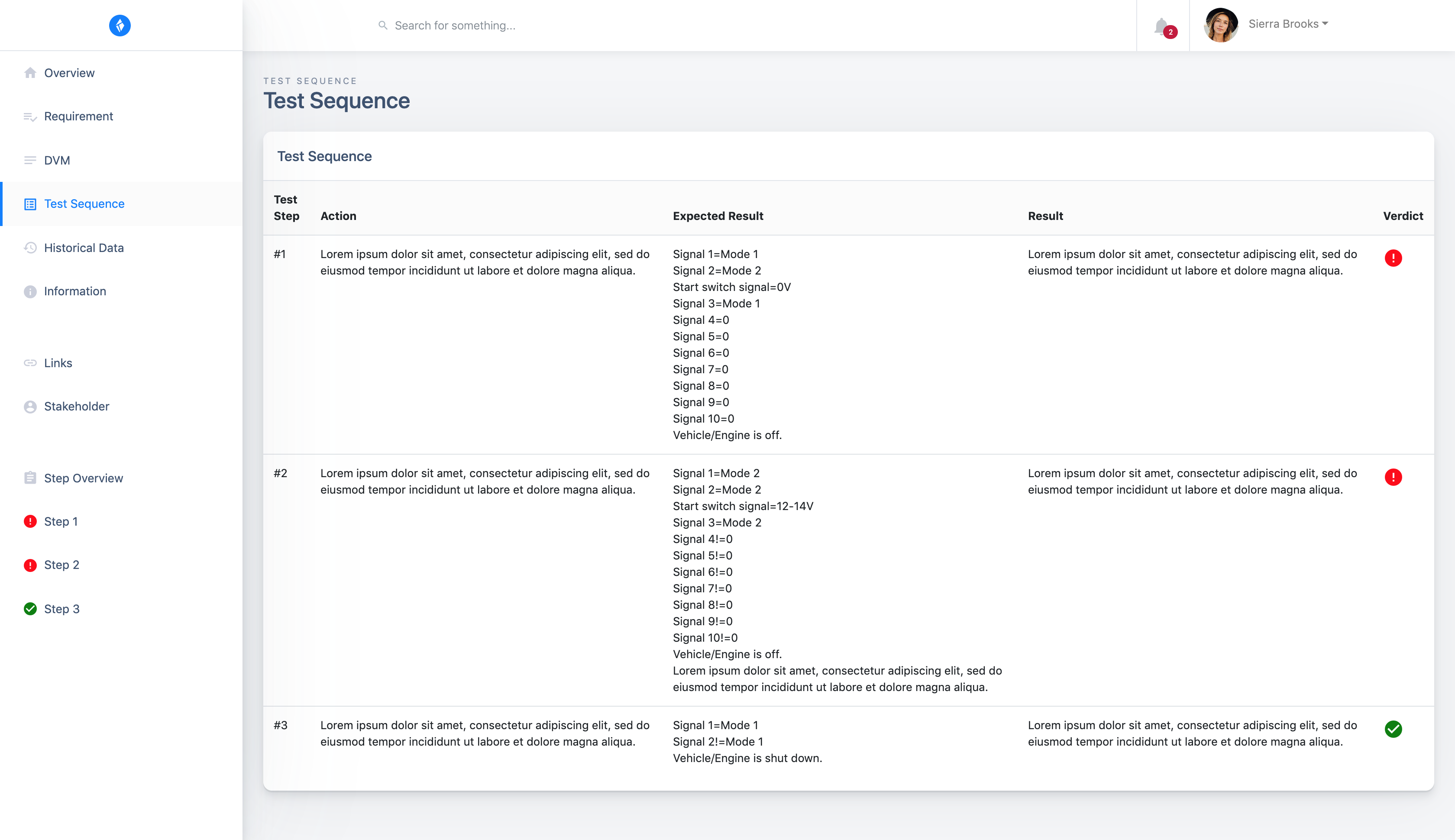}
    \caption{Final high-fidelity prototype of test sequence view.}
    \label{appendix:high2Prototype_testSequence}
\end{sidewaysfigure}

\begin{sidewaysfigure}
    \centering
    \includegraphics[width=.9\textheight]{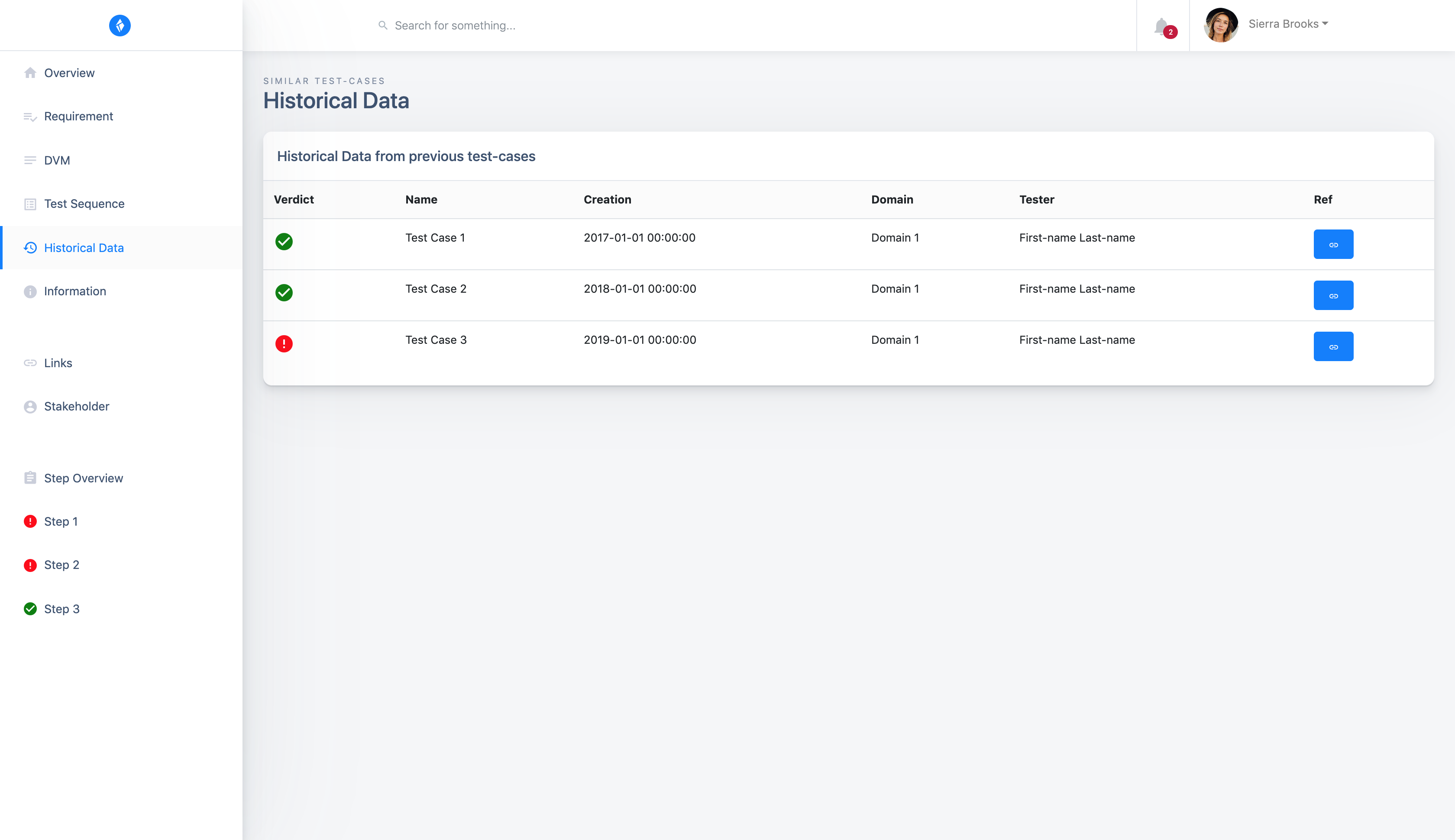}
    \caption{Final high-fidelity prototype of historical view.}
    \label{appendix:high2Prototype_historical}
\end{sidewaysfigure}

\begin{sidewaysfigure}
    \centering
    \includegraphics[width=.9\textheight]{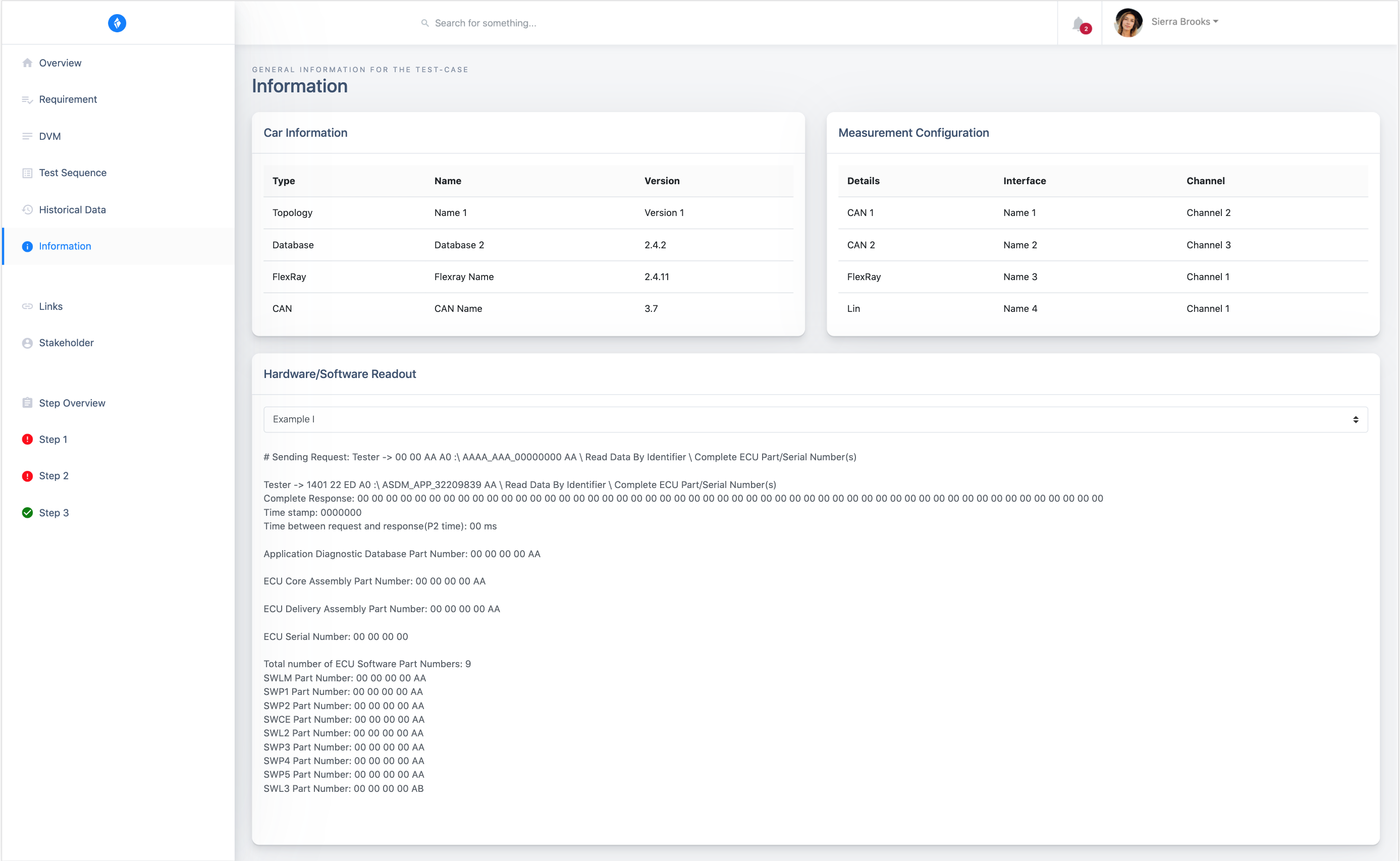}
    \caption{Final high-fidelity prototype of information view.}
    \label{appendix:high2Prototype_information}
\end{sidewaysfigure}

\begin{sidewaysfigure}
    \centering
    \includegraphics[width=.9\textheight]{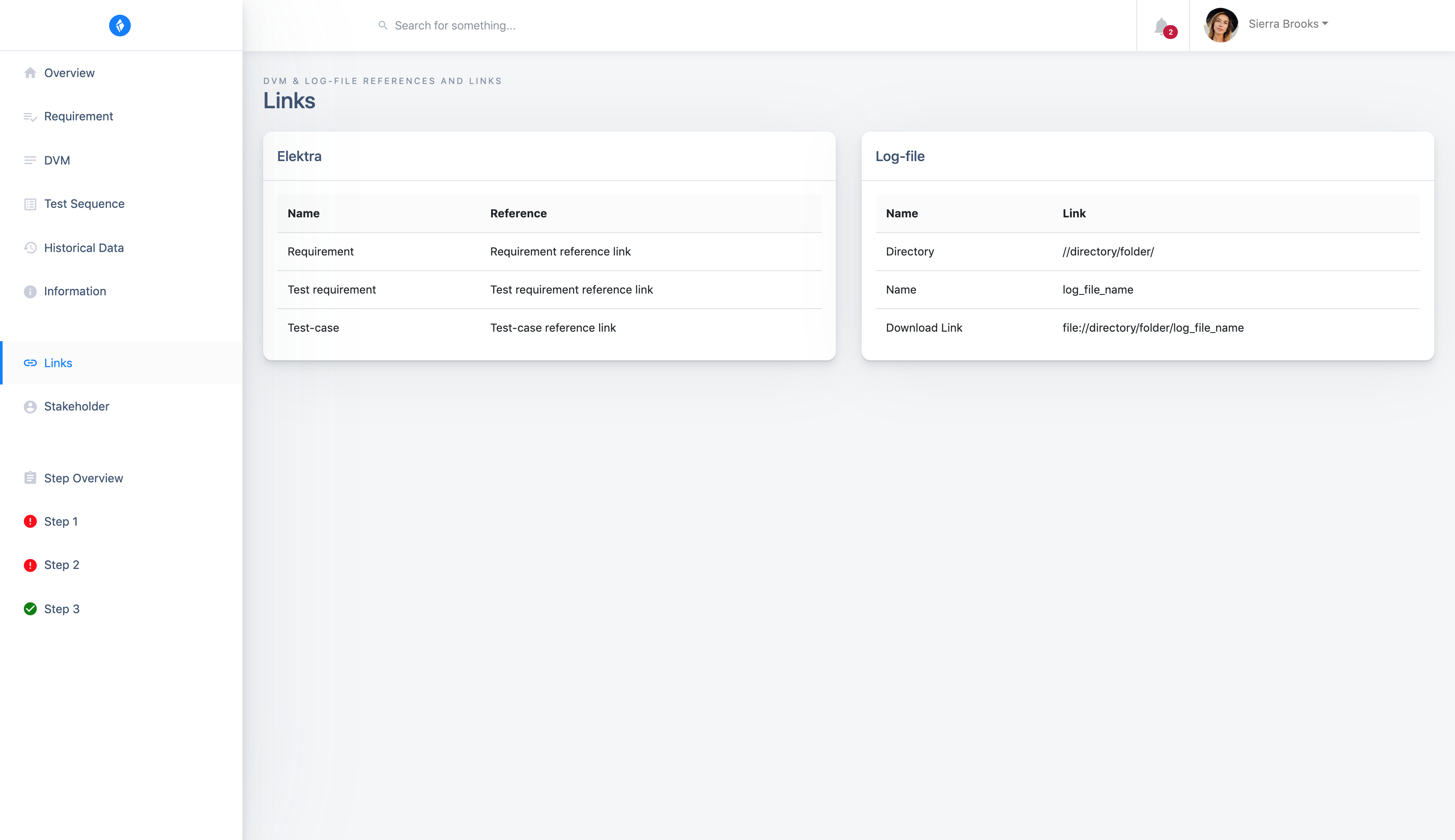}
    \caption{Final high-fidelity prototype of links view.}
    \label{appendix:high2Prototype_links}
\end{sidewaysfigure}

\begin{sidewaysfigure}
    \centering
    \includegraphics[width=.9\textheight]{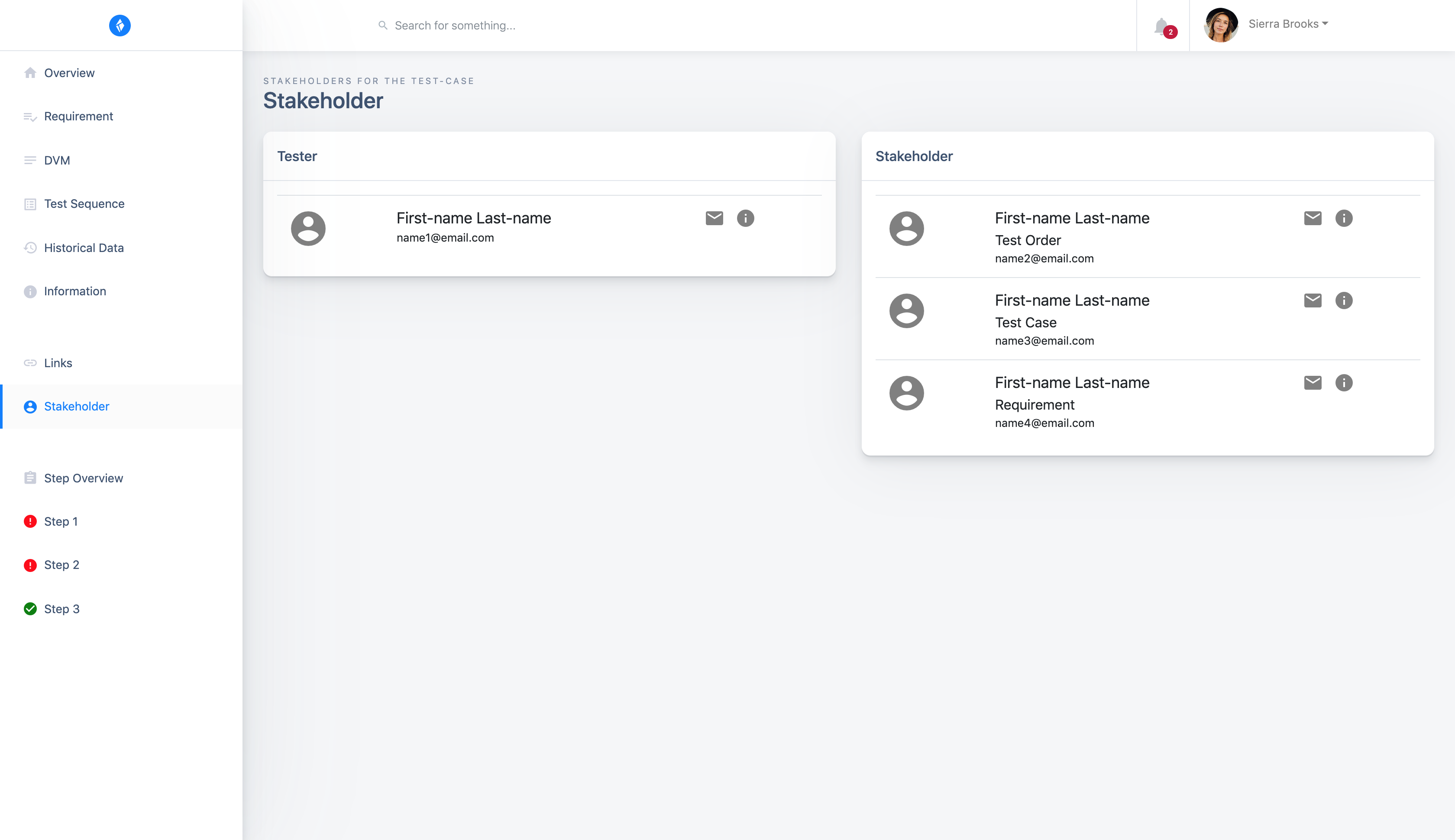}
    \caption{Final high-fidelity prototype of stakeholder view.}
    \label{appendix:high2Prototype_stakeholder}
\end{sidewaysfigure}

\begin{sidewaysfigure}
    \centering
    \includegraphics[width=.9\textheight]{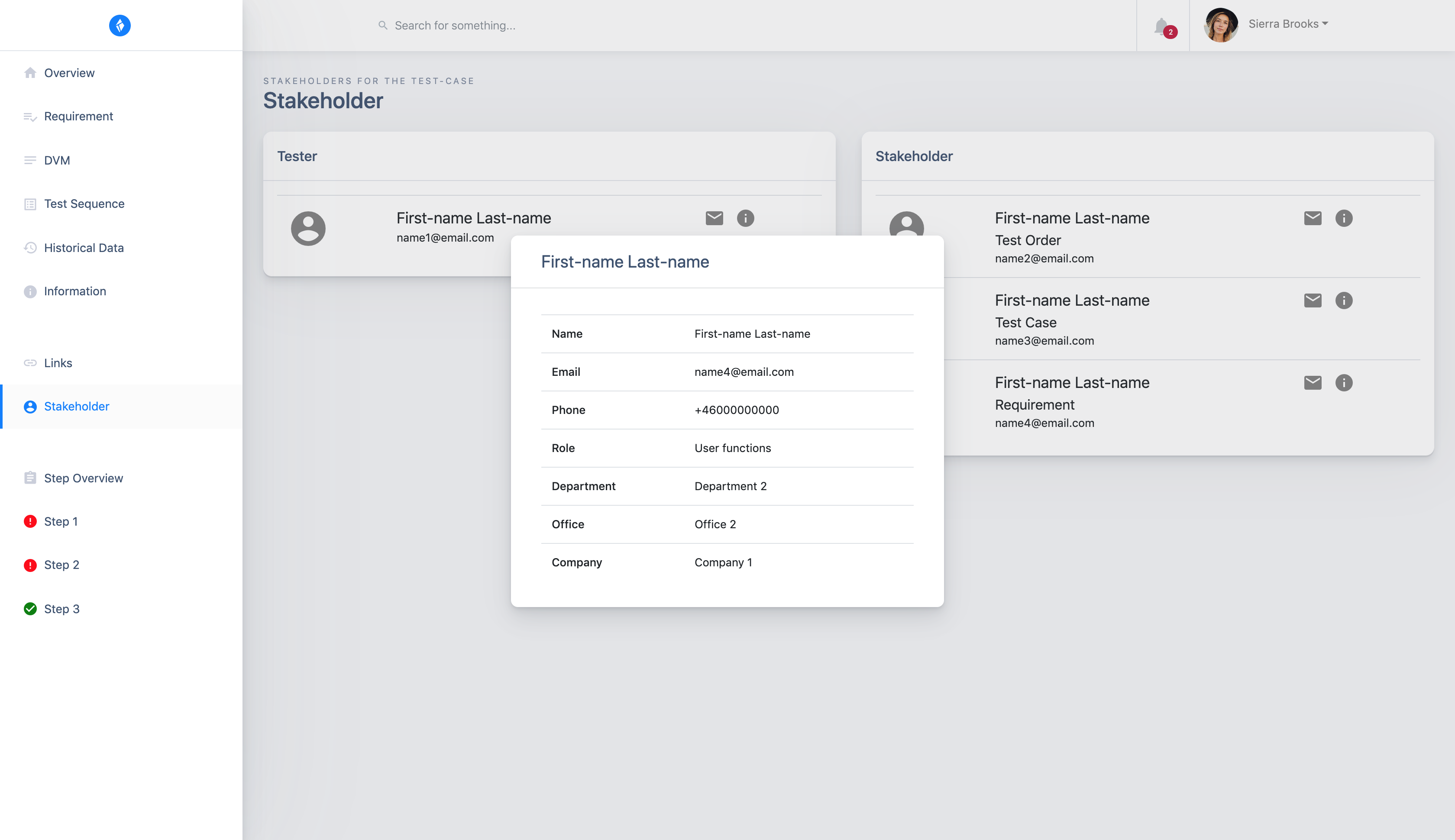}
    \caption{Final high-fidelity prototype of stakeholder sub-view.}
    \label{appendix:high2Prototype_stakeholderSubview}
\end{sidewaysfigure}

\begin{sidewaysfigure}
    \centering
    \includegraphics[width=.9\textheight]{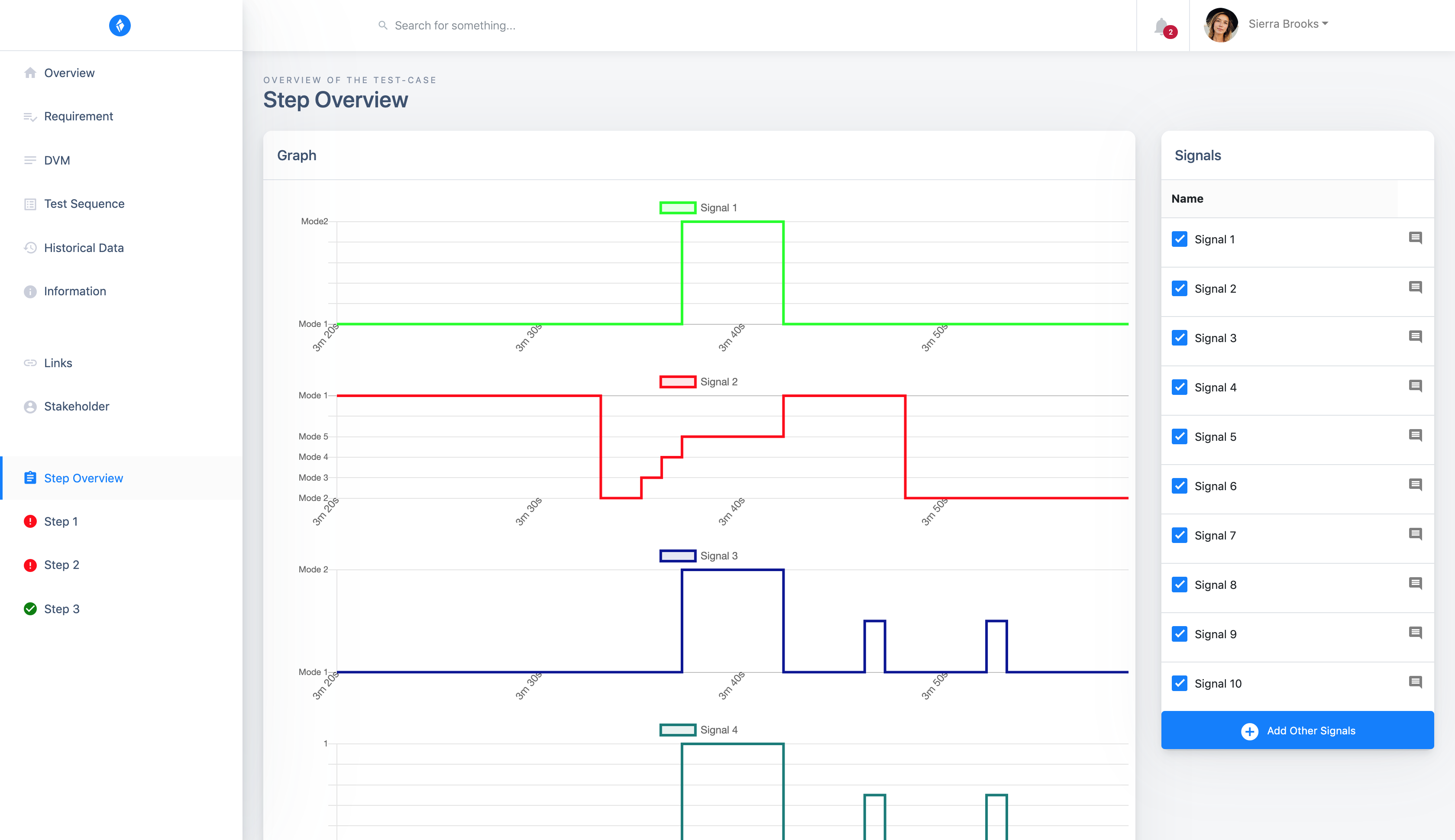}
    \caption{Final high-fidelity prototype of step overview view.}
    \label{appendix:high2Prototype_stepOverview}
\end{sidewaysfigure}

\begin{sidewaysfigure}
    \centering
    \includegraphics[width=.9\textheight]{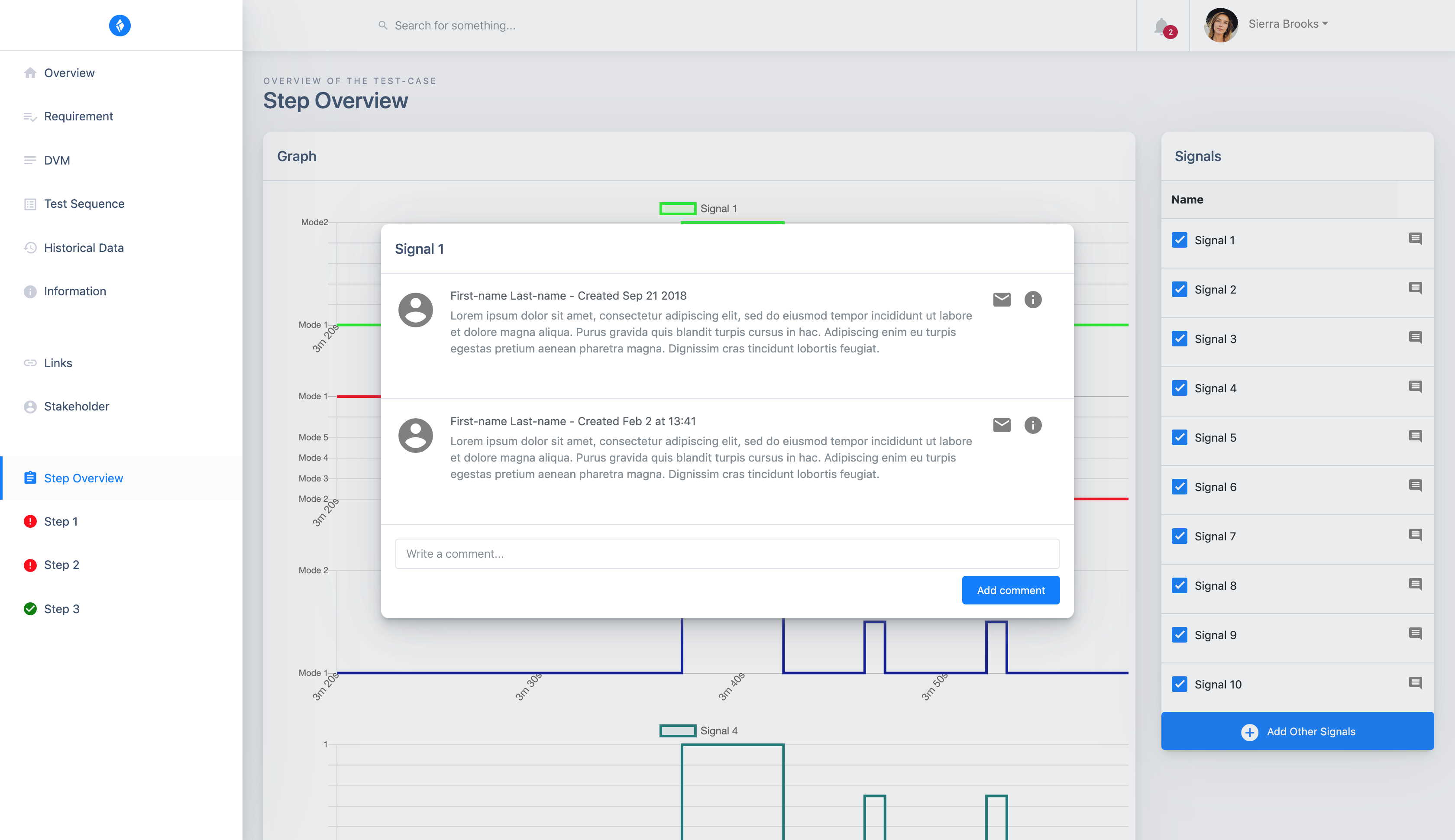}
    \caption{Final high-fidelity prototype of step signal comment view.}
    \label{appendix:high2Prototype_stepSignalCommentView}
\end{sidewaysfigure}

\begin{sidewaysfigure}
    \centering
    \includegraphics[width=.9\textheight]{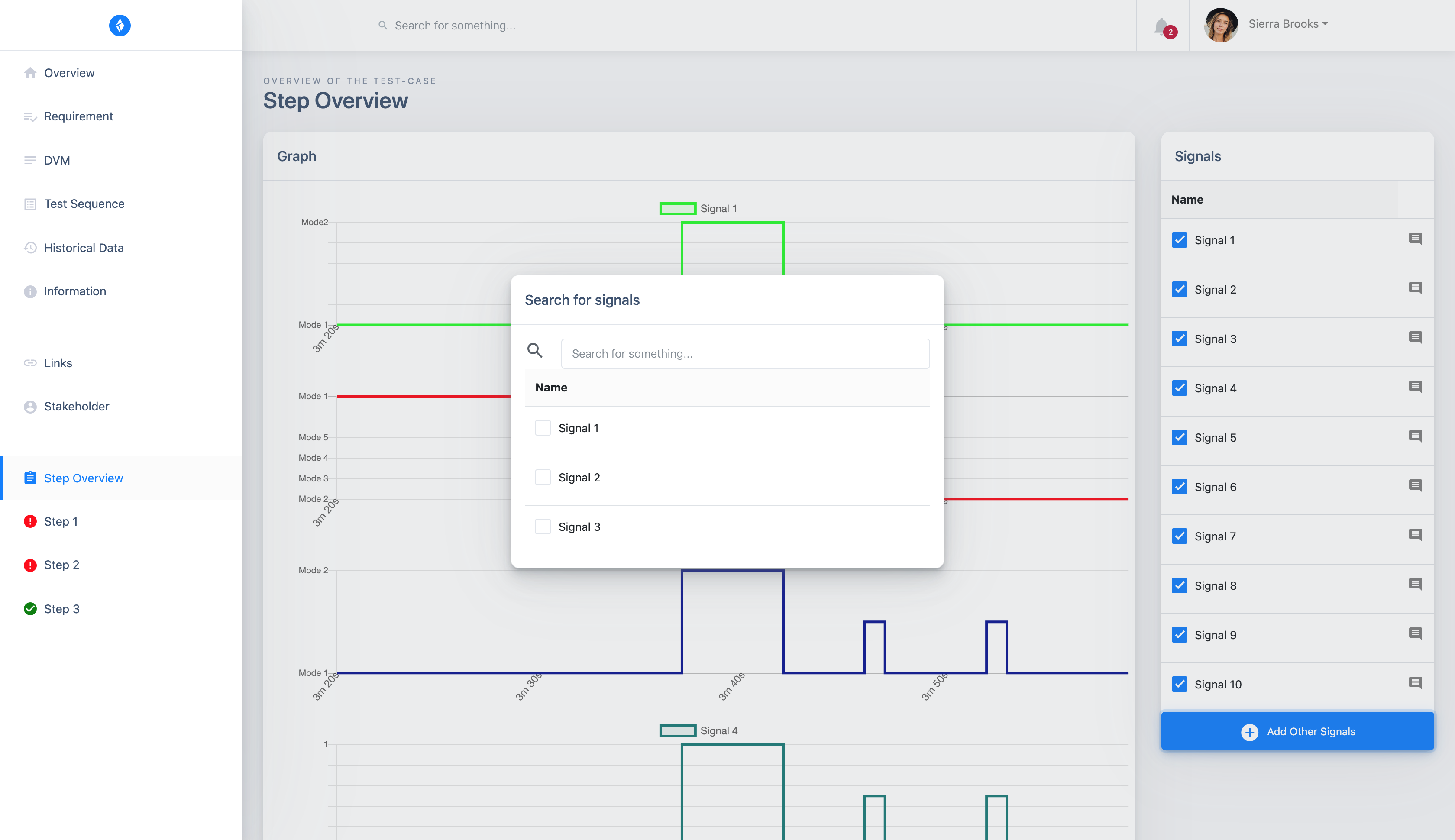}
    \caption{Final high-fidelity prototype of step signal search and add view.}
    \label{appendix:high2Prototype_stepSignalSearchView}
\end{sidewaysfigure}

\begin{sidewaysfigure}
    \centering
    \includegraphics[width=.9\textheight]{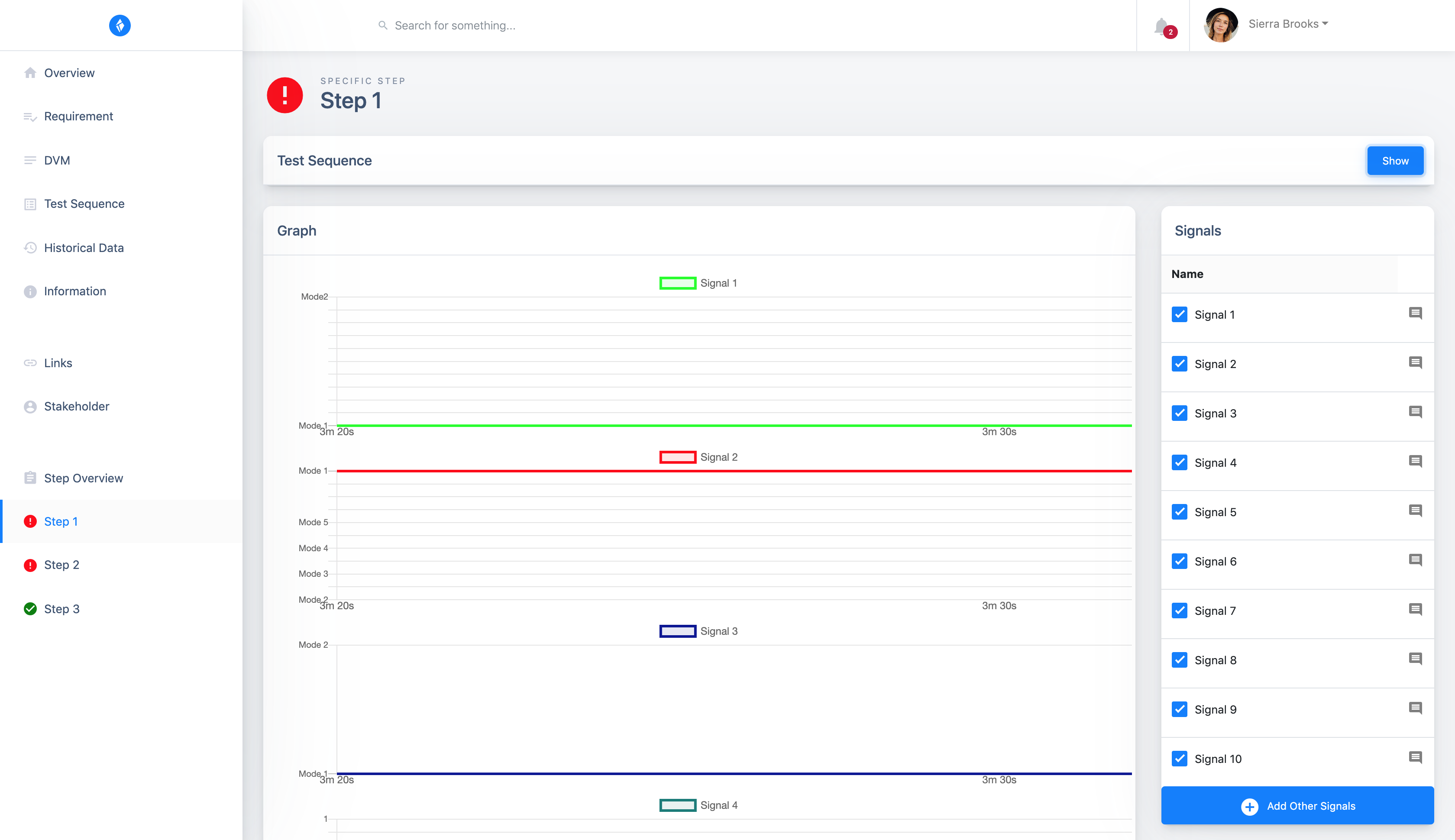}
    \caption{Final high-fidelity prototype of specific step view.}
    \label{appendix:high2Prototype_step}
\end{sidewaysfigure}

\begin{sidewaysfigure}
    \centering
    \includegraphics[width=.9\textheight]{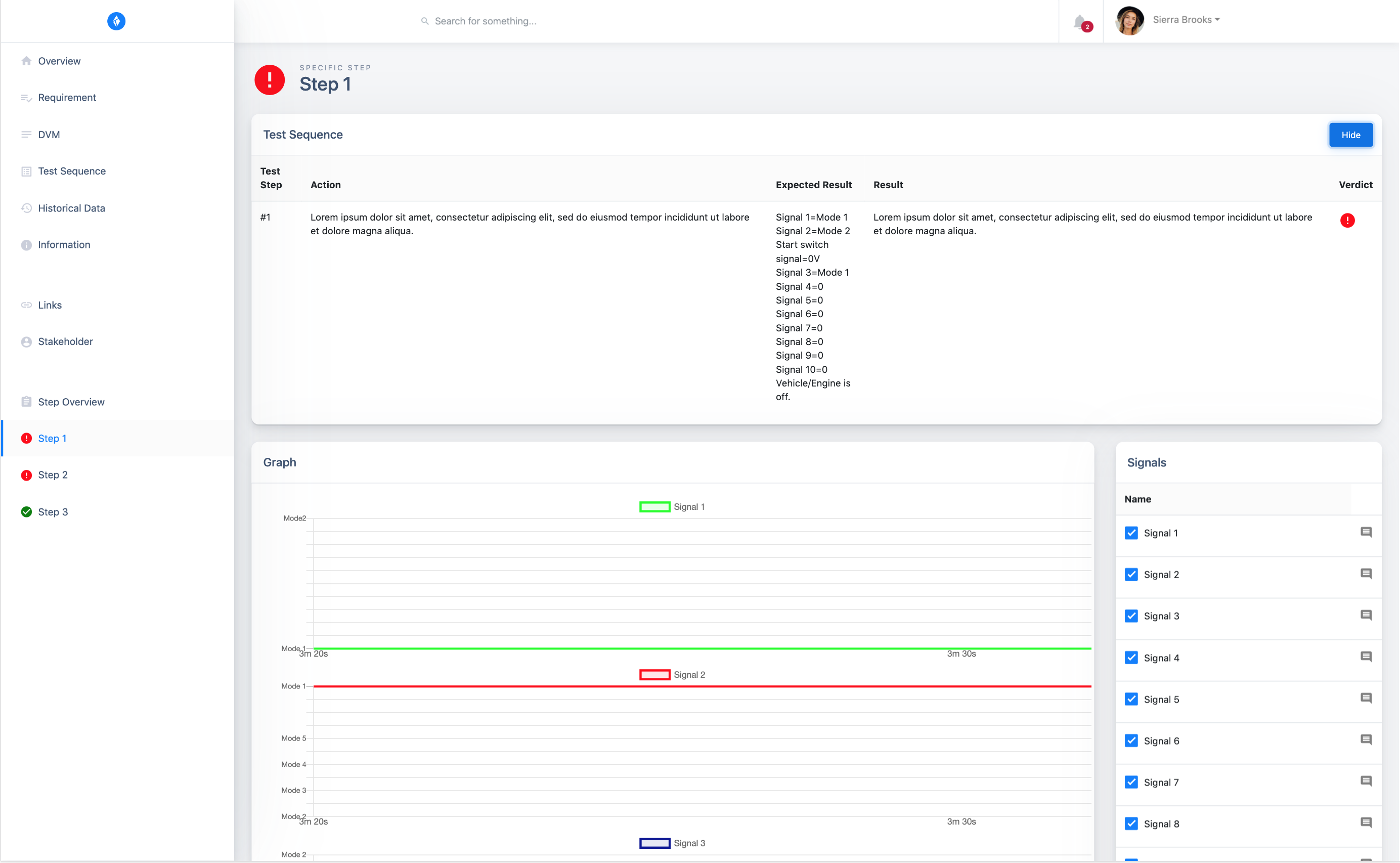}
    \caption{Final high-fidelity prototype of specific step view expanded.}
    \label{appendix:high2Prototype_stepExpanded}
\end{sidewaysfigure}

%% file: include/backmatter/Appendix_8.tex
\chapter{NASA Task Load Index Results for Current System / Old System}
\label{appendix:NasaTaskLoadIndexResultCurrentSystem}
NASA Task Load Index (TLX) result for each task collected from the usability test performed on the current system. Paper and pen approach was used with a TLX sheet\footnote{https://humansystems.arc.nasa.gov/groups/TLX/downloads/TLXScale.pdf} providing a 1-21 scale, which were multiple with 5 to give a 0-100\% scale. Weight for each demand was given by comparing each demand with each other where the subject determined which one were the more important one.

\hfill

\begin{table}[h!]
\centering
    \begin{tabular}{{ |l|c|c|c|c|c| }} 
        \hline
            \multicolumn{6}{|l|}{\textbf{Tester 1}} \\
        \hline
            \textbf{Demands} & \textbf{Task 1} & \textbf{Task 2} & \textbf{Task 3} & \textbf{Task 4} & \textbf{Weight} \\ 
        \hline
        \hline
            Mental Demand   & 60 & 75 & 85 & 10 & 5 \\
        \hline
            Physical Demand & 5 & 5 & 5 & 5 & 0 \\
        \hline
            Temporal Demand & 25 & 85 & 75 & 55 & 2 \\
        \hline
            Performance     & 10 & 75 & 25 & 55 & 3 \\
        \hline
            Effort          & 30 & 70 & 75 & 40 & 3 \\
        \hline
            Frustration     & 20 & 70 & 60 & 40 & 2 \\
        \hline
        \hline
            Product Sum     & 510 & 1120 & 995 & 525 & - \\
        \hline
            Score           & 34 & 74 & 66 & 35 & - \\
        \hline
    \end{tabular}
\end{table}

\begin{table}[h!]
\centering
    \begin{tabular}{{ |l|c|c|c|c|c| }} 
        \hline
            \multicolumn{6}{|l|}{\textbf{Tester 2}} \\
        \hline
            \textbf{Demands} & \textbf{Task 1} & \textbf{Task 2} & \textbf{Task 3} & \textbf{Task 4} & \textbf{Weight} \\ 
        \hline
        \hline
            Mental Demand   & 40 & 70 & 60 & 70 & 4 \\
        \hline
            Physical Demand & 5 & 5 & 5 & 5 & 0 \\
        \hline
            Temporal Demand & 45 & 60 & 65 & 65 & 2 \\
        \hline
            Performance     & 15 & 65 & 30 & 20 & 1 \\
        \hline
            Effort          & 70 & 80 & 65 & 40 & 5 \\
        \hline
            Frustration     & 60 & 85 & 35 & 60 & 3 \\
        \hline
        \hline
            Product Sum     & 795 & 1120 & 830 & 810 & - \\
        \hline
            Score           & 53 & 74 & 55 & 54 & - \\
        \hline
    \end{tabular}
\end{table}

\begin{table}[h!]
\centering
    \begin{tabular}{{ |l|c|c|c|c|c| }} 
        \hline
            \multicolumn{6}{|l|}{\textbf{Tester 3}} \\
        \hline
            \textbf{Demands} & \textbf{Task 1} & \textbf{Task 2} & \textbf{Task 3} & \textbf{Task 4} & \textbf{Weight} \\ 
        \hline
        \hline
            Mental Demand   & 65 & 70 & 40 & 75 & 4 \\
        \hline
            Physical Demand & 5 & 5 & 5 & 5 & 0 \\
        \hline
            Temporal Demand & 50 & 50 & 50 & 40 & 3 \\
        \hline
            Performance     & 20 & 25 & 20 & 40 & 2 \\
        \hline
            Effort          & 35 & 50 & 25 & 50 & 3 \\
        \hline
            Frustration     & 25 & 65 & 30 & 30 & 3 \\
        \hline
        \hline
            Product Sum     & 630 & 825 & 515 & 740 & - \\
        \hline
            Score           & 42 & 55 & 34 & 49 & - \\
        \hline
    \end{tabular}
\end{table}

\begin{table}[h!]
\centering
    \begin{tabular}{{ |l|c|c|c|c|c| }} 
        \hline
            \multicolumn{6}{|l|}{\textbf{Tester 4}} \\
        \hline
            \textbf{Demands} & \textbf{Task 1} & \textbf{Task 2} & \textbf{Task 3} & \textbf{Task 4} & \textbf{Weight} \\ 
        \hline
        \hline
            Mental Demand   & 65 & 20 & 80 & 70 & 4 \\
        \hline
            Physical Demand & 5 & 5 & 5 & 5 & 0 \\
        \hline
            Temporal Demand & 35 & 20 & 75 & 75 & 3 \\
        \hline
            Performance     & 30 & 25 & 15 & 10 & 1 \\
        \hline
            Effort          & 75 & 50 & 75 & 75 & 4 \\
        \hline
            Frustration     & 50 & 30 & 60 & 70 & 3 \\
        \hline
        \hline
            Product Sum     & 845 & 455 & 1040 & 1025 & - \\
        \hline
            Score           & 56 & 30 & 69 & 68 & - \\
        \hline
    \end{tabular}
\end{table}

\begin{table}[h!]
\centering
    \begin{tabular}{{ |l|c|c|c|c|c| }} 
        \hline
            \multicolumn{6}{|l|}{\textbf{Tester 5}} \\
        \hline
            \textbf{Demands} & \textbf{Task 1} & \textbf{Task 2} & \textbf{Task 3} & \textbf{Task 4} & \textbf{Weight} \\ 
        \hline
        \hline
            Mental Demand   & 80 & 60 & 20 & 45 & 5 \\
        \hline
            Physical Demand & 5 & 5 & 5 & 5 & 0 \\
        \hline
            Temporal Demand & 40 & 20 & 25 & 25 & 1 \\
        \hline
            Performance     & 20 & 25 & 15 & 30 & 2 \\
        \hline
            Effort          & 50 & 30 & 20 & 25 & 4 \\
        \hline
            Frustration     & 20 & 30 & 15 & 55 & 3 \\
        \hline
        \hline
            Product Sum     & 740 & 580 & 280 & 575 & - \\
        \hline
            Score           & 49 & 38 & 18 & 38 & - \\
        \hline
    \end{tabular}
\end{table}

\begin{table}[h!]
\centering
    \begin{tabular}{{ |l|c|c|c|c|c| }} 
        \hline
            \multicolumn{6}{|l|}{\textbf{Tester 6}} \\
        \hline
            \textbf{Demands} & \textbf{Task 1} & \textbf{Task 2} & \textbf{Task 3} & \textbf{Task 4} & \textbf{Weight} \\ 
        \hline
        \hline
            Mental Demand   & 55 & 60 & 55 & 45 & 4 \\
        \hline
            Physical Demand & 5 & 5 & 5 & 5 & 0 \\
        \hline
            Temporal Demand & 40 & 35 & 70 & 55 & 2 \\
        \hline
            Performance     & 20 & 20 & 40 & 50 & 3 \\
        \hline
            Effort          & 65 & 45 & 50 & 60 & 3 \\
        \hline
            Frustration     & 40 & 65 & 45 & 45 & 3 \\
        \hline
        \hline
            Product Sum     & 675 & 700 & 765 & 765 & - \\
        \hline
            Score           & 45 & 46 & 51 & 51 & - \\
        \hline
    \end{tabular}
\end{table}

%% file: include/backmatter/Appendix_9.tex
\chapter{NASA Task Load Index Results for New System}
\label{appendix:NasaTaskLoadIndexResultNewSystem}
NASA Task Load Index (TLX) result for each task collected from the usability test performed on the new system. Paper and pen approach was used with a TLX sheet\footnote{https://humansystems.arc.nasa.gov/groups/TLX/downloads/TLXScale.pdf} providing a 1-21 scale, which were multiple with 5 to give a 0-100\% scale. Weight for each demand was given by comparing each demand with each other where the subject determined which one were the more important one.

\hfill

\begin{table}[h!]
\centering
    \begin{tabular}{{ |l|c|c|c|c|c| }} 
        \hline
            \multicolumn{6}{|l|}{\textbf{Tester 1}} \\
        \hline
            \textbf{Demands} & \textbf{Task 1} & \textbf{Task 2} & \textbf{Task 3} & \textbf{Task 4} & \textbf{Weight} \\ 
        \hline
        \hline
            Mental Demand   & 30 & 5 & 65 & 10 & 3 \\
        \hline
            Physical Demand & 5 & 5 & 5 & 5 & 0 \\
        \hline
            Temporal Demand & 75 & 15 & 65 & 35 & 4 \\
        \hline
            Performance     & 25 & 20 & 15 & 5 & 2 \\
        \hline
            Effort          & 40 & 10 & 70 & 10 & 3 \\
        \hline
            Frustration     & 60 & 20 & 30 & 25 & 3 \\
        \hline
        \hline
            Product Sum     & 740 & 205 & 785 & 285 & - \\
        \hline
            Score           & 49 & 13 & 52 & 19 & - \\
        \hline
    \end{tabular}
\end{table}

\begin{table}[h!]
\centering
    \begin{tabular}{{ |l|c|c|c|c|c| }} 
        \hline
            \multicolumn{6}{|l|}{\textbf{Tester 2}} \\
        \hline
            \textbf{Demands} & \textbf{Task 1} & \textbf{Task 2} & \textbf{Task 3} & \textbf{Task 4} & \textbf{Weight} \\ 
        \hline
        \hline
            Mental Demand   & 20 & 10 & 30 & 45 & 4 \\
        \hline
            Physical Demand & 5 & 5 & 5 & 5 & 0 \\
        \hline
            Temporal Demand & 25 & 15 & 30 & 45 & 2 \\
        \hline
            Performance     & 15 & 5 & 15 & 15 & 1 \\
        \hline
            Effort          & 20 & 15 & 50 & 15 & 4 \\
        \hline
            Frustration     & 15 & 15 & 40 & 40 & 4 \\
        \hline
        \hline
            Product Sum     & 285 & 195 & 555 & 505 & - \\
        \hline
            Score           & 19 & 13 & 37 & 33 & - \\
        \hline
    \end{tabular}
\end{table}

\begin{table}[h!]
\centering
    \begin{tabular}{{ |l|c|c|c|c|c| }} 
        \hline
            \multicolumn{6}{|l|}{\textbf{Tester 3}} \\
        \hline
            \textbf{Demands} & \textbf{Task 1} & \textbf{Task 2} & \textbf{Task 3} & \textbf{Task 4} & \textbf{Weight} \\ 
        \hline
        \hline
            Mental Demand   & 10 & 15 & 30 & 20 & 5 \\
        \hline
            Physical Demand & 5 & 5 & 5 & 5 & 0 \\
        \hline
            Temporal Demand & 10 & 15 & 35 & 20 & 2 \\
        \hline
            Performance     & 10 & 15 & 15 & 20 & 2 \\
        \hline
            Effort          & 20 & 15 & 25 & 20 & 3 \\
        \hline
            Frustration     & 15 & 10 & 15 & 20 & 3 \\
        \hline
        \hline
            Product Sum     & 195 & 210 & 370 & 300 & - \\
        \hline
            Score           & 13 & 14 & 24 & 20 & - \\
        \hline
    \end{tabular}
\end{table}

\begin{table}[h!]
\centering
    \begin{tabular}{{ |l|c|c|c|c|c| }} 
        \hline
            \multicolumn{6}{|l|}{\textbf{Tester 4}} \\
        \hline
            \textbf{Demands} & \textbf{Task 1} & \textbf{Task 2} & \textbf{Task 3} & \textbf{Task 4} & \textbf{Weight} \\ 
        \hline
        \hline
            Mental Demand   & 30 & 20 & 70 & 35 & 4 \\
        \hline
            Physical Demand & 5 & 5 & 5 & 5 & 0 \\
        \hline
            Temporal Demand & 20 & 20 & 35 & 20 & 3 \\
        \hline
            Performance     & 15 & 25 & 35 & 10 & 2 \\
        \hline
            Effort          & 40 & 35 & 75 & 25 & 3 \\
        \hline
            Frustration     & 15 & 15 & 50 & 20 & 3 \\
        \hline
        \hline
            Product Sum     & 375 & 340 & 830 & 355 & - \\
        \hline
            Score           & 25 & 22 & 55 & 23 & - \\
        \hline
    \end{tabular}
\end{table}

\begin{table}[h!]
\centering
    \begin{tabular}{{ |l|c|c|c|c|c| }} 
        \hline
            \multicolumn{6}{|l|}{\textbf{Tester 5}} \\
        \hline
            \textbf{Demands} & \textbf{Task 1} & \textbf{Task 2} & \textbf{Task 3} & \textbf{Task 4} & \textbf{Weight} \\ 
        \hline
        \hline
            Mental Demand   & 10 & 25 & 20 & 20 & 3 \\
        \hline
            Physical Demand & 5 & 5 & 5 & 5 & 0 \\
        \hline
            Temporal Demand & 25 & 20 & 25 & 20 & 4 \\
        \hline
            Performance     & 10 & 25 & 15 & 15 & 2 \\
        \hline
            Effort          & 15 & 15 & 15 & 10 & 2 \\
        \hline
            Frustration     & 10 & 20 & 15 & 35 & 4 \\
        \hline
        \hline
            Product Sum     & 220 & 315 & 280 & 330 & - \\
        \hline
            Score           & 14 & 21 & 18 & 22 & - \\
        \hline
    \end{tabular}
\end{table}

\begin{table}[h!]
\centering
    \begin{tabular}{{ |l|c|c|c|c|c| }} 
        \hline
            \multicolumn{6}{|l|}{\textbf{Tester 6}} \\
        \hline
            \textbf{Demands} & \textbf{Task 1} & \textbf{Task 2} & \textbf{Task 3} & \textbf{Task 4} & \textbf{Weight} \\ 
        \hline
        \hline
            Mental Demand   & 15 & 20 & 40 & 30 & 4 \\
        \hline
            Physical Demand & 5 & 5 & 5 & 5 & 0 \\
        \hline
            Temporal Demand & 15 & 15 & 30 & 15 & 4 \\
        \hline
            Performance     & 15 & 20 & 40 & 10 & 3 \\
        \hline
            Effort          & 25 & 20 & 50 & 15 & 2 \\
        \hline
            Frustration     & 10 & 15 & 45 & 20 & 2 \\
        \hline
        \hline
            Product Sum     & 235 & 270 & 590 & 280 & - \\
        \hline
            Score           & 15 & 18 & 39 & 18 & - \\
        \hline
    \end{tabular}
\end{table}

%% file: include/backmatter/Appendix_11.tex
\chapter{Coded Usability Test Interviews}
\label{appendix:codedUsability}

\begin{sidewaysfigure}
    \centering
    \includegraphics[width=.9\textheight]{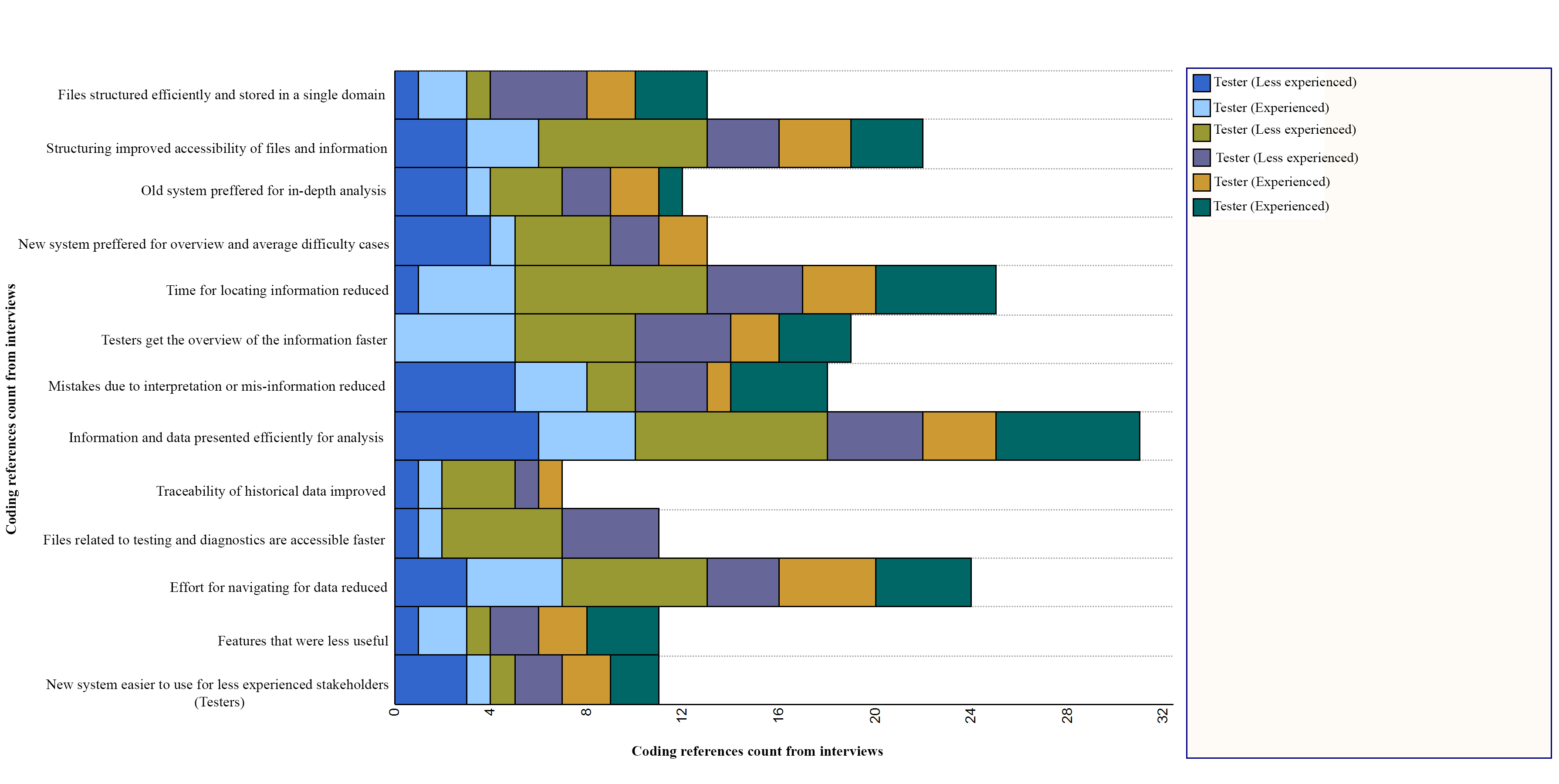}
    \caption{Findings from the usability test.}
    \label{appendix:codedInterviews:usabilityTest}
\end{sidewaysfigure}